\documentclass[12pt]{article}
\usepackage[utf8]{inputenc}
\usepackage{graphicx}
\usepackage{amsmath, amssymb,verbatim,bm, theorem}
\usepackage{hyperref}
\usepackage{setspace}
\usepackage{mathtools}
\usepackage[round]{natbib} 
\usepackage{kpfonts}
\usepackage{color}
\usepackage{multirow}
\usepackage{longtable,booktabs}
\usepackage{authblk}

\usepackage{latexsym}
\usepackage{marvosym}
\usepackage{pdflscape}
\usepackage{float,subcaption}
\usepackage{caption}
\usepackage{hvfloat}
\usepackage[title]{appendix}
\usepackage{longtable,booktabs}
\usepackage{chngcntr}

\DeclareMathOperator*{\argmin}{arg\,min}

\numberwithin{equation}{section}

\newtheorem{theorem}{Theorem}[section]
\newtheorem{lemma}{Lemma}[section] 
\newtheorem{assumption}{Assumption}[section] 
\newtheorem{defin}{Definition}[section] 
\newtheorem{corollary}{Corollary}[section] 

\newtheorem{remark}{Remark}[section]

\newcommand{\Dcon}{\stackrel{\mathcal{D}}{\rightarrow}}
\newcommand{\Pcon}{\stackrel{\mathcal{P}}{\rightarrow}}

\newcommand{\bK}{\boldsymbol{K}}
\newcommand{\btheta}{\boldsymbol{\theta}}

\newcommand{\bG}{\boldsymbol{G}}
\newcommand{\bV}{\boldsymbol{V}}
\newcommand{\bQ}{\boldsymbol{Q}}

\newcommand{\Ex}{\mathbb{E}}

\newcommand{\var}{ {\mathbb{V}\rm ar }}
\newcommand{\cov}{ {\mathbb{C}\rm ov}}
\newcommand{\corr}{ {\mathbb{C}\rm orr}}
\newcommand{\bs}{{\bm{s}}}
\newcommand{\bu}{{\bm{u}}}
\newcommand{\bx}{\bm{x}}
\newcommand{\by}{\bm{y}}
\newcommand{\bz}{\bm{z}}
\newcommand{\bw}{\bm{w}}

\DeclareMathOperator{\Tr}{Tr}

\newcommand*{\crosssymbol}{%
    \text{%
      \raisebox{1ex}{%
        \makebox[0pt][l]{%
          \rule[-.2pt]{.75ex}{.4pt}%
        }%
        \makebox[.75ex]{%
          \rule[-1ex]{.4pt}{1.5ex}%
        }%
      }%
    }%
}

\providecommand{\keywords}[1]
{
  \small	
  \textbf{Keywords: } #1
}

\title{On minimum contrast method for multivariate spatial point processes}

\author[1,2]{Lin Zhu \thanks{The first two authors have an equal contribution.}}  
\author[3]{Junho Yang \thanks{Correspondence to: Junho Yang, Institute of Statistical Science, No.128, Academia Road, Section 2, Nangang, Taipei 11529, Taiwan. \hspace{0.2em} Email: \url{junhoyang@stat.sinica.edu.tw}}}
\author[4]{Mikyoung Jun}
\author[5]{Scott Cook}
\affil[1]{Central University of Finance and Economics}
\affil[2]{China Fortune International Trust Co.,Ltd.}
\affil[3]{Institute of Statistical Science, Academia Sinica}
\affil[4]{Department of Mathematics, University of Houston}
\affil[5]{Department of Political Science, Texas A\&M University}

\date{\today}

\begin{document}

\maketitle

\begin{abstract}
Compared to widely used likelihood-based approaches, the minimum contrast (MC) method offers a computationally efficient method for estimation and inference of spatial point processes. These relative gains in computing time become more pronounced when analyzing complicated multivariate point process models. Despite this, there has been little exploration of the MC method for multivariate spatial point processes. Therefore, this article introduces a new MC method for parametric multivariate spatial point processes. A contrast function is computed based on the trace of the power of the difference between the conjectured $K$-function matrix and its nonparametric unbiased edge-corrected estimator. Under standard assumptions, we derive the asymptotic normality of our MC estimator. The performance of the proposed method is demonstrated through simulation studies of bivariate log-Gaussian Cox processes and five-variate product-shot-noise Cox processes.
\end{abstract}


\section{Introduction} \label{sec:intro}


In recent decades, with advancements in data collection methods, presence-only data have become increasingly common across various disciplines, and often include more than one ``type'' of outcome or event. Some applications can be found in epidemiology \citep{lawson2012bayesian}, neuroscience \citep{p:bad-14}, ecology \citep{p:was-16}, meteorology \citep{jun_et_al19}, and political science \citep{p:jun-coo-22}, to name a few. In statistical analysis, these multi-type locational data can be viewed as a realization of a multivariate spatial point process, and developing effective statistical tools to analyze multivariate spatial point processes is important for better understanding these multi-type event data in real-life applications.

Maximum likelihood estimation (MLE) is widely used to estimate and infer the parameters of spatial point processes. This is because the resulting MLE estimator has desirable large sample properties \citep{p:oga-78, p:hel-99}. However, one of the biggest challenges with MLE is that the likelihood function is, in general, analytically intractable. As such, MLE requires an appropriate approximation for the likelihood function. For example, in log-Gaussian Cox processes \citep[LGCP;][]{moller1998log}, the intractable likelihood function involves the expectation with respect to the logarithm of the latent intensity field. To approximate this expected value, and in turn, approximate the likelihood function, one can use Monte Carlo simulation as in \cite{jun_et_al19}.
Alternatively, Bayesian inferential methods \citep[e.g.,][]{rue2009approximate} can be implemented to obtain the approximation of the likelihood function of an LGCP model.

However, these likelihood approximation methods present challenges in application. 
First, for each alternative model considered, researchers need to re-specify the point process model and corresponding algorithms, which itself takes considerable time and effort. Second, as the model becomes increasingly complex or the number of computational grids increases, the aforementioned likelihood approximation methods are increasingly computationally intensive. Specifically, drawing each sample of model parameters and latent random fields on grids (in Monte Carlo simulations) or running long iterations to reach the convergence of the Markov chain (in Bayesian methods) is computationally demanding. To handle these computational issues, quasi-likelihood or composite likelihood approximation methods for point processes are proposed, such as \cite{p:jal-15} for Cox processes and \cite{p:raj-18} for Gibbs point processes.

As an alternative to these likelihood-based methods, Minimum Contrast (MC) estimation is a computationally efficient inferential method for spatial point processes. MC estimation selects the model parameters that minimize the discrepancy between the conjectured ``descriptor'' of point processes and its nonparametric estimator. For univariate point processes, a typical choice of a descriptor function is Ripley's $K$-function \citep{p:rip-76} or pair correlation function (PCF). This is because an analytic form of the $K$-function and PCF is available for many point process models, such as the shot noise Cox process \citep{p:mol-03}.

To elaborate, let $Q(\cdot; \btheta)$, $\btheta \in \Theta$, be a family of parametric descriptors of a stationary point process and let $\widehat{Q}(\cdot)$ be its estimator. Then, MC estimation aims to minimize the integrated distance between $Q(\cdot; \btheta)$ and $\widehat{Q}(\cdot)$ over the prespecified domain. For instance, if a univariate point process has the parametric $K$-function as its descriptor function, that is, $Q(h; \btheta) = K(h;\btheta)$ for $h \in [0,\infty)$, then the MC estimator is defined as
\begin{equation}\label{eq:U}
\widehat{\btheta} = \argmin_{\btheta \in \Theta} \mathcal{U}(\btheta) \quad \text{where} \quad
\mathcal{U}(\btheta) = \int_{0}^{R} w(h) \left\{ |K(h;\btheta)|^c - |\widehat{K}(h)|^c \right\}^2 dh.
\end{equation}
Here, $R$ is a positive range, $w(\cdot)$ is a non-negative weight function, $K(\cdot;\btheta)$ and $\widehat{K}(\cdot)$ are parametric $K$-functions and their nonparametric estimator (see Section \ref{sec:Kfcns} for precise definitions), and $c \in [0,\infty)$ is a non-negative power. Provided that $K(h;\btheta)$ has a known form and $\widehat{K}(h)$ can be easily evaluated for all $h \in [0,\infty)$, the MC estimator can be easily calculated using standard numerical optimization methods. As such, the MC method has been extensively used to analyze univariate point processes in applications \citep[e.g.,][]{moller1998log, p:mol-07, moller2010structured, p:dav-haz-13, cui2018vehicle}. There have also been developments in our theoretical understanding of these univariate point processes. \cite{p:hei-92} studied the asymptotic properties of the MC estimator for univariate homogeneous Poisson processes with $c=0.5$, and \cite{p:gua-07} extended the distribution theory of the MC estimator to a fairly large class of univariate point process models. When $c=1$, the MC procedure can be viewed as a special case of the second-order quasi-likelihood estimator in the sense of \cite{p:deng-14,p:deng-17}.

Despite the increasing use of the univariate MC method in applications, there is very little work on MC methods for multivariate point processes in both theory and application. This is because, unlike the univariate model, for multiple potentially related point patterns, one also needs to examine the cross-group interactions between the different types of point processes. Therefore, the descriptor function $Q(\cdot;\btheta)$ takes the form of a matrix-valued function. This requires the development of a new way to measure the discrepancy between these types of functions. Note several approaches have been suggested in the literature, for example, \cite{p:was-16} proposed the least squares estimation of the parametric multivariate LGCP model. Another general method to fit multivariate point processes is weighted composite likelihood as proposed by \cite{p:jal-15}. Even for these suggested approaches, however, there has been little theoretical justification, such as large sample properties, given for these estimators.

In this article, we extend existing research on MC estimation for multivariate point processes in at least three ways. First, we offer a new MC method to analyze a fairly large class of multivariate stationary point processes, which does not necessarily assume the specific structure of the point processes. Second, under the increasing-domain asymptotic framework, we derive the asymptotic normality of the MC estimator. Lastly, based on the asymptotic results, we propose a method to select the optimal control parameters of the MC method.

The rest of this article is organized as follows. In Section \ref{sec:joint}, we provide background on the spatial point process and define the higher-order joint intensity function of a multivariate spatial point process. In Section \ref{sec:Kfcns}, we introduce the marginal and cross $K$-functions and their nonparametric estimators. Using these $K$-function estimators, in Section \ref{sec:discrepancy}, we define a new discrepancy measure between the two matrix-valued functions. In Section \ref{sec:sampling}, we investigate the large sample properties of the MC estimator. In Section \ref{sec:practice}, we consider the practical application of the MC estimator, including the estimation of the asymptotic covariance matrix (Section \ref{sec:est_asymp_var}), selection of the optimal control parameters (Section \ref{sec:select_tune_par}), and constructing a confidence region (Section \ref{sec:CRs}).
In Section \ref{sec:simu}, we consider the bivariate LGCP and five-variate product-short-noise Cox Process \citep[PSNCP;][]{p:jal-15} models and investigate their finite sample properties.

Finally, auxiliary results, proofs, and additional simulations can be found in the Supplementary Materials (which we refer to as the Appendix hereafter). In particular, in Appendix \ref{sec:applic}, we apply our method to terrorism attacks in Nigeria.


\section{Joint intensity functions of multivariate point processes} \label{sec:joint}

In this section, we introduce the joint intensity function of $m$-tuple of point processes. To do so, we briefly review some background concepts that are used throughout the article. For greater detail on the mathematical presentations of point processes, we refer readers to \cite{b:dal-03}.

Let $X$ be a simple spatial point process defined on $\mathbb{R}^d$, where we observe the sample of $X$ in $D \subset \mathbb{R}^d$. For a bounded Borel set $E \subset D$, $N_X(E)$ denotes the random variable that counts the number of events of $X$ within $E$. For $n \in \mathbb{N}$, let $\lambda_n: D^{n} \mapsto \mathbb{R}$ be the $n$th-order intensity function (also known as the product density function) of $X$. Thus,
\begin{equation} \label{eq:intensity}
    \lambda_n(\bx_1,\cdots,\bx_n)=\lim_{|d\bx_1|,\cdots,|d\bx_n|\rightarrow 0} \frac{\mathbb{E}\left[ N_X(d\bx_1)\cdots N_X(d\bx_n)\right]}{|d\bx_1|\cdots|d\bx_n|},\quad (\bx_1, \cdots, \bx_n) \in D^{n,\neq},
\end{equation} and zero otherwise. 
Here, $D^{n,\neq}$ denotes the set of all $n$ pairwise distinct points in $D$, and for $i \in \{1,\dots,n\}$, $d\bx_i$ denotes the infinitesimal region in $D$ that contains $\bx_i \in D$, and $|d\bx_i|$ denotes the volume of $d\bx_i$. Under stationarity, we write $\lambda_1(\bx_1) = \lambda_{1}$ for all $\bx_1 \in D$ and
\begin{equation} \label{eq:lambda-stationary}
\lambda_n(\bx_1,\cdots,\bx_n) = \lambda_{n,\text{red}}(\bx_2-\bx_1,\cdots,\bx_n-\bx_1), \quad n \in \{2, 3, \cdots\},\quad 
\bx_1, \dots, \bx_n \in D.
\end{equation} We refer to $\lambda_{n,\text{red}}$ as the $n$-th order reduced intensity function of $X$. 

For a multivariate point process defined on the same probability space and sampling window, we can naturally extend the concept of $n$th-order intensity function and define the joint intensity function.
\begin{defin}[Joint intensity functions of multivariate point processes] \label{def:joint}
Let $\underline{X} = (X_{1}, \cdots, X_{m})$ be an $m$-variate spatial point process defined on the same probability space. Let $D \subset \mathbb{R}^d$ be the common sampling window of $\underline{X}$. Denote $\underline{n} = (n_{1}, \dots, n_{m})$ as a vector of non-negative integers and $\underline{\bx}_{i} = (\bx_{i,1}, \cdots, \bx_{i, n_{i}}) \in D^{n_i}$ for $i \in \{1, \dots, m\}$. 
Then, the joint intensity function of $\underline{X}$ of order $\underline{n}$, denoted by $\lambda_{\underline{n}}: D^N \mapsto \mathbb{R}$ (where $N = \sum_{i =1}^{m} n_{i}$), is defined as
\begin{equation} \label{eq:joint-intensity1}
\lambda_{\underline{n}} (\underline{\bx}_{1}, \dots, \underline{\bx}_{m}) = \lim_{|d\bx_{1,1}|,\cdots,|d\bx_{m,n_m}|\rightarrow 0} 
\frac{ \Ex \left[ \prod_{j=1}^{n_1} N_{X_{1}}(d \bx_{1,j}) \times \cdots \times \prod_{j=1}^{n_m} N_{X_{m}}(d \bx_{m,j}) \right]}{
\prod_{j=1}^{n_1} |d \bx_{1,j}| \times \cdots \times \prod_{j=1}^{n_m} |d \bx_{m,j}|}
\end{equation} for $(\bx_{i,1}, \cdots \bx_{i,n_i})\in D^{n_i,\neq}$, $i \in \{1, \dots,m\}$, and zero otherwise.
Following convention, we let $\prod_{1}^{0} = 1$. Sometimes, it will be necessary to consider the joint intensity of a subset of $\underline{X}$. In this case, for non-overlapping indices $\mathcal{I} = (i(1), \dots, i(k)) \subset \{1, \dots, m\}$, we denote
\begin{equation} \label{eq:joint-intensity2}
\lambda^{\mathcal{I}}_{\underline{n}^{}} (\underline{\bx}_{i(1)}, \dots, \underline{\bx}_{i(k)})
= \lim_{|d\bx_{i(1),1}|,\cdots,|d\bx_{i(k),n_{i(k)}}|\rightarrow 0} 
\frac{ \Ex \left[ \prod_{j=1}^{n_{i(1)}} N_{X_{i(1)}}(d \bx_{i(1),j})  
\times \cdots \times \prod_{j=1}^{n_{i(k)}} N_{X_{i(k)}}(d \bx_{i(k),j})
\right]}{
 \prod_{j=1}^{n_{i(1)}} |d \bx_{i(1),j}|
\times \cdots \times  \prod_{j=1}^{n_{i(k)}} |d \bx_{i(k),j}|
}
\end{equation}
 as the joint intensity function of $(X_{i(1)}, \dots, X_{i(k)})$ of order $\underline{n}^{\mathcal{I}} = (n_{i(1)}, \dots, n_{i(k)})$.
\end{defin}
If $\underline{X}$ is stationary, then analogous to (\ref{eq:lambda-stationary}), we have
\begin{equation} 
\lambda_{\underline{n}}^{} (\underline{\bx}_{1}, \dots, \underline{\bx}_{m}) 
= \lambda_{\underline{n}, red}^{} ( \bx_{1,2} - \bx_{1,1}, \cdots, \bx_{1,n_1}-\bx_{1,1}, \bx_{2,1} - \bx_{1,1}, \cdots,
\bx_{m, n_m} - \bx_{1,1}): D^{N-1} \mapsto \mathbb{R}_{},
\label{eq:joint-red}
\end{equation} where $N = \sum_{i=1}^{m} n_i$. 
We refer to $\lambda_{\underline{n}, \text{red}}$ as the reduced joint intensity function of $\underline{X}$ of order $\underline{n}$. For an index vector $\mathcal{I} = (i(1), \dots, i(k)) \subset \{1, \dots, m\}$, the reduced joint intensity function of the subset process $(X_{i(1)}, \dots, X_{i(k)})$, denoted by $\lambda_{\underline{n}, \text{red}}^{\mathcal{I}}$, can be defined similarly.

Finally, we introduce the joint cumulant intensity function, as we use this concept in the proof of the asymptotic normality of the target random variables (see Assumption \ref{assump:1}). By replacing the expectation with the cumulant in (\ref{eq:joint-intensity1}), we define the joint cumulant intensity function of $\underline{X}$ of order $\underline{n} = (n_1, \dots, n_m)$, denoted by $\gamma_{\underline{n}}(\underline{\bx}_{1}, \dots, \underline{\bx}_{m})$. Analogous to (\ref{eq:joint-red}), if $\underline{X}$ is stationary, then
\begin{equation} \label{eq:cumul-red}
\gamma_{\underline{n}}^{} (\underline{\bx}_{1}, \dots, \underline{\bx}_{m}) 
 = \gamma_{\underline{n}, red}^{} ( \bx_{1,2} - \bx_{1,1}, \dots, \bx_{1,n_1}-\bx_{1,1}, \bx_{2,1} - \bx_{1,1}, \dots.,
\bx_{m, n_m} - \bx_{1,1}).
\end{equation} 
We refer to $\gamma_{\underline{n}, \text{red}}$ as the reduced joint cumulant intensity function. For a subset process $\underline{X}^{\mathcal{I}} = (X_{i(1)}, \dots, X_{i(k)})$, we can define the joint cumulant intensity function of $\underline{X}^{\mathcal{I}}$, denoted by $\gamma^{\mathcal{I}}_{\underline{n}}$, and the reduced joint cumulant intensity function of $\underline{X}^{\mathcal{I}}$, denoted by $\gamma^{\mathcal{I}}_{\underline{n}, \text{red}}$, in the same manner.

\section{Minimum contrast for multivariate point processes}  \label{sec:MC}

In this section, we use the marginal and cross $K$-functions to formulate the minimum contrast method for multivariate point processes.
\subsection{The marginal and cross $K$-function} \label{sec:Kfcns}

The $K$-function is an important measure to quantify the second-order interaction between two points. Following the heuristic definition in \cite[page 615]{b:cre-93}, the $K$-function of a univariate stationary point process $X$ is defined as
\begin{equation}
    K(r) := \lambda_{1}^{-1} 
\mathbb{E} \Bigl \{ 
\text{Number of distinct points within distance $r$ of a given point}
 \Bigr \},~~ r \in [0,\infty), 
\label{eq:K(r)}
\end{equation}
where $\lambda_{1}$ is the homogeneous first-order intensity of $X$. A formal formulation of $K$-function based on the second-order reduced Palm measure can be found in \cite{p:rip-76}.

Given a realization of $X$ sampled within $D_n \subset \mathbb{R}^d$ for $n \in \mathbb{N}$, a naive estimator of $K(r)$ is
\begin{equation}\label{eq:empirical_K0}
\widehat{K}_{0,n}(r)=
\left( |D_n|{\hat{\lambda}_{1,n}}^2 \right)^{-1} \sum_{\bx, \by \in X} \textbf{1}_{\{0<\|\bx-\by\|\leq r\}}, \quad r \in [0,\infty),
\quad n \in \mathbb{N},
\end{equation}
where $\hat{\lambda}_{1,n} = N_X(D_n)/|D_n|$ is an unbiased estimator of the first-order intensity of $X$, $\mathbf{1}_{\{ \}}$ is the indicator function, and $\| \cdot \|$ is the Euclidean norm. It is well-known that $\widehat{K}_{0,n}(r)$ is negatively biased since it neglects the undetected observations near the boundary of $D_n$. 
To ameliorate the boundary issue, we consider the edge-corrected estimator of $K(r)$, which can be written in the following general form:
\begin{equation}\label{eq:empirical_K}
    \widehat{K}_{n}(r)=\left( |D_n| \hat{\lambda}_{1,n}^2 \right)^{-1} \sum_{\bx, \by \in X} b(\bx, \by)
\textbf{1}_{\{0<\|\bx-\by\|\leq r\}}, \quad r \in [0,\infty), \quad n \in \mathbb{N}.
\end{equation} 
Here, $b(\cdot, \cdot)$ is an edge-correction factor which depends on the observation domain $D_n$ and radius $r$. In particular, we assume that $b(\cdot, \cdot)$ belongs to one of the two commonly used edge-correction factors:
\begin{itemize}
\item[(i)] Translation correction: 
we let $b_{1}(\bx, \by) = |D_n| / |D_n \cap (D + \bx - \by)|$, where for a set $A \subset \mathbb{R}^d$ and a point $\bx \in \mathbb{R}^d$, $A + \bx$ is defined as $\{\by: \by = \boldsymbol{a} + \bx, \boldsymbol{a} \in A\}$.
\item[(ii)] Minus Sampling correction: 
we let $b_{2}(\bx, \by) = \{ |D_n| / |D_{0n} \} \mathbf{1}_{\{\bx \in D_{0n}\}}$, where
\begin{equation} \label{eq:D0}
D_{0n} = \{ \bx \in D_n : \inf_{\by \notin D_n} \|\bx - \by\| > r \}.
\end{equation}  For example, if $D_n = B_{d}(\bx, s)$, a ball in $\mathbb{R}^d$ with the center $\bx \in \mathbb{R}^d$ and radius $s>r$, then $D_{0n} = B_{d}(\bx, s-r)$.
\end{itemize}
If we further assume that the underlying point process $X$ is isotropic, then we consider the above two edge-correction factors \textit{plus} the following edge-correction factor:
\begin{itemize}
\item[(iii)] 
Ripley's edge-correction \citep{p:rip-76}: we let $b_{3}(\bx, \by) = \lambda_{d-1} \{ B_{d}(\bx, \|\bx-\by\|)\} / \lambda_{d-1} \{ B_{d}(\bx, \|\bx-\by\|) \cap D_n\}$, where $\lambda_{d-1}\{\cdot\}$ is the $(d-1)$-dimensional surface measure.
\end{itemize} 
We note that $b \in \{b_{1}, b_{2}\}$ (or $b \in \{b_{1}, b_{2}, b_{3}\}$, assuming $X$ is isotropic) yields $\hat{\lambda}_{1,n}^2 \widehat{K}_n(r)$ is an unbiased estimator of $\lambda_{1}^2 K(r)$ for all $n \in \mathbb{N}$.

For a bivariate stationary point process, \cite{hanisch1979formulas} extended the definition of (\ref{eq:K(r)}) and defined the cross $K$-function of $(X_i, X_j)$ as
\begin{eqnarray}
    K_{ij}(r) &=& (\lambda^{(j)}_{1})^{-1} \mathbb{E}
\Bigl \{ 
\text{Number of (distinct) points of $X_j$} \nonumber \\
&& \qquad \qquad \qquad \text{within distance $r$ of a given point of $X_i$}
 \Bigr \},\quad r \in [0,\infty), 
\label{eq:K12(r)}
\end{eqnarray}
where $\lambda^{(j)}_{1}$ is the (homogeneous) first-order intensity of $X_j$. 
The nonparametric edge-corrected estimator of $K_{ij}(r)$ within the sampling window $D_n \subset \mathbb{R}^d$ is
\begin{equation}\label{eq:empirical_cross_K}
    \widehat{K}_{ij,n}(r)=\left( |D_n|\hat{\lambda}_{1,n}^{(i)} \hat{\lambda}_{1,n}^{(j)} \right)^{-1} \sum_{\bx \in X_i} \sum_{\by \in X_j}  b(\bx, \by)  \textbf{1}_{\{\|\bx-\by\|\leq r\}}, \quad r \in [0,\infty), \quad n \in \mathbb{N},
\end{equation}
where $\hat{\lambda}_{1,n}^{(i)} = N_{X_i}(D_n)/|D_n|$ is the estimator of the marginal first-order intensity function of $X_i$, and $\hat{\lambda}_{1,n}^{(j)}$ is defined similarly for $X_j$.


\subsection{The discrepancy measure and minimum contrast estimator} \label{sec:discrepancy}

Now, our aim is to fit a parametric multivariate stationary spatial point process model using the marginal and cross $K$-functions from the observations.  To establish the increasing domain asymptotic framework, we define a sequence of sampling windows $\{D_n\}_{n \in \mathbb{N}}$ in $\mathbb{R}^d$ such that $D_1 \subset D_2 \subset \dots$ and $\lim_{n \rightarrow \infty}|D_n| = \infty$. 

Before introducing our discrepancy measure, we fix the following terms: Let
 \begin{equation} \label{eq:Kmatrix}
\bK_{} (r; \btheta) = [K_{ij}(r;\btheta)]_{i,j=1}^{m}, \quad r \in [0,\infty), \quad \btheta \in \Theta,
\end{equation} 
be the family of $m \times m$ matrix-valued parametric $K$-functions, where for $i,j \in \{1, \dots,m\}$,
$K_{ij}(\cdot, \btheta)$ is the conjectured marginal (if $i=j$) or the cross (if $i \neq j$) $K$-function
defined as in (\ref{eq:K(r)}) and (\ref{eq:K12(r)}), respectively. An estimator of $\bK(r; \btheta)$ based on the data sampled within $D_n$ is
\begin{equation} \label{eq:Knr}
\widehat{\bK}_n (r) = [\widehat{K}_{ij,n}(r)]_{i,j=1}^{m}, \quad r \in [0,\infty), \quad n \in \mathbb{N},
\end{equation} 
where for $i,j \in \{1, \dots,m\}$, $\widehat{K}_{ij,n}(r)$ is the nonparametric edge-corrected estimator of the true marginal (if $i=j$) and cross (if $i\neq j$) $K$-function, defined as in (\ref{eq:empirical_K}) and (\ref{eq:empirical_cross_K}), respectively. We note that $\widehat{\bK}_n (r)$ is slightly biased for the ``true'' $K$-function matrix due to the additional randomness in the estimation of the first-order intensities. Therefore, we define the scaled parametric $K$-function matrix (which we refer to as the $Q$-function matrix hereafter) by
\begin{equation} \label{eq:Q}
\bQ_{} (r; \btheta) = \text{Diag} ( \lambda_{1}^{(1)}, \dots, \lambda_{1}^{(m)}) \bK_{} (r; \btheta) \text{Diag} ( \lambda_{1}^{(1)}, \dots, \lambda_{1}^{(m)}), \quad r \in [0,\infty),
\end{equation} 
where for $i \in \{1, \dots,m\}$, $\lambda_{1}^{(i)} = \lambda_1^{(i)}(\btheta)$ denotes the first-order intensity of $X_i$, and $\text{Diag}$ denotes a diagonal matrix. Therefore, a nonparametric unbiased estimator of the true $\bQ$-function is
\begin{equation} \label{eq:Qn}
\widehat{\bQ}_{n}(r) = \text{Diag}(\hat{\lambda}_{1,n}^{(1)}, \dots, \hat{\lambda}_{1,n}^{(m)}) \widehat{\bK}_{n}(r) \text{Diag}(\hat{\lambda}_{1,n}^{(1)}, \dots, \hat{\lambda}_{1,n}^{(m)}), \quad r \in [0,\infty), \quad n \in \mathbb{N}.
\end{equation} 
With this notation, we propose our discrepancy measure between $\bQ$ and $\widehat{\bQ}_n$ as
\begin{equation} 
\label{eq:multiK_discrep_diff_power}
U_{n}(\btheta) = \int_{0}^{R} w(h) \Tr \left[ \left(\bQ(h;\btheta)_{}^{\circ C} - \widehat{\bQ}_{n}(h)^{\circ C} \right)
\left(\bQ(h;\btheta)_{}^{\circ C} - \widehat{\bQ}_{n}(h)^{\circ C} \right)^{\top}\right] dh, 
\quad n \in \mathbb{N}, \quad \btheta \in \Theta.
\end{equation} 
Here, $R$ is a positive range, $w(\cdot)$ is a non-negative weight function with $\sup_{h} w(h) <\infty$, $\Tr$ denotes the trace operator, $C = [c_{i,j}]_{i,j=1}^{m}$ is a symmetric matrix with positive entries, and $A^{\circ C} = [a_{i,j}^{c_{i,j}}]_{i,j=1}^{m}$ for $A = [a_{i,j}]_{i,j=1}^{m}$ is the Hadamard power of the matrix $A$ to the power $C$. 
For computational purposes, using the standard matrix norm identity, $U_{n}(\btheta)$ can be equivalently written as
\begin{equation} \label{eq:multiK_discrep_diff_power2}
U_{n}(\btheta) = \sum_{i,j=1}^{m}
\int_{0}^{R} w(h) \left\{ [\bQ_{}(h;\btheta)]_{i,j}^{c_{i,j}} - [\widehat{\bQ}_{n}(h)]_{i,j}^{c_{i,j}} \right\}^2 dh, \quad n \in \mathbb{N},
\end{equation} 
where for a matrix $A$, $[A]_{i,j}$ denotes the $(i,j)$-th element of $A$. It is worth mentioning that when $m=1$ (univariate case), then the form of $U_n(\btheta)$ is nearly identical to the discrepancy measure considered in \cite{p:gua-07}. One difference is that the authors of \cite{p:gua-07} considered the $Q$-function without edge-correction. See Appendix \ref{sec:comparison} for the moment comparison between the two $Q$-function estimators.


Lastly, using the above discrepancy measure, our proposed minimum contrast estimator for multivariate point processes is
\begin{equation} \label{eq:thetahat}
\widehat{\btheta}_n = \arg \min_{\btheta \in \Theta} U_n(\btheta), \qquad n \in \mathbb{N}.
\end{equation} 

\section{Sampling properties of the minimum contrast estimator} \label{sec:sampling} 

In this section, we study the large sample properties of the minimum contrast estimator defined as in (\ref{eq:thetahat}).
We note that in practical scenario, one can consider the Riemann sum approximation
\begin{eqnarray} 
U_{n}^{\crosssymbol}(\btheta) &=& \sum_{k=1}^{n_0} w(\frac{R k}{n_0}) \Tr \left[ \left(\bQ(\frac{R k}{n_0};\btheta)_{}^{\circ C} - \widehat{\bQ}_{n}(\frac{R k}{n_0})^{\circ C} \right)
\left(\bQ(\frac{R k}{n_0};\btheta)_{}^{\circ C} - \widehat{\bQ}_{n}(\frac{R k}{n_0})^{\circ C} \right)^{\top}\right] \nonumber \\
&=&  \sum_{k=1}^{n_0} \sum_{i,j=1}^{m} w(\frac{R k}{n_0}) \left\{ [\bQ_{}(\frac{R k}{n_0};\btheta)]_{i,j}^{c_{i,j}} - [\widehat{\bQ}_{n}(\frac{R k}{n_0})]_{i,j}^{c_{i,j}} \right\}^2\quad \text{for} \quad n_0 \in \mathbb{N}
\label{eq:multiK_discrep}
\end{eqnarray} 
to obtain the feasible criterion of $U_n(\btheta)$. However, in this article, we will not study the asymptotic behavior of $\btheta_n^{\crosssymbol} = \arg \min_{\btheta \in \Theta} U_{n}^{\crosssymbol}(\btheta)$.


\subsection{Assumptions} \label{sec:assum}

To derive the sampling properties of $\widehat{\btheta}_n$, we assume that the increasing sequence of sampling windows $\{D_n\}$ in $\mathbb{R}^d$ are \textit{convex averaging windows} (c.a.w.) that satisfy
\begin{equation} \label{eq:caw}
D_n \text{ is convex}, \quad |D_n| \propto n^{d}, \quad \text{and} \quad \lambda_{d-1}(\partial D_n) \propto n^{d-1},
\end{equation}
where $a_n \propto b_n$ means that there exists $C>0$ such that $C^{-1} < \inf_{n \in \mathbb{N}} |a_n|/|b_n| \leq \sup_{n \in \mathbb{N}} |a_n|/|b_n| < C$. Condition (\ref{eq:caw}) implies that $D_n$ grows ``regularly'' in all coordinates of $\mathbb{R}^d$, which is important in our theoretical development. For example, (\ref{eq:caw}) implies that $\{D_n\}$ is regular in the sense of \cite{p:ngu-79} and for an arbitrary fixed $r>0$, $\lim_{n\rightarrow \infty} |D_{0n}|/|D_{n}|=1$ and $|D_n \cap D_{0n}^{\text{c}}|/|D_n| \propto n^{-1}$ as $n \rightarrow \infty$, where $D_{0n}$ is defined as in (\ref{eq:D0}). See \cite{p:hei-08}. These properties are used to calculate the asymptotic covariance matrix of $\bQ$-function estimators. See Appendices \ref{sec:comparison} and \ref{sec:sigma}.


The next set of assumptions is on the higher-order structure and the $\alpha$-mixing condition (cf. \cite{p:ros-56}) of the point processes. For $E_i, E_j \subset \mathbb{R}^d$, $E_i \cong E_j$ means $E_i$ and $E_j$ are congruent. For compact and convex subsets $E_i, E_j \subset \mathbb{R}^d$, let $d(E_{i}, E_{j})=\inf_{\bx_i \in E_{i}, \bx_j \in E_{j}} \|\bx_i - \bx_j\|_{\infty}$, where $\|\bx\|_{\infty}$ is the $\ell_\infty$-norm (max norm). Then, the $\alpha$-mixing coefficient is defined as
\begin{eqnarray}
\alpha_{p}^{}(k) &=& \sup_{A_{i}, A_j, E_i, E_j}  \Bigl\{\left| P(A_{i} \cap A_{j}) - P(A_{i}) P(A_{j}) \right|: \nonumber \\
&& \qquad A_{i} \in \mathcal{F}_{}(E_{i}), A_{j} \in \mathcal{F}_{}(E_{j}), E_i \cong E_j, |E_{i}|=|E_{j}| \leq p, d(E_{i}, E_{j}) \geq k \Bigr\},~~p,k \in (0,\infty).
 \label{eq:alpha}
\end{eqnarray} 
Here, $\mathcal{F}_{}(E)$ denotes the $\sigma$-field generated by the superposition of $\underline{X}$ in $E_{} \subset \mathbb{R}^d$ and the supremum is taken over all compact and convex congruent subsets $E_i$ and $E_j$.
\begin{assumption} \label{assump:1}
$\underline{X} = (X_1, \dots, X_m)$ is a simple, stationary, and ergodic multivariate point process on $\mathbb{R}^d$ that satisfies the following two conditions:
\begin{itemize}
\item[(i)]
Let $\ell \geq 4$. The reduced joint cumulant intensity of $(X_{i(1)}, \dots, X_{i(k)})$ of order $\underline{n} = (n_{1}, \dots, n_{k})$ defined as in (\ref{eq:cumul-red}) is well-defined for any non-overlapping subset $(i(1), \dots, i(k)) \subset \{1, \dots,m\}$ and positive order $\underline{n} = (n_{1}, \dots, n_{k})$ such that $\sum_{i=1}^{k} n_i \leq \ell$. Moreover, if $2 \leq \sum_{i=1}^{k} n_i \leq \ell$, then the reduced joint cumulant intensity functions are absolutely integrable on $\mathbb{R}^{N-1}$, where $N = \sum_{i=1}^{k} n_i$.

\item[(ii)] 
There exists $\varepsilon > 0$ such that $\sup_{p \in (0,\infty)} \alpha_{p}(k)/\max(p,1) = O(k^{-d-\varepsilon})$ as $k \rightarrow \infty$.
\end{itemize} 
\end{assumption}
We require Assumption \ref{assump:1}(i) (for $\ell=4$)
to show the convergence of the asymptotic covariance between the two empirical processes $[\widehat{\bG}_{n}(h)]_{a,b}$ and $[\widehat{\bG}_{n}(h)]_{c,d}$ ($a,b,c,d \in \{1, \dots,m\}$), where
\begin{equation} \label{eq:Gn}
[\widehat{\bG}_{n}(h)]_{i,j} =
|D_n|^{1/2} \left\{ [\widehat{\bQ}_{n}(h)]_{i,j} - \Ex [\widehat{\bQ}_{n}(h)]_{i,j} \right\},
 \quad i,j \in \{1, \dots,m\} \quad \text{and} \quad h \in [0,\infty).
\end{equation} 
Assumption \ref{assump:1}(ii) together with
 \begin{equation} \label{eq:momentQ1}
    \max\limits_{1\leq i,j \leq m}\sup\limits_n  \mathbb{E} |[\widehat{\bG}_{n}(h)]_{i,j}|^{2+\delta}  <\infty \quad \text{for some} \quad \delta>0
\end{equation} 
are used to show the multivariate CLT for $\{ [\widehat{\bG}_{n}(h)]_{i,j}\}_{i,j=1}^{m}$.
When we choose $b \in \{b_1, b_2, b_3\}$, then a sufficient condition for (\ref{eq:momentQ1}) (for $\delta = 1$) to hold is Assumption \ref{assump:1}(i) for $\ell=6$. A proof for the sufficiency is similar to those in \cite{p:jol-78}, Theorems 2 and 3. 
We refer interested readers to \cite{p:yang-24}, Section 4, for various (univariate) point process models that satisfy Assumption \ref{assump:1}.

The last set of assumptions is on the parameter space. For $i,j \in \{1, \dots,m\}$, let
\begin{equation}
[\nabla_{\btheta}\bQ(h;\btheta)]_{i,j} = \frac{\partial}{\partial \btheta} [\bQ(h;\btheta)]_{i,j}
\quad \text{and} \quad
[\nabla_{\btheta}^2 \bQ(h;\btheta)]_{i,j} = \frac{\partial^2}{\partial \btheta \partial \btheta^{\top}} [\bQ(h;\btheta)]_{i,j}
\label{eq:Q-deriv}
\end{equation} 
be the gradient and Hessian of $ [\bQ(h;\btheta)]_{i,j}$, respectively. Further, let 
\begin{equation} 
[\widehat{\bQ}_{1n}(h)]_{i,j} 
= \int_{0}^{h} w(u) [\widehat{\bQ}_n(u)]_{i,j} [\bQ(h;\btheta_0)]_{i,j}^{2c_{i,j}-2} [\nabla_{\btheta} \bQ(u;\btheta_0)]_{i,j} du, \quad h \in [0,\infty),
\label{eq:Q1n}
\end{equation} where $\btheta_0 \in \Theta$ is the true parameter of $\underline{X}$. 
 
\begin{assumption} \label{assump:3}
\begin{itemize}
\item[(i)] The parameter space $\Theta \subset \mathbb{R}^{p}$ is convex and compact, $\btheta \rightarrow \bQ(\cdot;\btheta)$ is continuous and injective, and the true parameter $\btheta_{0}$ lies in the interior of $\Theta$.  

\item[(ii)] For $i,j \in \{1, \dots,m\}$, $[\nabla_{\btheta}^2 \bQ(h;\btheta)]_{i,j}$ defined as in (\ref{eq:Q-deriv}) exists and is continuous with respect to $\btheta \in \Theta$. 

\item[(iii)] For $i,j \in \{1, \dots,m\}$, let $[\widehat{\bQ}_{1n}(r)]_{i,j}$ be defined as in (\ref{eq:Q1n}). Then, there exists $\delta >0$ such that
\begin{equation*}
    \max\limits_{1\leq i,j \leq m}\sup\limits_n  \mathbb{E}  \Bigl\|\sqrt{|D_n|} \Bigl\{[\widehat{\bQ}_{1n}(h)]_{i,j} - \mathbb{E}\left[ [\widehat{\bQ}_{1n}(h)]_{i,j} \right] \Bigr\} \Bigr\|_{}^{2+\delta} <\infty.
\end{equation*}
\end{itemize} 
\end{assumption}


\subsection{Asymptotic results} \label{sec:asymp}

In this section, we state our main asymptotic results. The first theorem addresses the asymptotic joint normality of $\{ [\widehat{\bQ}_n(h)]_{i,j}\}_{i,j=1}^{m}$. We first note that for $b \in \{b_1,b_2\}$ (or, $b\in \{b_1, b_2, b_3\}$, assuming $\underline{X}$ is isotropic), $[\widehat{\bQ}_n(h)]_{i,j}$ is unbiased in the sense that $\Ex [\widehat{\bQ}_n(h)]_{i,j} = [\bQ(h;\btheta_0)]_{i,j}$. Therefore, recalling (\ref{eq:Gn}), we have
\begin{equation} \label{eq:Gn2} 
[\widehat{\bG}_{n}(h)]_{i,j} =
|D_n|^{1/2} \left\{ [\widehat{\bQ}_{n}(h)]_{i,j} -[\bQ_{}(h;\btheta_0)]_{i,j} \right\},
 \quad i,j \in \{1, \dots,m\} \quad \text{and} \quad h \in [0,\infty).
\end{equation} 
Since we take into account the asymmetric edge-correction factors, $\widehat{\bG}_{n}(\cdot)$ may not be symmetric. Therefore, we consider the full entries of $\widehat{\bG}_n$ and define
\begin{equation} \label{eq:vecGn}
\text{vec}(\widehat{\bG}_{n}(h)) = ( [\widehat{\bG}_{n}(h)]_{i,j})_{1\leq i, j \leq m}, \quad h \in [0,\infty),
\end{equation}
as the $m^2$-dimensional vectorization of $\widehat{\bG}_{n}(h)$. The following theorem shows the asymptotic normality of $\text{vec}(\widehat{\bG}_{n}(h))$.

\begin{theorem} \label{thm:Jasymp}
Let $\underline{X} = (X_1, \dots, X_m)$ be a multivariate stationary point process that satisfies Assumption \ref{assump:1}(i) (for $\ell=2$). Moreover, we assume that the increasing sequence of sampling windows $\{D_n\}$ in $\mathbb{R}^d$ is c.a.w., and the edge-correction factor is such that $b \in \{b_1, b_2\}$ (or $b \in \{b_1, b_2, b_3\}$, assuming $\underline{X}$ is isotropic).
Then, for fixed $R>0$ and $i,j \in \{1, \dots,m\}$,
\begin{equation}\label{eq:unifJ}
\sup_{0 \leq h \leq R} |D_n|^{-1/2} \left|  [\widehat{\bG}_{n}(h)]_{i,j} \right| = 
\sup_{0 \leq h \leq R} \left| [\widehat{\bQ}_{n}(h)]_{i,j} -[\bQ_{}(h;\btheta_0)]_{i,j} \right|
 \rightarrow 0 \quad \text{almost surely}
\end{equation} as $n \rightarrow \infty$.
Furthermore, under Assumptions \ref{assump:1}(i) (for $\ell=4$), (ii), and (\ref{eq:momentQ1}), we have 
\begin{equation}\label{eq:asympJ}
\emph{vec}(\widehat{\bG}_{n}(h))  \Dcon \mathcal{N}(\bm{0}_{m^2}, \Sigma(h;\btheta_{0})), \quad h \in [0,\infty),
\end{equation} 
where $\Dcon$ denotes weak convergence and $\mathcal{N}$ represents the multivariate normal distribution. An expression for the asymptotic covariance matrix $\Sigma(h;\btheta_{0})$ can be found in Appendix \ref{sec:sigma}.
\end{theorem}
\noindent \textit{Proof}. See Appendix \ref{proof:thm1}. \hfill $\Box$ 


\vspace{0.5em}

The next theorem addresses the asymptotic normality of the minimum contrast estimator. To obtain the asymptotic covariance matrix of $\widehat{\btheta}_n$, we define the following two quantities:
\begin{equation} \label{eq:Btheta}
    B(\btheta_{0}) 
= \sum_{i,j=1}^{m} c_{i,j}^2 \int_{0}^{R}  w(h)
[\bQ(h;\btheta_0)]_{i,j}^{2c_{i,j}-2} 
\left\{ [\nabla_{\btheta}\bQ(h;\btheta_{0})]_{i,j} \right\}
\left\{ [\nabla_{\btheta}\bQ(h;\btheta_{0})]_{i,j} \right\}^{\top}  dh
\end{equation} and
\begin{eqnarray} 
 S(\btheta_{0}) &=& \sum_{i_{1},j_{1},i_{2},j_{2}=1}^{m} c_{i_1,j_1}^2 c_{i_2,j_2}^2 \int_{0}^{R} 
\int_{0}^{R}  w(s) w(h) \sigma^2_{(i_{1},j_{1}: i_{2},j_{2})} (s,h) \nonumber \\
&& \qquad \qquad \qquad \qquad \qquad \times    \left\{ 
[\bQ(s;\btheta_0)]_{i_{1},j_{1}}^{2c_{i_1,j_1}-2} \right\}
\left\{
[\bQ(h;\btheta_0)]_{i_{2},j_{2}}^{2c_{i_2,j_2}-2}
 \right\} \nonumber \\
 &&\qquad \qquad \qquad \qquad \qquad \times  
\left\{ [\nabla_{\btheta}\bQ(s;\btheta_{0})]_{i_1,j_1} \right\}
\left\{ [\nabla_{\btheta}\bQ(h;\btheta_{0})]_{i_2,j_2} \right\}^{\top} 
dsdh, 
\label{eq:Stheta}
\end{eqnarray} where for $i_1,j_1,i_2,j_2 \in \{1, \dots,m\} $,
\begin{equation} \label{eq:sigma}
\sigma^2_{(i_{1},j_{1}: i_{2},j_{2})} (s,h) = 
\lim_{n \rightarrow \infty}  \cov  \left\{ [\widehat{\bG}_n(s)]_{i_{1},j_{1}}, [\widehat{\bG}_n(h)]_{i_{2},j_{2}} \right\},
 \quad s,h \in [0,\infty).
\end{equation} 
Under Assumption \ref{assump:1}(i) (for $\ell=4$), the limit of the right-hand side of (\ref{eq:sigma}) exists and is finite for all indices $i_1,j_1,i_2,j_2 \in \{1, \dots,m\}$ and fixed $s,h \geq 0$. An exact expression of $\sigma^2_{(i_{1},j_{1}: i_{2},j_{2})} (s,h)$ can be derived using a similar technique to calculate an expression of $\Sigma(h;\boldsymbol{\theta}_{0})$ in Appendix \ref{sec:sigma}, but the form is more complicated.

Using this notation, we show the asymptotic normality of $\widehat{\boldsymbol{\theta}}_n$.

\begin{theorem} \label{thm:normal_c} 
Let $\underline{X} = (X_1, \dots, X_m)$ be a multivariate stationary point process that satisfies Assumptions \ref{assump:1} (for $\ell=4$), \ref{assump:3}(i), and (\ref{eq:momentQ1}). Moreover, we assume that the increasing sequence of sampling windows $\{D_n\}$ in $\mathbb{R}^d$ is c.a.w. and the edge-correction factor is such that $b \in \{b_1, b_2\}$ (or $b \in \{b_1, b_2, b_3\}$, assuming $\underline{X}$ is isotropic).
Then, $\widehat{\boldsymbol{\theta}}_n$ defined as in (\ref{eq:thetahat}) uniquely exists and
\begin{equation} \label{eq:theta-consist}
\widehat{\btheta}_n \rightarrow \btheta_0 \qquad \text{almost surely~~as~~} n \rightarrow \infty.
\end{equation}
We further assume Assumptions \ref{assump:3}(ii) and (iii) hold and $B(\btheta_{0})$ defined as in (\ref{eq:Btheta}) is invertible. Then,
\begin{equation} \label{eq:theta-normal}
\sqrt{|D_n|} (\widehat{\btheta}_n-\btheta_{0}) \Dcon \mathcal{N}(\bm{0}_p, B(\btheta_{0})^{-1} S(\btheta_{0}) B(\btheta_{0})^{-1}).
\end{equation}
\end{theorem}
\noindent \textit{Proof}. See Appendix \ref{proof:thm2}. \hfill $\Box$

\begin{remark}
The univariate analogous results for Theorem \ref{thm:normal_c} were proved in \cite{p:gua-07}, Theorem 4. However, they only showed the consistency of $\widehat{\boldsymbol{\theta}}_n$. As far as we are aware, almost sure convergence of the MC estimator, even for the univariate case.
\end{remark}

\begin{remark}[Application to the homogeneity testing] \label{rmk:two-sample}
As a direct application of our asymptotic results, let $\{\bx_{ij}: i \in \{1, \dots, m\}, j \in \{1, \dots, n_{1i}\}\}$ and $\{\by_{ij}: i \in \{1, \dots, m\}, j \in \{1, \dots, n_{2i}\}\}$ be two independent configurations of $m$-variate spatial point patterns on $D_n$ sampled from distributions $\mathcal{F}_{\boldsymbol{\theta}_1}$ and $\mathcal{F}_{\boldsymbol{\theta}_2}$, respectively, where $\boldsymbol{\theta}_1, \boldsymbol{\theta}_2 \in \Theta \subset \mathbb{R}^p$. Now, we are interested in whether these configurations are sampled from the same distribution or not. That is, we want to test the hypotheses $H_{0}: \boldsymbol{\theta}_1 = \boldsymbol{\theta}_2$ versus $H_{A}: \boldsymbol{\theta}_1 \neq \boldsymbol{\theta}_2$. Let $\widehat{\boldsymbol{\theta}}_{1,n}$ and $\widehat{\boldsymbol{\theta}}_{2,n}$ be the MC estimators based on the point patterns $\{\bx_{ij}\}$ and $\{\by_{ij}\}$, respectively. For $k \in \{1,2\}$, let $\widehat{\Sigma}_{k,n}$ be the consistent estimator of the asymptotic covariance matrix of $|D_n|^{-1/2}\widehat{\boldsymbol{\theta}}_{k,n}$ (one example of the consistent estimator can be found in Section \ref{sec:est_asymp_var} below). Then, from Theorem \ref{thm:normal_c}, under the null of homogeneity,
$T_n = |D_n| ( \widehat{\btheta}_{1,n} - \widehat{\btheta}_{2,n})^\top ( \widehat{\Sigma}_{1,n} + \widehat{\Sigma}_{2,n})^{-1} ( \widehat{\btheta}_{1,n} - \widehat{\btheta}_{2,n})$ converges weakly to a chi-squared distribution with degrees of freedom $p$. 
Under the alternative, as $n \rightarrow \infty$, $T_n$ converges to the non-central chi-squared distribution with divergent mean. Thus, $T_n$ has statistical power.
\end{remark}

\section{Practical considerations} \label{sec:practice}

\subsection{Estimator of the asymptotic covariance matrix} \label{sec:est_asymp_var}

In this section, our aim is to estimate the asymptotic covariance matrix of $\widehat{\btheta}_n$. 
Recall (\ref{eq:theta-normal}),
\begin{equation} \label{eq:sigmatheta}
\lim_{n \rightarrow \infty}|D_n| \var \widehat{\btheta}_n 
= B(\btheta_{0})^{-1} S(\btheta_{0}) B(\btheta_{0})^{-1} =: \Sigma(\btheta_0).
\end{equation} 
We first estimate $B(\btheta)$ defined as in (\ref{eq:Btheta}). From its definition, provided that $\bQ(\cdot; \btheta)$ and $\nabla_{\btheta} \bQ(\cdot; \btheta)$ have known expressions, $\bQ(\cdot; \btheta_0)$ and $\nabla_{\btheta} \bQ(\cdot; \btheta_0)$ can be easily estimated by replacing $\btheta_0$ with its estimator $\widehat{\btheta}_n$. Therefore, this gives a natural estimator of $B(\btheta_{0})$ and $B(\btheta_{0})^{-1}$, denoted by $B(\widehat{\btheta}_n)$ and $B(\widehat{\btheta}_n)^{-1}$, respectively.
Next, to estimate $S(\btheta_{0})$, we use a Monte Carlo method, which we will describe below.

Recall (\ref{eq:Stheta}), (\ref{eq:sigma}), and (\ref{eq:Gn}). It is easily seen that $S(\btheta_{0}) =  \lim \limits_{n \rightarrow \infty} \var \left\{ \bV_n(\btheta_0) \right\}$,
where
\begin{eqnarray}
\bV_n(\btheta_0) &=& \sqrt{|D_n|} \sum_{i,j=1}^{m} c_{i,j}^2
\int_{0}^{R} \bigg\{ [\widehat{\bQ}_n(h)]_{i_{},j_{}} - [\bQ(h;\btheta_0)]_{i,j} \bigg\} \bigg\{ [\bQ(h;\btheta_0)]_{i_{},j_{}}^{2c_{i,j}-2} \bigg\} 
\nonumber \\
&& \qquad \qquad \qquad \qquad \times
 \bigg\{ [\nabla_{\btheta}\bQ^{}(h;\btheta_{0})]_{i,j} \bigg\}dh. 
\label{eq:Vn}
\end{eqnarray}
To generate the Monte Carlo samples of $\bV_n(\btheta_0)$, we simulate the multivariate point process $\underline{X}$ from the fitted model based on $\widehat{\btheta}_n$. For each simulation, we calculate an estimator of $\bV_n(\btheta_0)$ by replacing $\btheta_0$ with $\widehat{\btheta}_n$ in (\ref{eq:Vn}). Therefore, an estimator of $S(\btheta_{0})$, denoted $\widehat{S}_n(\btheta_{0})$, can be obtained using the sample variance of the Monte Carlo samples of $\bV_n(\widehat{\btheta}_n)$. Under the assumptions stated in Theorem \ref{thm:normal_c}, it can be shown that $B(\widehat{\btheta}_n)$ and $\widehat{S}_n(\btheta_{0})$ are both consistent. Therefore, our final consistent estimator of the asymptotic covariance matrix of $\widehat{\btheta}_n$ is
\begin{equation} \label{eq:sigmahat}
\widehat{\Sigma}_n(\btheta_0) = 
\{ B(\widehat{\btheta}_n)\}^{-1} \{ \widehat{S}_n(\btheta_{0})\} \{  B(\widehat{\btheta}_n)\}^{-1}, \qquad n \in \mathbb{N}.
\end{equation}

\begin{remark}[Alternative estimation methods] \label{rmk:alt-cov}
As pointed out by the two referees, there are alternative approaches to estimate $B(\btheta_0)$ and $S(\btheta_0)$. Firstly, $B(\btheta_0)$ can be estimated using Monte Carlo samples of $B(\btheta_0)$ from simulated spatial point patterns of $\underline{X}$ based on $\widehat{\btheta}_n$. Secondly, $S(\btheta_0)$ can be estimated using a subsampling method as described in \cite{p:bis-19} (see also \cite{p:yang-24}, Appendix H).
In detail, we evaluate subsamples of $\bV_n(\widehat{\btheta}_n)$, denoted $\bV_n^{(\boldsymbol{k})}(\widehat{\btheta}_n)$, where the sampling window is the subregions of $D_n$ of the form $D_n^{(\boldsymbol{k})} = \boldsymbol{k} + [-a_n,a_n]^d$, $\boldsymbol{k} \in \mathbb{Z}^d$. Here, $\{a_n\}$ is an increasing sequence of positive numbers that satisfies $\lim_{n\rightarrow \infty} a_n / n = 0$. We then estimate $S(\btheta_0)$ using the subsampling variance of $\bV_n^{(\boldsymbol{k})}(\widehat{\btheta}_n)$. Under appropriate moment and mixing conditions such as conditions ($\mathcal{S}$1)–($\mathcal{S}$6) in \cite{p:bis-19}, one can show the consistency of the subsampling variance estimator.
Further details, including the sampling properties and empirical studies of the alternative estimations of $B(\btheta_0)$ and $S(\btheta_0)$, will not be considered in this study.
\end{remark}

\subsection{Selection of the optimal control parameters} \label{sec:select_tune_par}

Selecting the control parameters is a challenging task in the MC method, even for the univariate case. For a univariate point process with a discrepancy measure $\mathcal{U}(\btheta)$ defined as in (\ref{eq:U}), \cite{diggle2003statistical} suggested some empirical rules on the choice of the control parameters. However, these choices are \textit{ad hoc}. Moreover, to best of our knowledge, there is no existing work on the selection of control parameters for the MC method applied to multivariate point processes.

Now, using the asymptotic variance estimator, we propose a data-driven criterion for selecting control parameters in multivariate point processes. Recall (\ref{eq:multiK_discrep_diff_power}) and (\ref{eq:theta-normal}), where the discrepancy function $U_n(\btheta)$ and the asymptotic covariance matrix of $\widehat{\btheta}_n$ depend on the weight $w(\cdot)$, the range $R$, and the power matrix $C = [c_{i,j}]_{i,j=1}^{m}$. Both $w(\cdot)$ and $C$ control the fluctuation of $\widehat{\bQ}_n$. For simplicity, we fix $w(h) \equiv 1$ and allow $C$ and $R$ to vary. It is worth noting that $w(\cdot)$ could be selected using the subsampling method proposed in \cite{p:bis-19}, but this method is not considered in our study. 

By fixing the weight function unity, we propose the following criterion for selecting control parameters $C$ and $R$:
\begin{equation} \label{eq:CR}
(C_{opt}, R_{opt}) = \argmin_{C, R} \det \Sigma(\btheta_0),
\end{equation} 
where $\Sigma(\btheta_0)$ is the asymptotic covariance matrix of $\widehat{\btheta}_n$ defined as in (\ref{eq:sigmatheta}). We refer to (\ref{eq:CR}) as ``optimal'' in the sense that for a fixed level, $(C_{opt}, R_{opt})$ provides the smallest volume of the confidence ellipse. In practical scenarios, our final feasible criterion for selecting optimal control parameters is
\begin{equation} \label{eq:CR2}
(\widetilde{C}_{opt}, \widetilde{R}_{opt}) = \argmin_{C,R} \det \widehat{\Sigma}_n(\btheta_0),
\end{equation} where $\widehat{\Sigma}_n(\btheta_0)$ is defined as in (\ref{eq:sigmahat}) and the minimum is taken over the finite grids of $(C,R)$.

\subsection{Constructing confidence regions} \label{sec:CRs}
Let $(\widetilde{C}_{opt}, \widetilde{R}_{opt})$ be the optimal control parameters as in (\ref{eq:CR}), and let $\widetilde{\Sigma}_n$ be the asymptotic covariance estimator corresponding to $(\widetilde{C}_{opt}, \widetilde{R}_{opt})$. Then, by Theorem \ref{thm:normal_c}, an asymptotic $(1-\alpha)$ confidence ellipsoid using the optimal parameters $(\widetilde{C}_{opt}, \widetilde{R}_{opt})$ is given by
\begin{equation} \label{eq:Conf1}
\big\{\btheta \in \mathbb{R}^p : T(\btheta) \leq \chi^2_{p}(1-\alpha)\big\},
\end{equation}
where $T(\btheta) = |D_n| (\widehat{\btheta}_{n} - \btheta)^\top (\widetilde{\Sigma}_n)^{-1} (\widehat{\btheta}_{n} - \btheta)$, and $\chi^2_{p}(1-\alpha)$ is the $(1-\alpha)$th quantile of the chi-squared distribution with degrees of freedom $p$. 
However, as referees have pointed out, using (\ref{eq:Conf1}) as a confidence region may produce low coverage probabilities. Indeed, in our bivariate LGCP simulation study in Section \ref{sec:sim-model} below, we encountered this issue even for a large sampling window size $D_n = [-15,15]^2$. A possible explanation for this phenomenon is that since $\widetilde{\Sigma}_n$ yields the smallest determinant among all asymptotic covariance estimators, using $(\widetilde{C}_{opt}, \widetilde{R}_{opt})$ to calculate the asymptotic covariance estimator may underestimate the true asymptotic covariance corresponding to $(\widetilde{C}_{opt}, \widetilde{R}_{opt})$.

As a remedy, we consider a simulation-based confidence region to solve the low coverage probability issue in finite samples. Let $\widehat{\btheta}_n$ be the MC estimator calculated based on the optimal control parameters $(\widetilde{C}_{opt}, \widetilde{R}_{opt})$. Next, we simulate \textit{i.i.d.} replications of multivariate point patterns from the fitted parametric model. Then, we estimate the simulation-based asymptotic covariance matrix by
$\widetilde{\Sigma}_n^{\star} = (|D_n| B)^{-1} \sum_{i=1}^{B} (\widehat{\btheta}_n^{(i)} - \overline{\widehat{\btheta}_n}) (\widehat{\btheta}_n^{(i)} - \overline{\widehat{\btheta}_n})^{\top}$,
where $B$ is the number of replications, $\widehat{\btheta}_n^{(i)}$ is the MC estimator derived from the $i$-th replication, and $\overline{\widehat{\btheta}_n} = B^{-1} \sum_{i=1}^{B} \widehat{\btheta}_n^{(i)}$. Our final simulation-based $(1-\alpha)$ confidence ellipsoid is
\begin{equation} \label{eq:Conf2}
\big\{\btheta \in \mathbb{R}^p: T^{\star}(\btheta) \leq \chi^2_{p}(1-\alpha)\big\},
\end{equation} 
where $T^{\star}(\btheta) = |D_n| (\widehat{\btheta}_{n} - \btheta)^\top (\widetilde{\Sigma}_n^{\star})^{-1} (\widehat{\btheta}_{n} - \btheta)$. In the same bivariate LGCP simulations, we observed that using (\ref{eq:Conf2}) as a confidence region successfully recovers the $(1-\alpha)$ coverage rate for all window sizes. Therefore, for practical purposes, we recommend using (\ref{eq:Conf2}) to construct a confidence region.


\section{Simulations} \label{sec:simu}

To validate our theoretical results and assess the finite sample performance, we conduct simulation studies for bivariate LGCP and five-variate PSNCP models. Supplementary simulation results can also be found in Appendix \ref{appen:fig}.
 
\subsection{The bivariate LGCP model} \label{sec:sim-model}

For the data-generating process, we first consider the bivariate LGCP model, which is also used to fit the real data in Appendix \ref{sec:applic}. Let $\underline{X} = (X_1, X_2)$ be a stationary LGCP on $\mathbb{R}^2$ driven by the latent intensity field $\underline{\Lambda}(\bs) = (\Lambda_1(\bs), \Lambda_2(\bs)) = (\exp(Y_1(\bs)), \exp(Y_{2}(\bs)))$, $\bs \in \mathbb{R}^2$, where $Y_i(\cdot)$ is a stationary Gaussian random field with the parameter restriction $\mathbb{E} [ e^{Y_i(\bs)}] = 1$. To formulate the joint distribution of $Y_1$ and $Y_2$, we consider the following additive structure:
\begin{equation}
\begin{pmatrix}
Y_1(\bs) \\ Y_2(\bs)
\end{pmatrix}=
\begin{pmatrix}
\mu_{Y_1} \\ \mu_{Y_2}
\end{pmatrix}
+
\begin{pmatrix}
1 & 0 & 1 \\ 0 & 1 & b
\end{pmatrix}
\begin{pmatrix}
Z_1(\bs) \\ Z_2(\bs) \\ Z_3(\bs)
\end{pmatrix}
= 
\begin{pmatrix}
\mu_{Y_1} + Z_1(\bs) + Z_3(\bs) \\ \mu_{Y_2} + Z_2(\bs) + b Z_3(\bs)
\end{pmatrix}, \quad \bs \in \mathbb{R}^2,
\label{eq:LMC}
\end{equation}
where $b \in \{-1,1\}$ indicates the positive ($b=1$) or negative ($b=-1$) correlation between $Y_1$ and $Y_2$, and $\{Z_i\}_{i=1}^{3}$ are mean zero independent Gaussian processes on $\mathbb{R}^2$ with isotropic exponential covariance functions. Therefore,
\begin{equation} \label{eq:exp-cov}
\cov \{ Z_i (\bs_1), Z_j(\bs_2) \} = \sigma_{Z_i}^2 \exp(- \|\bs_1 - \bs_2\| / \phi_{Z_i})\textbf{1}_{i=j}, \quad \bs_1, \bs_2 \in \mathbb{R}^2, \quad i,j \in \{1,2,3\}.
\end{equation}
Here, for $i \in \{1,2,3\}$, $\sigma_{Z_i}$ and $\phi_{Z_i}$ are the positive \textit{scale} and \textit{range} parameters of the covariance function of $Z_i(\cdot)$. We note that the bivariate LGCP under consideration satisfies Assumption \ref{assump:1}. See Lemma \ref{lemma:LGCP} in the Appendix. Using \cite{b:cre-93}, Equations (8.3.32) and (8.6.10), and \cite{moller1998log}, Equation (4), the entries of $\bK(r;\btheta)$ are given by
\begin{equation}
[\bK(r;\btheta)]_{i,j}  = 2 \pi \int_{0}^{r} h \exp(C_{ij}(h;\btheta)) dh, \quad r \in [0,\infty), \quad i,j \in \{1,2\},
\label{eq:LGCP-Qr}
\end{equation} 
where $\btheta = (\sigma_{Z_1}, \phi_{Z_1},\sigma_{Z_2}, \phi_{Z_2},\sigma_{Z_3}, \phi_{Z_3})^\top$ is the set of parameter of interest of our approach, and \\
$ C_{ij}(h;\bm{\theta}) = \cov\{\log\Lambda_i(\bs_1), \log\Lambda_j(\bs_2)\}$ for $h = \|\bs_1-\bs_2\|$. To further investigate the correlation between $X_1$ and $X_2$, let 
\begin{equation*}
\rho = \rho(\btheta) = \corr\{ \log \Lambda_1(\textbf{0}), \log \Lambda_2(\textbf{0})\}
\end{equation*}
be the cross-correlation coefficient of $X_1$ and $X_2$. Explicit expressions for $C_{ij}(\cdot;\btheta)$ and $\rho$ in terms of the model parameters can be found in Appendix \ref{appen:explicit}. The $Q$-functions $[\bQ(r;\btheta)]_{i,j}$ for $i,j \in \{1,2\}$ can be calculated using (\ref{eq:LGCP-Qr}) and (\ref{eq:Q}). Throughout the simulation, we assume that the first-order intensities are known to be equal to one. Thus, when utilizing the $Q$-function, the set of parameters of interest remains the same as that used for the $K$-function.

Now, we consider the aforementioned bivariate LGCP model with four different combinations of the true model parameters, denoted as (M1)--(M4), as displayed in Table \ref{tab:true_par_LMC}. For each model, we let $b$ take values of either 1 or -1. Therefore, (M1)--(M4) allow flexible modeling of bivariate point patterns, including weak cross-correlation (corresponding to (M1)) to strong cross-correlation (corresponding to (M4)), as well as positive ($b=1$) and negative ($b=-1$) correlations. Figure \ref{fig:pp_simu_biv_lgcp_density} in the Appendix illustrates a realization of each model.
\begin{table}[h]    
    \centering
    \begin{tabular}{c|rrrrrr|c} 
    \hline
        Model
        &$\sigma_{Z_1}$ 
        & $\phi_{Z_1}$ 
        &$\sigma_{Z_2}$ 
        & $\phi_{Z_2}$
        &$\sigma_{Z_3}$ 
        & $\phi_{Z_3}$
        & $\rho$ \\
    \hline \hline
      (M1)  & 1 & 0.5 & 0.8 & 1 & 0.4 & 1.5 & $\pm$ 0.166\\
      (M2) & 0.8 & 0.5 & 0.6 & 1 & 0.5 & 1.5 & $\pm$ 0.339 \\
      (M3) & 0.7 & 0.5 & 0.4 & 1.3 & 0.6 & 1 & $\pm$ 0.541 \\
      (M4)  & 0.5 & 0.5 & 0.4 & 1.3 & 0.8 & 1 & $\pm$ 0.758\\
      \hline
    \end{tabular}
\caption{
Four different data generating processes of the parametric bivariate LGCP model. Note that the reported cross-correlation coefficient $\rho$ in the column is calculated based on the model parameters and $b \in \{-1,1\}$. See (\ref{eq:rho}) in the Appendix.
}
 \label{tab:true_par_LMC}
\end{table}
For each model, we generate the bivariate point patterns on the window $D=[-\text{WL}/2,\text{WL}/2]^2$ of window length $\text{WL} \in \{10,20,30\}$. Therefore, the expected numbers of points within the sampling windows are 100, 400, and 900, respectively. For each simulation, we evaluate the parameter estimators and the correlation coefficient $\rho$ using two different methods: the minimum contrast estimator (MC; see (\ref{eq:thetahat})) and the Bayesian inferential method for LGCP using the Metropolis-adjusted Langevin algorithm \citep[BI; see][]{moller1998log, taylor2015bayesian}, with the latter serving as our benchmark.

\subsubsection{Processing the MC and BI estimators}

For MC, we use Ripley's edge-correction (which corresponds to $b_3(\bx, \by)$ in Section \ref{sec:Kfcns}) to evaluate the $Q$-function matrix estimator. Ripley's edge-correction is known to outperform Translation correction or Minus sampling correction in simulations (cf. \citep{p:dog-89}, page 565). To numerically approximate the discrepancy measure of the MC method, we use a Riemann sum approximation with 512 equally-spaced grids as in (\ref{eq:multiK_discrep}). When selecting the optimal control parameters for MC, we set the weight function as a unit constant and assume common powers $c_{i,j} = c$ for $i, j \in \{1,2\}$ in (\ref{eq:multiK_discrep}). This allows us to avoid introducing excessive control parameters. Finally, we select the optimal control parameter $(c,R)$ using the method in Section \ref{sec:select_tune_par} on the grids $c \in \{0.1, 0.2, 0.3, 0.4, 0.5\}$ and $R$ from 1 to $0.35 \times \text{WL}$ with increment $0.025 \times \text{WL}$. For BI, we implement the \texttt{lgcp} package in R CRAN \citep{taylor2015bayesian} to compute the parameter estimates using a 64 $\times$ 64 computational grid for all values of WL, and the Markov chain in BI runs for $3.1 \times 10^6$ iterations. Lastly, for each model, we generate only 50 simulations for BI due to its extensive computational cost, whereas we generate 500 simulations for MC.

\subsection{Results} \label{sec:results}

\subsubsection{Computation time} \label{sec:compute}

Table \ref{tab:comp-time} presents the average computing time per simulation for evaluating the two estimators for model (M1). Here, we vary the sign of correlation, the maximum range $R$ (for MC only), and the window length (WL). The average computing times for models (M2)--(M4) exhibit a similar pattern so we omit those tables.
\begin{table}[h!]
\small
\begin{center}
\begin{tabular}{c|ccrrr}
\hline
Estimator & Correlation & $R$ & $\text{WL}=10$ & $\text{WL}=20$ & $\text{WL}=30$  \\
 \hline \hline
\multirow{4}{*}{MC}
& \multirow{2}{*}{Negative} 
& 0.15WL  &1.50 &3.91 &9.73 \\
& & 0.35WL &2.39 &9.04 &36.67 \\
\cmidrule(lr){2-6}
& \multirow{2}{*}{Positive} 
& 0.15WL  &1.78 &5.02 &12.09 \\
& & 0.35WL &2.73 &10.17 &33.03 \\
\cmidrule(lr){1-6}
\multirow{2}{*}{BI}
& Negative & -- &656.66 &658.77 &661.65 \\
& Positive & -- &596.00 &586.32 &598.42 \\ 
\hline 
\end{tabular}
\end{center}
\caption{
Average computing time per simulation (unit: minute) for MC and BI estimator from (M1). Here, we use the common power $c=0.2$ for MC. Estimation is done on a parallel computing cluster in R using a server with a 56-processor Intel Xeon CPU E5-2680 v4 2.40GHz.
} 
\label{tab:comp-time}
\end{table}
For the bivariate LGCP models under consideration, there is a significant difference in computation time between the MC and BI methods. The fastest computing time for BI is about 657 minutes for negative correlation and 586 minutes for positive correlation, whereas the computing time for MC remains under 36 minutes across all settings. Even when employing grid search in the MC method, which will be elaborated in the next section, the computation time for the MC estimator remains much faster than that of BI. Please refer to Table \ref{tab:selection-add-Q} in the Appendix for details.

Now, we discuss the effects of the control parameters on the computation time of MC and BI. As expected, for MC, a larger range $R$ results in longer computing times. As WL increases, the computing time of MC also increases due to the increase in the number of observations. However, BI estimation does not seem to be affected by the WL. Regarding the sign of correlation, except for the case of $(\text{WL}, R) = (30, 0.35\text{WL})$, computation time of the MC estimation for the negatively correlated models seems to be slightly faster than those of the positively correlated models, whereas the trend is opposite for the BI.

\subsubsection{\label{sec:opt_control_par}Optimal control parameters for MC}

In an effort to obtain more accurate parameter estimates for MC, we employ a grid search method to select the optimal control parameters $(c, R)$. To do so, we first fix the window $D=[-5,5]^2$ and determine the appropriate size of Monte Carlo samples. Table \ref{tab:selection-add-Q} in the Appendix summarizes the selected optimal control parameters $(c, R)$ based on two independent realizations (Experiment I and II) for each model, along with different numbers of Monte Carlo samples of $\widehat{\Sigma}_n(\btheta_0)$ in (\ref{eq:sigmahat}). Additional information such as the log determinant of $\widehat{\Sigma}_n(\btheta_0)$ and the total computing time can also be found in the same table.

From Table \ref{tab:selection-add-Q} in the Appendix, we observe that using 300 or more Monte Carlo samples leads to consistent selection of optimal control parameters across different realizations. However, with only 100 Monte Carlo samples, the selected optimal parameters from two different experiments lack consistency. This suggests that using 300 Monte Carlo samples is often sufficient for selecting control parameters, providing robust results without unnecessary additional computation.

Next, Table \ref{tab:select_c_max_simu_Q_mu0_diffWL} below presents the optimal parameter $(c, R)$ based on 300 Monte Carlo samples for each model and window size. We observe that the optimal parameter sets $(c, R)$ are consistent across two different realizations (Experiment I and II). For models (M1) and (M4) with WL=10, the optimal control parameters are stable across sign of correlations. However, interestingly, for models (M2) and (M3) with WL=10, these values vary between positive and negative correlated models. As the window length increases (corresponding to WL$ \in \{20,30\}$), the effect of correlation becomes more prominent. For instance, for models (M1), (M3), and (M4) with WL=30, the difference in the selected maximum range $R$ between positively and negatively correlated models exceeds 3 units. Specifically, positively correlated models tend to choose larger $R$ values compared to those of negatively correlated models. Moreover, when WL$\in\{20,30\}$, models with larger absolute cross-correlation ($|\rho|$) tend to choose smaller $c$ values. For example, for models (M3) and (M4), the optimal $c$ is less than or equal to 0.3. This observation may be due to the fact that larger $|\rho|$, which induces more fluctuation in the cross $K$-function, necessitating a smaller $c$ to mitigate this fluctuation.  However, we do not have a theoretical justification of this founding.
\begin{table}[h!]
\small
\begin{center}
\begin{tabular}{c|cccc}
\hline
Model & Correlation & WL=10 & WL=20 & WL=30  \\ \hline \hline
\multirow{2}{*}{(M1)} &  Negative &(0.5, 2.00) &(0.2, 5.00) &(0.5, 2.50)  \\
  &Positive &(0.5, 2.50) &(0.5, 6.00) &(0.5, 7.00)   \\
\hline
\multirow{2}{*}{(M2)} & Negative &(0.5, 3.50) &(0.5, 2.00) &(0.5, 3.25)   \\
  & Positive &(0.4, 2.50) &(0.5, 4.50) &(0.5, 3.25)  \\
\hline
\multirow{2}{*}{(M3)} &  Negative &(0.5, 1.00) &(0.2, 5.50) &(0.2, 6.25)   \\
  & Positive &(0.4, 2.50) &(0.3, 6.50) &(0.1, 10.00)  \\
 \hline
\multirow{2}{*}{(M4)} & Negative &(0.5, 3.50) &(0.3, 7.00) &(0.3, 5.50)  \\
  & Positive &(0.5, 3.50) &(0.1, 5.50) &(0.2, 8.50)  \\
 \hline
\end{tabular}
\caption{
Optimal control parameters $(c,R)$ for different models and different window sizes. Here, we use 300 Monte Carlo samples to estimate the asymptotic covariance matrix. 
} \label{tab:select_c_max_simu_Q_mu0_diffWL} 
\end{center}
\end{table}

\subsubsection{Parameter accuracy} \label{sec:LGCP-accuracy}

Now, we assess the accuracy of both the MC and BI estimators. When implementing the MC method, we consider two variants of the MC estimators: one with the optimal control parameters obtained in Table \ref{tab:select_c_max_simu_Q_mu0_diffWL} (referred to as ``MC\_opt'') and another using fixed control parameters $(c, R) = (0.2, 0.15\text{WL})$, where WL $\in \{10, 20, 30\}$ is the window length (referred to as ``MC\_fix''). This allows us to assess the improvements to the MC approach from using our grid search method for control parameter selection.

\begin{figure}[h!]
    \centering
    \includegraphics[width=0.97\linewidth]{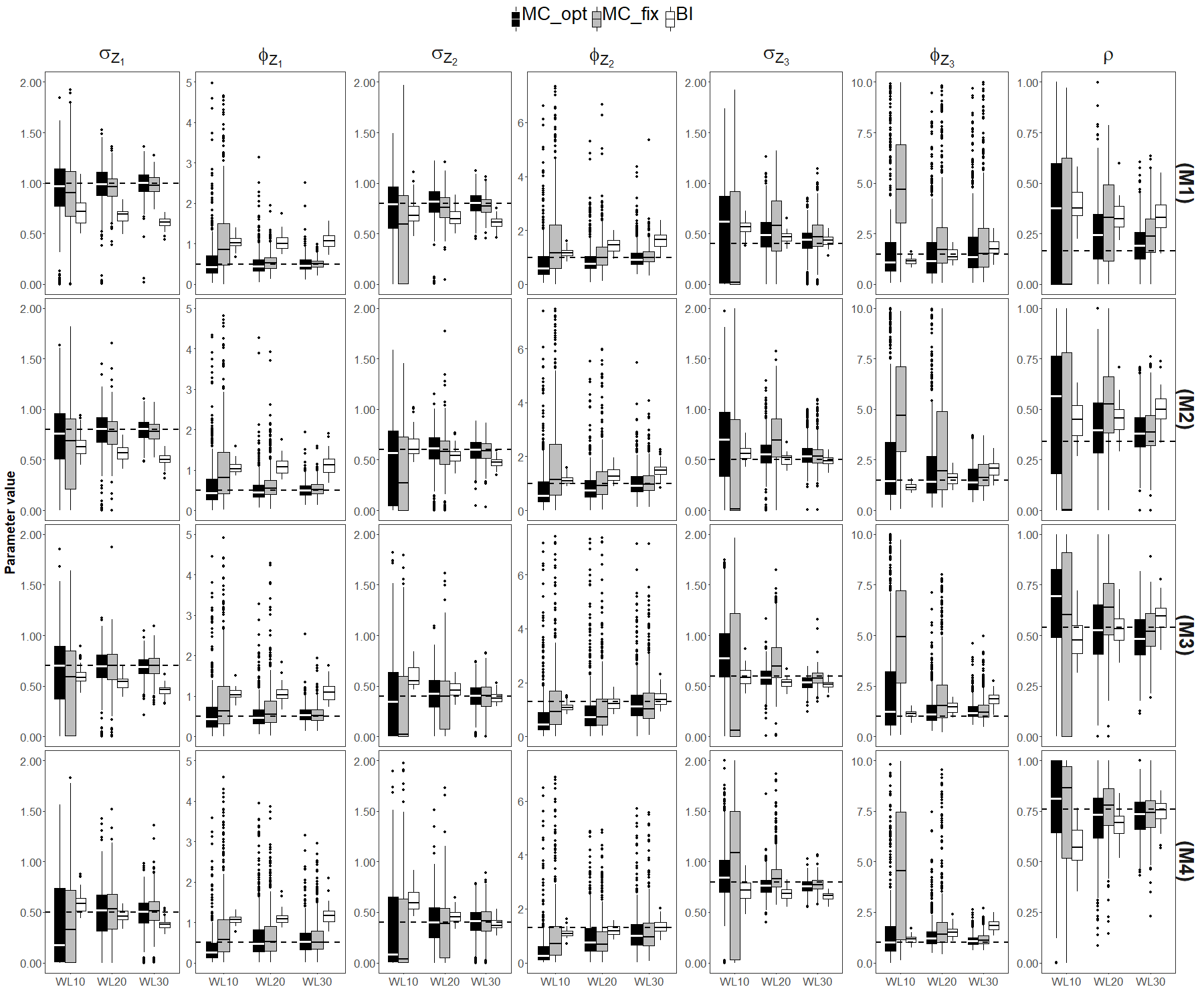}
    \caption{
Boxplot of parameter estimates from MC estimators (``MC\_opt'' and ``MC\_fix'') and BI estimator for positively correlated models (different rows) and different window sizes (``WL10'', ``WL20'' and ``WL30'' in each panel). True parameter values are marked with horizontal dashed lines.}
\label{fig:boxplot_MCM_BI_diffWL_pos}
\end{figure}

Figure \ref{fig:boxplot_MCM_BI_diffWL_pos} displays boxplots of the parameter estimates from the three estimators for positively correlated models. The corresponding boxplots for negatively correlated models can be found in Figure \ref{fig:boxplot_MCM_BI_diffWL_neg} in the Appendix. The mean absolute error (MAE), standard deviation (SD), and root mean squared error (RMSE) of the three estimators are also summarized in Tables \ref{tab:bias-M1}--\ref{tab:bias-M4} in the Appendix.

From Figure \ref{fig:boxplot_MCM_BI_diffWL_pos} (see also Figure \ref{fig:boxplot_MCM_BI_diffWL_neg} in the Appendix), the BI estimator has the smallest SD across all models and window sizes. However, for some parameters (e.g., $\sigma_{Z_1}$ and $\phi_{Z_1}$ in positively correlated models), significant biases are present, resulting in the largest MAD and RMSE for these parameters. This may indicate the difficulty of obtaining reliable estimators for these parameters based on the current computational grids and the length of Markov chains. We mention that the biases of the BI method are lessened when using finer (256 $\times$ 256) computational grids, as compared to the coarser (64 $\times$ 64) grids reported here. However, BI estimation on the finer grids takes approximately five times longer than on the coarse grids, and it still has larger MAD and RMSE values than the MC estimators. 

Moving on, we focus on a variant of the MC estimators. Compared to the MC estimator results using the fixed control parameter, we observe a clear improvement in terms of MAE, SD, and RMSE for all models and window sizes when using the optimal control parameter to calculate the MC estimator. These gains are more significant for smaller sampling windows, as the variance of $\widehat{\btheta}_n$ is $|D_n|^{-1} \Sigma(\btheta_0) + o(|D_n|^{-1})$ as $n \rightarrow \infty$, and smaller $|D_n|$ implies a larger difference in the variances between ``MC\_opt'' and ``MC\_fix''. Considering that obtaining reliable parameter estimates is one of the top priorities in the estimation procedure, implementing the optimal control parameter selection in our MC estimation outweighs its disadvantage of an additional computation time caused by the grid search method.

Lastly, for the smallest window (WL=10), the MC estimators have larger SDs as compared to those of the BI estimator, especially for the scale parameters $\sigma_{Z_i}$. This is because our estimator is constructed based on the nonparametric estimator of the $K$-function matrix, which may require more observed points to obtain reliable estimator results (an average number of points in WL=10 is 100 for each type). We observe that as the WL increases, SDs of all parameters of the MC estimators decrease and become comparable to those of the BI estimator. Furthermore, as window size increases, the RMSE of both MC estimators in all parameter settings tends to converge to zero, supporting the consistency result of the MC estimator as shown in Theorem \ref{thm:normal_c}. For the largest window (WL=30), boxplots of the scale parameters ($\sigma_{Z_i}$'s) exhibit symmetric distributions with less than 3\% of outliers for all models, suggesting the asymptotic Gaussianity of the MC estimator. However, when evaluating the range parameters ($\phi_{Z_i}$'s), even for the large sampling window, the distributions are slightly right-skewed with the presence of up to 5\% of outliers. This may indicate an overestimation of range parameters, especially for those with large true range parameter values.


\subsection{Utilizing the MC method for multivariate PSNCP model} \label{sec:PSNCP}

In addition to the bivariate LGCP models above, we also implement our MC method to the multivariate PSNCP model which offers a flexible approach to modeling spatial point patterns characterized by clustered intra-specific interactions alongside positive or negative inter-specific interactions. Specifically, we consider the five-variate PSNCP model $\underline{X}=(X_1,\ldots,X_5)$ on $\mathbb{R}^2$, driven by the latent intensity field $\underline{\Lambda}(\bx) = (\Lambda_1(\bx), \ldots, \Lambda_5(\bx))$, where each $\Lambda_i(\bx)$ satisfies $\Lambda_i(\bx) = S_i(\bx)F_i(\bx)$ for $\bx \in \mathbb{R}^2$ and $i \in \{1, \ldots, 5\}$. Here, the shot-noise fields $S_i(\bx)$ are given by $S_i(\bx) = \kappa_i^{-1} \sum_{\by \in \Phi_i} k(\|\bx - \by\|)$, where $\Phi_i$ is a homogeneous Poisson point process with intensity $\kappa_i \in (0,\infty)$, and $k(r)$ is the Gaussian kernel function defined as
\begin{equation}\label{eq:g_kernel}
k(r)=\frac{1}{2\pi \omega^2}\exp(-r^2/(2\omega^2)), \quad r \in \mathbb{R}.
\end{equation} 
$S_i(\cdot)$ accounts for the clustering within the $i$-th point process. Next, the compound fields in PSNCP model are given by
\begin{equation}\label{eq:compound_field}
    F_i(\bx) = \exp\left(\sum_{\ell \neq i} \dfrac{\kappa_l\xi_{li}}{k(0)} \right) ~~ \prod_{l\neq i} \prod_{\by\in \Phi_l} \left\{ 1+\xi_{li}\widetilde{k}(\|\bx-\by\|)\right\}, \quad i \in \{1, \dots, 5\}.
\end{equation}
Here, $\xi_{li} \in (-1,\infty)$ and $\widetilde{k}(r) = k(r)/ k(0)$, with $k(r)$ defined as in (\ref{eq:g_kernel}). The coefficients $\{\xi_{ij}\}$ in (\ref{eq:compound_field}) indicate whether $X_i$ is clustered around ($\xi_{ij} > 0$) or repelled by ($\xi_{ij} < 0$) the latent process $\Phi_j$. The compound field $F_i(\cdot)$ captures the combined effect of all other processes on the $i$-th process. We note that the analytic expressions for the marginal and cross PCFs are given in \cite{p:jal-15}, Equation (10). Thus, straightforward computation of $K$- and $Q$-function matrices are available for multivariate PSNCP.

For the true data-generating process, we assume that the first-order intensities $\lambda_1^{(i)}$ are known and equal to one for all $i$. Furthermore, for the interest of parsimony, we use common parameters $\kappa_i = \kappa = 0.2$ and $\omega_i^2 = \eta = 0.25$ in the shot-noise fields $S_i(\bx)$ and sparse interaction matrix $\{\xi_{ij}\}$:
\begin{equation*}
\xi_{12}=\xi_{21}= \alpha_1 = 0.7, \quad
\xi_{1,5}=\xi_{5,1}=\alpha_2 = -0.8, \quad 
\xi_{3,5}=\xi_{5,3}= \alpha_3 = 0.3, \quad
\xi_{45}=\xi_{54}= \alpha_4 = 0.5,
\end{equation*} 
and $\xi_{i,j}=0$ otherwise. Therefore, the parameters of interest are $\btheta=(\kappa,\eta,\alpha_1,\alpha_2,\alpha_3,\alpha_4)$.

Lastly, in simulations, we generate 500 five-variate spatial point patterns from the above model and calculate two parameter estimators using (1) the MC method and (2) the weighted composite likelihood (WCL) method as considered in \cite{p:jal-15}.

\subsection{Results}

Analogous to Section \ref{sec:opt_control_par}, we use 300 Monte Carlo samples to select the optimal control parameters of the MC estimator. The selected optimal control parameters are $(c,R)=(0.3,1.75)$ for $D=[-5,5]^2$; $(c,R)=(0.5,1.00)$ for $D=[-10,10]^2$; and $(c,R)=(0.5,1.00)$ for $D=[-15,15]^2$.

\begin{figure}[h!]
    \centering
    \includegraphics[width=0.97\linewidth]{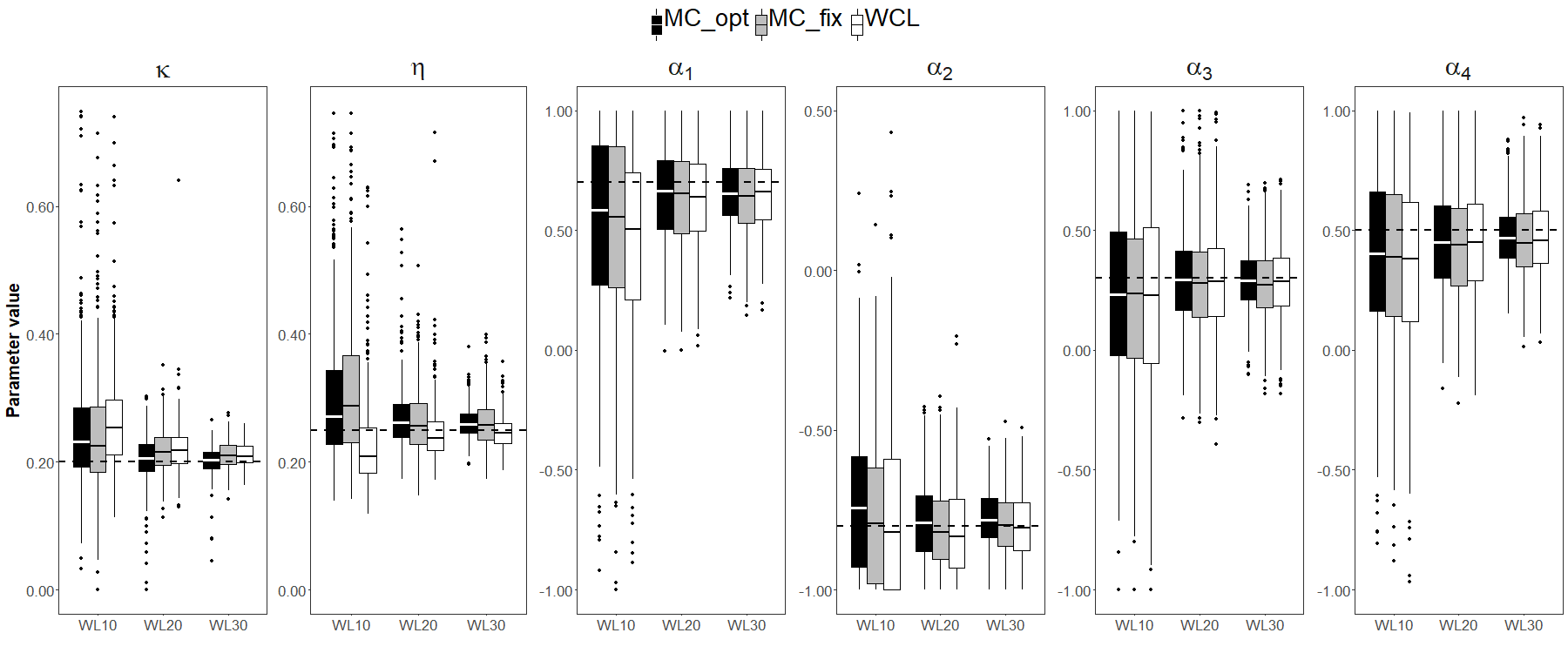}
    \caption{
Boxplot of parameter estimates from MC estimators (``MC\_opt'' and ``MC\_fix'') and WCL estimator for different window sizes (``WL10'', ``WL20'' and ``WL30'' in each panel). True parameter values are marked with horizontal dashed lines. }
\label{fig:boxplot_MCM_WCL_PSNCP_diffWL_pos}
\end{figure}

In Figure \ref{fig:boxplot_MCM_WCL_PSNCP_diffWL_pos}, we present boxplots of the parameter estimates from the 500 simulations. For the MC estimator, we consider both the optimal control parameters (referred to as ''MC\_opt'') and fixed control parameters $(c,R) = (0.2, 0.15\text{WL})$ (referred to as ''MC\_fix''). Results for the computation time and evaluation metrics (MAE, SD, and RMSE) for the MC and WCL estimators are reported in Table \ref{tab:comp-time-est-PSNCP} in the Appendix.

Regarding computational time, we observe that for the five-variate PSNCP model under consideration, our MC method (both ''MC\_opt'' and ''MC\_fix'') is faster than the WCL method across all window sizes. As discussed in Section \ref{sec:compute}, the computing time of MC increases as the window size increases, while the time required for the WCL estimator is not impacted by the sampling window. 

Turning to parameter estimation of the MC (either type) and WCL estimators, we observe that both perform quite well across all window sizes. Specifically, the biases and standard errors of both estimators tend to zero as WL increases, and for a large window size (WL=30), boxplots are symmetric and centered around the true values. These results indicate that both the MC and WCL estimators yield satisfactory large sample properties. Now, we compare their accuracies. As discussed in Section \ref{sec:LGCP-accuracy}, the performance of ''MC\_opt'' is superior to ''MC\_fix'' for all parameters and windows, although the distinction between the two MC estimators is not as clear as in the bivariate LGCP models considered in Section \ref{sec:sim-model}. Interestingly, for parameters $\kappa$ and $\eta$ that correpond to the shot-noise fields $S_{i}(\cdot)$, our estimator (both ''MC\_opt'' and ''MC\_fix'') has smaller RMSE than the WCL estimator across all window sizes. However, the WCL method performs better (in terms of RMSE) for the parameters in the interaction matrix ($\alpha_1, \dots, \alpha_4$).


\section{Concluding remarks and discussions} \label{sec:conclusion} 

In this article, we propose a new inferential method for multivariate stationary spatial point processes by minimizing the contrast (MC) between the matrix-valued scaled $K$-function and its nonparametric edge-corrected estimator. When the model is correctly specified, the resulting MC estimator has satisfactory large sample properties. These enable us to conduct various statistical inferences on multivariate spatial point processes, such as tests for the homogeneity of multivariate spatial point patterns, as discussed in Remark \ref{rmk:two-sample}. Moreover, the proposed method is computationally efficient, and the form of the asymptotic covariance matrix of the MC estimator provides insight into the selection of optimal control parameters in the discrepancy measure.

From the results of our simulations, we believe that our method could serve as a valuable alternative to the BI or WCL methods for analyzing multivariate spatial point processes. The significantly faster computing speed is the chief advantage of our method, enabling researchers to obtain initial values, analyze large samples, and evaluate numerous complex point process models efficiently. Moreover, it is intriguing that under certain parameter settings, our estimator outperforms those in the BI or WCL methods. Currently, we lack a theoretical basis for explaining the relative efficiency of these estimators, but this appears to be a good avenue for future research.

However, implementing our estimator for large-dimensional multivariate models requires some caution. Note that the ``full'' parameters in the $m$-variate PSNCP model as in Section \ref{sec:PSNCP} increase with order $O(m^2)$. Therefore, even for $m=5$, numerical calculation of the MC estimator for the full model may not be feasible without regularization methods. This limitation also applies to the BI and WCL methods when analyzing large-dimensional multivariate models. Indeed, we observe that both MC and WCL estimators become highly unstable when the number of parameters in the five-variate PSNCP exceeds 10. In such situations, it may be necessary to consider regularization methods as discussed in \cite{p:cho-20, p:hes-22}. Details of the regularized MC method for multivariate spatial point processes will be explored in future research.

Lastly, we discuss two possible extensions of our study. Firstly, the emphasis of this paper is on using the scaled $K$-function matrix to construct a discrepancy function. However, we believe that similar arguments as in Section \ref{sec:asymp} (also, Appendices \ref{sec:sigma} and \ref{sec:proof}) can be applied to derive the asymptotic normality of the MC estimator based on the PCF matrix $G(\bx) = [g_{i,j}(\bx)]_{i,j=1}^{m}$. Please see \cite{p:bis-22}, Section 4.1 for some technical results when implementing the PCF. Secondly, in practical scenarios, the joint stationarity assumption is often too stringent. Therefore, we may relax this assumption and consider the MC estimator for second-order intensity reweighted stationary 
\citep[SOIRS;][]{p:bad-00} processes. Indeed, \cite{p:was-16} proposed a least squares estimation of the multivariate SOIRS LGCP process. See also \cite{p:chu-22} for the quasi-likelihood approach to fitting multivariate SOIRS processes and \cite{p:gua-09, p:waa-09} for theoretical developments.

\section*{Acknowledgement}
JY's research was supported from the National Science and Technology Council, Taiwan (grant 110-2118-M-001-014-MY3). MJ and SC acknowledge support by NSF DMS-1925119 and DMS-2123247. MJ also acknowledges support by NIH P42ES027704. The authors also wish to thank the three anonymous referees and editors for their valuable comments and corrections, which have greatly improved the article in all aspects.

\appendix
\counterwithin{figure}{section}
\counterwithin{table}{section}
\counterwithin{equation}{section}


\section{A comparison between the two $Q$-function estimators} \label{sec:comparison}

Recall (\ref{eq:empirical_K0}) and (\ref{eq:empirical_K}). For $i,j \in \{1, \dots,m\}$, we define the two marginal ($i=j$) and cross ($i\neq j$) $Q$-function estimators as
\begin{equation}\label{eq:Q0}
    [\widehat{\bQ}_{0n}(r)]_{i,j}=|D_n|^{-1} \sum_{\bx \in X_i} \sum_{\by \in X_j} \textbf{1}_{\{0<\|\bx-\by\|\leq r\}}
\end{equation} and
\begin{equation}\label{eq:Qedge}
    [\widehat{\bQ}_{n}(r)]_{i,j}= |D_{n}|^{-1} \sum_{\bx \in X_i} \sum_{\by \in X_j} b(\bx,\by) \textbf{1}_{\{0<\|\bx-\by\|\leq r\}}, \quad r\geq 0, \quad n\in \mathbb{N},
\end{equation} where $b(\cdot,\cdot)$ is an edge-correction factor.
The following theorem addresses the first and second moment bounds for $[\widehat{\bQ}_{0n}(r)]_{i,j}$ and $[\widehat{\bQ}_{n}(r)]_{i,j}$. The proof technique is almost identical to that in \cite{p:bis-22}, Theorems 3.5 and 4.1, so we omit the details.

\begin{theorem} \label{theorem:edge}
Let $\underline{X} = (X_1, \dots, X_m)$ be a multivariate stationary point process that satisfies Assumption \ref{assump:1}(i) (for $\ell=4$). Moreover, we assume that the increasing sequence of sampling windows $\{D_n\}$ in $\mathbb{R}^d$ is c.a.w., and the edge-correction factor is such that $b \in \{b_1, b_2\}$ (or $b \in \{b_1, b_2, b_3\}$, assuming $\underline{X}$ is isotropic). Then, for $i,j \in \{1, \dots, m\}$, the following three assertions hold:
\begin{eqnarray} 
&& \Ex[ [\widehat{\bQ}_{0n}(r)]_{i,j} - [\widehat{\bQ}_{n}(r)]_{i,j}] = O(n^{-1}), \label{eq:edge-compare1} \\
&& 
\var\{ [\widehat{\bQ}_{0n}(r)]_{i,j} -[\widehat{\bQ}_{n}(r)]_{i,j}\} = O(n^{-1}|D_n|^{-1}) = O(n^{-d-1}),  \label{eq:edge-compare2} \\
&& 
\var\{[\widehat{\bQ}_{n}(r)]_{i,j}\} =  O(|D_n|^{-1}) = O(n^{-d}).
 \label{eq:edge-compare3}
\end{eqnarray}
\end{theorem}
 As a corollary, we show that the asymptotic covariance matrix of $\text{vec}(\widehat{\bG}_{n}(h))$ as in (\ref{eq:vecGn}) is equal to that of a vectorization of non-edge corrected counterparts
\begin{equation} \label{eq:G0n} 
[\widehat{\bG}_{0n}(h)]_{i,j} =
|D_n|^{1/2} \left\{ [\widehat{\bQ}_{0n}(h)]_{i,j} -\Ex [\widehat{\bQ}_{0n}(h)]_{i,j} \right\},
 \quad i,j \in \{1, \dots,m\}, \quad h \geq 0.
\end{equation}

\begin{corollary} \label{coro:edge}
Suppose the same set of Assumptions and notation as in Theorem \ref{theorem:edge} holds. For $i,j \in \{1, \dots,m\}$, let $[\widehat{\bG}_{n}(h)]_{i,j}$ and $[\widehat{\bG}_{0n}(h)]_{i,j}$ are defined as in (\ref{eq:Gn2}) and (\ref{eq:G0n}), respectively. Then,
\begin{equation}
\lim_{n\rightarrow \infty} \left| \cov \left\{
[\widehat{\bG}_{n}(h)]_{a,b} , [\widehat{\bG}_{n}(h)]_{c,d}
 \right\} - \cov \left\{
[\widehat{\bG}_{0n}(h)]_{a,b} , [\widehat{\bG}_{0n}(h)]_{c,d}
 \right\}\right| = 0, \quad a,b,c,d \in \{1,\dots,m\}.
\end{equation}
\end{corollary}

\noindent \textit{Proof}. By triangular inequality and Cauchy-Schwarz inequality, we have
\begin{eqnarray}
&& |\cov (X_1, Y_1) - \cov(X_2, Y_2)| \nonumber \\
&&\quad \leq
|\cov (X_1-X_2, Y_1)| +|\cov (X_2, Y_1-Y_2)| \nonumber \\
&& \quad \leq
 \{ \var (X_1-X_2)\}^{1/2}   \{ \var (Y_1)\}^{1/2} +  \{ \var (Y_1-Y_2)\}^{1/2}  \{ \var (X_2)\}^{1/2}.
\label{eq:cauchy1}
\end{eqnarray} Let $X_1 = [\widehat{\bG}_{n}(h)]_{a,b}$, $X_2 = [\widehat{\bG}_{0n}(h)]_{a,b}$, 
$Y_1 = [\widehat{\bG}_{n}(h)]_{c,d}$, and
$Y_2 = [\widehat{\bG}_{0n}(h)]_{c,d}$. Then, from (\ref{eq:edge-compare2}), we have
\begin{equation} \label{eq:cov1}
\var (X_1-X_2) = |D_n| \var\{[\widehat{\bQ}_{0n}(h)]_{a,b} - [\widehat{\bQ}_{n}(h)]_{a,b}\} = O(n^{-1}), \quad n\rightarrow \infty.
\end{equation} Similarly, $\var (Y_1-Y_2) = O(n^{-1})$ as $n\rightarrow \infty$. Moreover, from (\ref{eq:edge-compare2}) and (\ref{eq:edge-compare3}), we have
\begin{equation} \label{eq:cov2}
\var X_{2} \leq 2( \var (X_2-X_1) + \var(X_1)) = O(1), \quad n\rightarrow \infty.
\end{equation} Similarly, we have $\var Y_1 = O(1)$ as $n\rightarrow \infty$. Substitute (\ref{eq:cov1}) and (\ref{eq:cov2}) into (\ref{eq:cauchy1}) gives
\begin{eqnarray*}
\left|\cov \{
[\widehat{\bG}_{n}(h)]_{a,b} , [\widehat{\bG}_{n}(h)]_{c,d}
 - 
\cov \{
[\widehat{\bG}_{0n}(h)]_{a,b} , [\widehat{\bG}_{0n}(h)]_{c,d}
\right| = O(n^{-1/2}), \quad a,b,c,d \in \{1,\dots,m\}
\end{eqnarray*} as $n \rightarrow \infty$. Thus, we get the desired results.
\hfill $\Box$

\section{Expression for the asymptotic covariance matrix}  \label{sec:sigma}

In this section, we provide an expression for the asymptotic covariance matrix of $\text{vec}(\widehat{\bG}_{n}(h))$ in terms of the joint intensity functions of the underlying point process $\underline{X}$. Let
\begin{equation} \label{eq:sigma2}
\sigma^2_{(a,b:c,d)} (h) = 
\lim_{n \rightarrow \infty}  \cov  \left\{ [\widehat{\bG}_n(h)]_{a,b}, [\widehat{\bG}_n(h)]_{c,d} \right\}.
\quad a,b,c,d \in \{1, \dots,m\},
\end{equation} 
Due to the asymmetry of the edge-correction, $\widehat{\bG}_{n}(h)$ may not be symmetric. Thus, calculations of $\sigma^2_{(a,b:c,d)} (h)$ are cumbersome. As a remedy, for $i,j \in \{1, \dots, m\}$, let
$[\widehat{\bQ}_{0n}(h)]_{i,j}$ be a nonparametric estimator of $[\bQ(h;\btheta_0)]_{i,j}$, but without the edge-correction as in (\ref{eq:Q0}). Let
\begin{equation} \label{eq:sigmatilde2}
\eta^2_{(a,b:c,d)} (h) = 
\lim_{n \rightarrow \infty}  \cov  \left\{ [\widehat{\bG}_{0n}(h)]_{a,b}, [\widehat{\bG}_{0n}(h)]_{c,d} \right\},
\quad a,b,c,d \in \{1, \dots,m\},
\end{equation} where $[\widehat{\bG}_{0n}(h)]_{i,j}$ is an empirical process of $[\bQ(h;\btheta_0)]_{i,j}$ defined as in (\ref{eq:G0n}). Then, in Corollary \ref{coro:edge}, we show
\begin{equation} \label{eq:sigma-identity}
\sigma^2_{(a,b:c,d)} (h) = \eta^2_{(a,b:c,d)} (h), \quad a,b,c,d \in \{1, \dots,m\}.
\end{equation}
One advantage of using $\eta^2_{(a,b:c,d)} (h)$ over $\sigma^2_{(a,b:c,d)} (h)$ is that $\widehat{\bG}_{0n}(h)$ is symmetric. Therefore, the number of different cases to consider in the expression $\eta^2_{(a,b:c,d)} (h)$ is significantly reduced. Below, we provide the complete list of expressions for $\sigma^2_{(a,b:c,d)} (h)$.
\begin{theorem} \label{thm:asymp-var}
Let $\underline{X} = (X_1, \dots, X_m)$ be a multivariate stationary point process that satisfies Assumption \ref{assump:1}(i) (for $\ell=4$). Moreover, we assume that the increasing sequence of sampling windows $\{D_n\}$ in $\mathbb{R}^d$ is c.a.w. and the edge-correction factor is such that $b \in \{b_1, b_2\}$ (or $b \in \{b_1, b_2, b_3\}$, assuming $\underline{X}$ is isotropic). Then, $\sigma^2_{(a,b:c,d)} (h)$ is well-defined for all $a, b, c, d \in \{1, \dots, m\}$ and we have
\begin{equation} \label{eq:sigma-sym}
\sigma^2_{(a,b:c,d)} (h) = \sigma^2_{(b,a:c,d)} (h) = \sigma^2_{(a,b:d,c)} (h) = \sigma^2_{(b,a:d,c)} (h),
\quad a,b,c,d \in \{1, \dots,m\}.
\end{equation}
Let $i,j,k,\ell$ be the distinct indices of $\{1, \dots,m\}$ (If $m <4$, then, we select at most $m$ distinct indices) and $I()$ be the indicator function.
Then, using (\ref{eq:sigma-sym}), we have seven distinct expressions for $\sigma^2_{(a,b:c,d)} (h)$, which we will list below.

\noindent {\bf [case1]}: $(a,b)=(c,d)=(i,i)$;
\begin{eqnarray*}
&& \sigma^2_{(i,i:i,i)} (h) \\
&&\quad = \iiint I(\|\bu_{1}\|\leq h) I(\|\bu_{3}\|\leq h) \left\{ \lambda^{(i)}_{4,red} (\bu_{1},\bu_{2},\bu_{2}+\bu_{3}) - \lambda^{(i)}_{2,red} (\bu_{1})\lambda^{(i)}_{2,red}(\bu_{3}) \right\} 
d\bu_{1}d\bu_{2}d\bu_{3} \\
&&\quad + 4\iint I(\|\bu_{1}\|\leq h) I(\|\bu_{2}\|\leq h) \lambda^{(i)}_{3,red} (\bu_{1},\bu_{2}) d\bu_{1}d\bu_{2} +
2 \int I(\|\bu_{1}\|\leq h) \lambda^{(i)}_{2,red} (\bu_{1}) d\bu_{1}.
\end{eqnarray*}

\noindent {\bf [case2]}: $(a,b)=(c,d)=(i,j)$;
\begin{eqnarray*}
&& \sigma^2_{(i,j:i,j)} (h)\\
&&\quad = \iiint I(\|\bu_{1}\|\leq h) I(\|\bu_{3}\|\leq h) \left\{ \lambda^{(i,j)}_{2,2} 
(\bu_{2},\bu_{1},\bu_{2}+\bu_{3}) - \lambda^{(i,j)}_{1,1,red} (\bu_{1})\lambda^{(i,j)}_{1,1,red}(\bu_{3}) \right\} d\bu_{1}d\bu_{2}d\bu_{3}\\
&&\quad + \iint I(\|\bu_{1}\|\leq h) I(\|\bu_{2}\|\leq h) \left\{ \lambda^{(i,j)}_{2,1,red} (\bu_{1},\bu_1 + \bu_{2}) + \lambda^{(j,i)}_{1,2,red} (\bu_{1},\bu_{2}) \right\} d\bu_{1}d\bu_{2} \\
&&\quad +  \int I(\|\bu_{1}\|\leq h) \lambda^{(i,j)}_{1,1,red} (\bu_{1}) d\bu_{1}.
\end{eqnarray*}

\noindent {\bf [case3]}: $(a,b)=(i,i), (c,d) = (j,j)$;
\begin{eqnarray*}
&& \sigma^2_{(i,i:j,j)} (h) \\
&& \quad  = \iiint I(\|\bu_{1}\|\leq h) I(\|\bu_{3}\|\leq h) \left\{ \lambda^{(i,j)}_{2,2,red} (\bu_{1},\bu_{2},\bu_{2}+\bu_{3}) - \lambda^{(i,j)}_{1,1,red} (\bu_{1})\lambda^{(i,j)}_{1,1,red}(\bu_{3}) \right\} 
d\bu_{1}d\bu_{2}d\bu_{3}
\end{eqnarray*}

\noindent {\bf [case4]}: $(a,b)=(i,i), (c,d) = (i,j)$;
\begin{eqnarray*}
&& \sigma^2_{(i,i:i,j)} (h) \\
&& \quad =  \iiint I(\|\bu_{1}\|\leq h) I(\|\bu_{3}\|\leq h) \left\{ \lambda^{(i,j)}_{3,1,red} 
(\bu_{1},\bu_{2},\bu_{2}+\bu_{3}) - \lambda^{(i)}_{2,red} (\bu_{1})\lambda^{(i,j)}_{1,1,red}(\bu_{3}) \right\} d\bu_{1}d\bu_{2}d\bu_{3}\\
&&\quad + \iint I(\|\bu_{1}\|\leq r) I(\|\bu_{2}\|\leq h) \left\{
\lambda^{(i,j)}_{2,1,red} (\bu_{1},\bu_{1}+\bu_{2})
+\lambda^{(i,j)}_{2,1,red} (\bu_{1},\bu_{2})\right\}
 d\bu_{1}d\bu_{2}.
\end{eqnarray*}

\noindent {\bf [case5]}: $(a,b)=(i,i), (c,d)=(j,k)$;
\begin{eqnarray*}
&&\sigma^2_{(i,i:j,k)} (h) \\
&&\quad = \iiint I(\|\bu_{1}\|\leq h) I(\|\bu_{3}\|\leq h) \left\{ \lambda^{(i,j,k)}_{2,1,1,red} 
(\bu_{1},\bu_{2},\bu_{2}+\bu_{3}) - \lambda^{(i)}_{2,red} (\bu_{1})\lambda^{(j,k)}_{1,1,red}(\bu_{3}) \right\} d\bu_{1}d\bu_{2}d\bu_{3}
\end{eqnarray*}

\noindent {\bf [case6]}: $(a,b)=(i,j), (c,d)=(i,k)$;
\begin{eqnarray*}
&&\sigma^2_{(i,j:i,k)} (h) \\
&&\quad = \iiint I(\|\bu_{1}\|\leq h) I(\|\bu_{3}\|\leq h) \left\{ \lambda^{\{i,j,k\}}_{2,1,1,red} 
(\bu_{2},\bu_{1},\bu_{2}+\bu_3) - \lambda^{(i,j)}_{1,1,red} (\bu_{1})\lambda^{(i,k)}_{1,1,red}(\bu_{3}) \right\} d\bu_{1}d\bu_{2}d\bu_{3} \\
&&\quad + \iint I(\|\bu_{1}\|\leq h) I(\|\bu_{2}\|\leq h) \lambda^{(i,j,k)}_{1,1,1,red} (\bu_{1},\bu_{2}) d\bu_{1}d\bu_{2}.
\end{eqnarray*}

\noindent {\bf [case7]}: $(a,b)=(i,j), (c,d)=(k,\ell)$;
\begin{eqnarray*}
&& \sigma^2_{(i,j:i,k)} (h) \\
&&\quad =\iiint I(\|\bu_{1}\|\leq h) I(\|\bu_{3}\|\leq h) \left\{ \lambda^{(i,j,k,\ell)}_{1,1,1,1,red} 
(\bu_{1},\bu_{2},\bu_{2}+\bu_{3}) - \lambda^{(i,j)}_{1,1} (\bu_{1})\lambda^{(k,\ell)}_{1,1,red}(\bu_{3}) \right\} d\bu_{1}d\bu_{2}d\bu_{3}.
\end{eqnarray*}

\end{theorem}

\noindent \textit{Proof}. 
Under (\ref{eq:caw}) and Assumption \ref{assump:1}, for $i,j \in \{1, \dots, m\}$, it is straightforward that the limit of $\var [\widehat{\bG}_n(h)]_{i,j}$ finitely exists. Therefore, using the identity $\cov(X, Y) = \frac{1}{2} \left( \var \{X + Y\} - \var X - \var Y \right)$, $\sigma^2_{(a,b:c,d)} (h)$ also exists for all $a, b, c, d \in \{1, \dots, m\}$. Showing (\ref{eq:sigma-sym}) is a direct consequence of (\ref{eq:sigma-identity}) and the symmetry of $\widehat{\bG}_{0n}(h)$.

Next, due to (\ref{eq:sigma-identity}), it is enough to calculate $\eta^2_{(a,b:c,d)} (h)$ instead of $\sigma^2_{(a,b:c,d)} (h)$. As indexed by {\bf [case 1]}--{\bf [case 7]} in the statement of the theorem, there are seven different expressions for $\eta^2_{(a,b:c,d)} (h)$. Since {\bf [case 1]} is shown in \cite{p:gua-07}, Appendix B, we only show {\bf [case 2]}. The cases {\bf [case 3]}--{\bf [case 7]} can be derived in a similar way. Let $\phi(\bx, \by) = I(\|\bx - \by\| \leq h)$, and we simplify the notation $\bx \in X_{i}$ to $\bx \in [i]$ for $i \in \{1, \dots, m\}$. Using (\ref{eq:sigma2}) and (\ref{eq:Q0}), for $i \neq j$, we have
\begin{eqnarray*}
\cov  \left\{ [\widehat{\bG}_{0n}(h)]_{i,j}, [\widehat{\bG}_{0n}(h)]_{i,j} \right\}
  &=& |D_{n}|^{-1}   \cov  \left\{ \sum_{\bx \in [i], \by \in [j]} \phi(\bx,\by), 
\sum_{\bz \in [i], \bw \in [j]} \phi(\bz,\bw)
\right\} \\
&=&|D_{n}|^{-1} \Ex \left[ \sum_{\bx, \bz \in [i]; ~ \by, \bw \in [j]}
\phi(\bx,\by)\phi(\bz,\bw) \right] \\
&& -  |D_{n}|^{-1}  \Ex\left[ \sum_{\bx \in [i], ~ \by \in [j]} \phi(\bx,\by)] \right] \Ex \left[  \sum_{\bz \in [i], ~ \bw \in [j]} \phi(\bz,\bw) \right].
\end{eqnarray*}
Since $\underline{X}$ is simple, $P\left( \bx \in [i] ~~ \&~~ \bx \in [j] \right) = 0$. Therefore, the above can be decomposed as
\begin{equation} \label{eq:case2-decom}
\cov  \left\{ [\widehat{\bG}_{0n}(h)]_{i,j}, [\widehat{\bG}_{0n}(h)]_{i,j} \right\}
= A_{1} +A_{2} +A_{3} +A_{4},
\end{equation} where
\begin{eqnarray*}
&& A_{1} = |D_{n}|^{-1} \Ex \left[ \sum_{\bx\in [i], \by\in [j]} \phi(\bx,\by)^2 \right], \quad
A_{2} = |D_{n}|^{-1}  \Ex \left[ \sum_{\bx\neq \bz\in [i], \by\in [j]}\phi(\bx,\by)\phi(\bz,\by) \right], \\
&& A_{3} = |D_{n}|^{-1} \Ex \left[ \sum_{\bx\in [i], \by \neq \bw\in [j]} \phi(\bx,\by)\phi(\bx,\bw)\right], \quad \text{and} \\
&& A_{4} = |D_{n}|^{-1} \left\{ \Ex \left[ \sum_{\bx\neq \bz\in [i], \by\neq \bw\in [j]} \phi(\bx,\by)\phi(\bz,\bw) \right]
- \Ex\left[ \sum_{\bx \in [i], ~ \by \in [j]} \phi(\bx,\by)] \right] \Ex \left[  \sum_{\bz \in [i], ~ \bw \in [j]} \phi(\bz,\bw) \right]
\right\}.
\end{eqnarray*} 
To represent each term above in integral form, we use the celebrated Campbell's formula which states that
\begin{equation} \label{eq:joint-intensity}
\Ex \left[  \sum_{\bx_{1,1}, \dots, \bx_{1,n_1} \in X_1}^{\neq} \cdots \sum_{\bx_{m,1}, \dots, \bx_{m, n_m} \in X_m }^{\neq} 
g(\underline{\bx}) \right] =
\int_{D^{N}} g(\underline{\bx}) \lambda_{\underline{n}}^{} (\underline{\bx}) \mu(d\underline{\bx})
\end{equation} 
For any non-negative measurable function $g: D^{N} \mapsto \mathbb{R}$ ($N = n_1 + \dots + n_m$), where $\sum_{\bx_{1}, \cdots, \bx_{n} \in X}^{\neq}$ is the sum over the $n$ pairwise distinct points $\bx_{1}, \cdots, \bx_{n}$ in $X$ and $\mu(d\underline{\bx})$ is a Lebesgue measure on $\mathbb{R}^N$. Using (\ref{eq:joint-intensity}) and since $\phi(\bx, \by)^2 = \phi(\bx, \by)$, $A_1$ can be written as
\begin{equation*}
A_{1} =|D_{n}|^{-1} \Ex \left[ \sum_{\bx\in [i], \by\in [j]} \phi(\bx,\by)^2 \right] = 
|D_{n}|^{-1} \iint_{D_{n}^2} I(\|\bx-\by\|\leq h) \lambda^{(i,j)}_{1,1} (\bx,\by) d\bx d\by.
\end{equation*} 
Using change of variables: $\bu_{1} = \by-\bx$ and $\bu_2 = \bx$ and since $\lambda^{(i,j)}_{1,1} (\bx,\by) = 
\lambda^{(i,j)}_{1,1,red} (\by-\bx)$,
\begin{eqnarray*}
A_{1} &=& |D_{n}|^{-1} \int_{D_{n}} \left( \int_{D_{n}-\bu_2} I(\|\bu_{1}\| \leq h) \lambda^{(i,j)}_{1,1,red} (\bu_{1}) d\bu_1 \right) d\bu_{2} \\
&=& |D_{n}|^{-1} \int_{D_{0n}} (\sim)  d\bu_{2} + |D_{n}|^{-1} \int_{E_n} (\sim) d\bu_{2},
\end{eqnarray*}
where $D_n - \bu_2 = \{\bu_1 - \bu_2 : \bu_1 \in D_n\}$, $D_{0n}$ is an inner window of $D_n$ of depth $h$ as in (\ref{eq:D0}), and $E_n = D_{0n}^{\text{c}} \cap D_n$.

Next, we calculate each term in $A_1$. When $\bu_2 \in D_{0n}$, then $B_d(\bu_2, h) \subset D_n$. Thus,
\begin{equation*}
\int_{D_{n}-\bu_2} I( \|\bu_{1}\| \leq h) \lambda^{(i,j)}_{1,1,red} (\bu_{1}) d\bu_1
= \int_{\mathbb{R}^d} I(\|\bu_{1}\| \leq h) \lambda^{(i,j)}_{1,1,red} (\bu_{1}) d\bu_1.
\end{equation*} 
Therefore, the first integral in $A_1$ is
\begin{equation} 
|D_{n}|^{-1} \int_{D_{0n}} \int_{D_{n}-\bu_2} I(\|\bu_{1}\| \leq h) \lambda^{(i,j)}_{1,1,red} (\bu_{1})  d\bu_1 d\bu_2  = \frac{|D_{0n}|}{|D_n|} \int_{} I(\|\bu_{1}\| \leq h) \lambda^{(i,j)}_{1,1,red} (\bu_{1}) d\bu_1.
\label{eq:A11}
\end{equation} The second integral in $A_2$ is bounded with
\begin{eqnarray} 
&& |D_{n}|^{-1} \left| \int_{E_n} \int_{D_{n}-\bu_2} I(\|\bu_{1}\| \leq h) \lambda^{(i,j)}_{1,1,red} (\bu_{1})  d\bu_1  d\bu_{2} \right| \nonumber \\
&& \quad \leq |D_{n}|^{-1} \sup_{\|\bu\|\leq h} |\lambda^{(i,j)}_{1,1,red} (\bu_{})|  \int_{E_n} \int_{} I(\|\bu_{1}\| \leq h) \bu_1 \bu_2 = O\left( h^{d} \frac{|E_n|}{|D_n|}\right) = O(n^{-1})
\label{eq:A12}
\end{eqnarray} as $n\rightarrow \infty$,
 where the first identity is due to Assumption \ref{assump:1}(i) and the second identity is due to (\ref{eq:caw}). Combining (\ref{eq:A11}), (\ref{eq:A12}), and using that $\lim_{n \rightarrow \infty} |D_{0n}|/|D_n| =1$ due to (\ref{eq:caw}), we have
\begin{equation*}
\lim_{n \rightarrow \infty} A_{1} = \int_{} I(\|\bu_{1}\| \leq h) \lambda^{(i,j)}_{1,1,red} (\bu_{1}) d\bu_1.
\end{equation*}
Similarly, we can show 
\begin{eqnarray*}
\lim_{n \rightarrow \infty} A_{2} &=& \iint I(\|\bu_{1}\|\leq h) I(\|\bu_{2}\|\leq h) \lambda^{(i,j)}_{2,1,red} (\bu_{1},\bu_{1} + \bu_{2}) d\bu_{1}d\bu_{2}, \\
\lim_{n \rightarrow \infty} A_{3} &=&  \iint I(\|\bu_{1}\|\leq h) I(\|\bu_{2}\|\leq h) \lambda^{(i,j)}_{1,2,red} (\bu_{1},\bu_{2}) d\bu_{1}d\bu_{2}.
\end{eqnarray*}
Evaluation of $A_4$ is slightly different. Again, using Campbell's formula, we have
\begin{eqnarray*}
A_4 &=& |D_n|^{-1}  \iiiint_{D_n^4} 
I(\|\bx -\by \|\leq h) I(\|\bz -\bw \|\leq h)  \\
&&  \quad \quad  \times \left\{
 \lambda^{(i,j)}_{2,2} (\bx,\bz,\by,\bw) -  \lambda^{(i,j)}_{1,1} (\bx,\by) \lambda^{(i,j)}_{1,1} (\bz,\bw)
\right\} d\bx d\by d\bz d\bw.
\end{eqnarray*} By using the following change of variables: $\bu_1 = \by-\bx$, $\bu_2 = \bz-\bx$, $\bu_3 = \bw-\bz$, and 
$\bu_4 = \bx$, we have,
\begin{eqnarray*}
&& A_4 = |D_n|^{-1}  \int_{D_n} \int_{D_n-D_n} \int_{D_n-\bu_4}
I(\|\bu_1 \|\leq h) I(\| \bu_3 \|\leq h)  \\
&&  \quad \times \left(
  \int_{D_n-\bu_3-\bu_4} \left\{ \lambda^{(i,j)}_{2,2,red} (\bu_2,\bu_1,\bu_2+\bu_3) -  \lambda^{(i,j)}_{1,1,red} (\bu_1) \lambda^{(i,j)}_{1,1,red} (\bu_3) \right\} d\bu_2
\right)  d\bu_1 d\bu_3 d\bu_4.
\end{eqnarray*} 
It is not straightforward that the above integral exists. However, in Lemma \ref{lemma:cuexpansion}, we show that $\lambda^{(i,j)}_{2,2,\text{red}} (\bu_2, \bu_1, \bu_2 + \bu_3) - \lambda^{(i,j)}_{1,1,\text{red}} (\bu_1) \lambda^{(i,j)}_{1,1,\text{red}} (\bu_3)$ can be written as a sum of the reduced joint cumulant intensities where each term contains $\bu_2$. Then, the absolute integrability of the above integral follows from Assumption \ref{assump:1}(i) (for $\ell=4$). Therefore, we can apply Fubini's theorem. Using the similar techniques applied for the representation of $A_1$, we can show that
\begin{eqnarray*}
&& \lim_{n \rightarrow \infty} A_{4}  = \iiint
I(\|\bu_1 \|\leq h) I(\| \bu_3 \|\leq h) \nonumber \\
&&\quad \quad \quad \times
 \left\{ \lambda^{(i,j)}_{2,2,red} (\bu_2,\bu_1,\bu_2+\bu_3) -  \lambda^{(i,j)}_{1,1,red} (\bu_1) \lambda^{(i,j)}_{1,1,red} (\bu_3) \right\}  d\bu_2 d\bu_1 d\bu_3.
\end{eqnarray*} All together with (\ref{eq:case2-decom}), we prove {\bf [case 2]}. Thus, proves the theorem.
\hfill $\Box$



\section{Proof of Theorems \ref{thm:Jasymp} and \ref{thm:normal_c}} \label{sec:proof}

\subsection{Proof of Theorem \ref{thm:Jasymp}} \label{proof:thm1}

We first show (\ref{eq:unifJ}). Recall (\ref{eq:Qn}). We have
\begin{equation} \label{eq:Qn2}
[\widehat{\bQ}_{n}(h)]_{i,j} = \frac{1}{|D_{n}|} \sum_{\bx \in X_i, \by \in X_j} b(\bx,\by)
\textbf{1}_{\{0<\|\bx-\by\|\leq r\}} \quad \text{a.s.} \quad  h\geq 0.
\end{equation} Under ergodicity of $\underline{X}$,
we can apply \cite{p:ngu-79}, Theorem 1 and obtain
\begin{equation*}
\lim_{n \rightarrow \infty} [\widehat{\bQ}_{n}(h)]_{i,j}  = \Ex [\widehat{\bQ}_{n}(h)]_{i,j} = [\bQ(h;\btheta_0)]_{i,j}
 \quad \text{a.s.} \quad i,j \in \{1,\dots,m\} , ~~ h\geq 0.
\end{equation*} 
The identity above is due to the unbiasedness of $[\widehat{\bQ}_{n}(h)]_{i,j}$ when using the appropriate edge-correction factor. Since both $\widehat{\bQ}_{n}(h)$ and $[\bQ(h;\btheta_0)]_{i,j}$ are positive and increasing functions of $h \in [0, \infty)$, the uniform almost sure convergence of the above on $h \in [0, R]$ can be obtained by applying standard techniques as in \cite{b:kal-21}, Proposition 5.24. Therefore, (\ref{eq:unifJ}) follows.

Next, we show (\ref{eq:asympJ}). Techniques to prove (\ref{eq:asympJ}) are similar to those in \cite{guan2004nonparametric}, Theorem 1, and \cite{p:paw-09}, Section 3, so we only sketch the proof. Recall (\ref{eq:Gn2}). To show the central limit theorem of $\text{vec}(\widehat{\bG}_{n}(h))$, we use the so-called sub-block technique. Let $\{D_{\ell(n)}^{i}: 1 \leq i \leq k_n\}$ be the $k_n$ number of non-overlapping subcubes of $D_n$ with side length $\ell(n) = n^{\beta}$, where
\begin{equation}\label{eq:alpha0}
\beta \in (2d/(2d+\varepsilon), 1).
\end{equation} Here, $\varepsilon>0$ is from Assumption \ref{assump:1}(ii). Since $\{D_{\ell(n)}^{i}: 1\leq i \leq k_n\}$ are non-overlapping, we have
\begin{equation*}
\left| \cup_{i=1}^{k_n} D_{\ell(n)}^{i} \right| = k_n \ell(n)^{d} = k_n n^{d\beta} \leq |D_n| \leq Cn^{d}.
\end{equation*} Here, the last inequality is due to (\ref{eq:caw}). Therefore, we have
\begin{equation} \label{eq:kn}
k_n = O(n^{d(1-\beta)}) \quad \text{as} \quad n \rightarrow \infty.
\end{equation}
Next, let $\{D_{m(n)}^{i} : 1 \leq i \leq k_n\}$ be the subcubes of $\{D_{\ell(n)}^{i} : 1 \leq i \leq k_n\}$, where for $i \in \{1, \dots, k_n\}$, $D_{m(n)}^{i}$ is a subcube of $D_{\ell(n)}^{i}$ with the same center and side length
\begin{equation} \label{eq:mn}
m(n) = n^{\beta} - n^{\eta} < n^{\beta} = \ell(n) \quad \text{for some} \quad \eta \in (2d/(2d+\varepsilon), \beta).
\end{equation} Then, we have
\begin{equation} \label{eq:distmn}
d(D_{\ell(n)}^{i}, D_{m(n)}^{j}) \leq n^{\eta} \quad \text{for} \quad i \neq j \in \{1,\dots,k_n\}.
\end{equation} 
For $i \in \{1, \dots, k_n\}$, let $\widehat{\bQ}_n^{(i)}(h)$, $h \geq 0$, be a nonparametric edge-corrected estimator of $\bQ(h; \btheta_0)$ defined similarly to (\ref{eq:Qn}), but within the sampling window $D_{m(n)}^{i}$. For $i \in \{1, \dots, k_n\}$, let
\begin{equation} \label{eq:Gni}
[\widehat{\bG}_n^{(i)}(h)]_{p,q} = |D_{m(n)}^{i}|^{1/2} \left\{ [\widehat{\bQ}_n^{(i)}(h)]_{p,q} - [\bQ(h;\btheta_0)]_{p,q} \right\},
\quad p,q \in \{1,\dots,m\}, \quad h\geq 0
\end{equation} and 
\begin{equation} \label{eq:Tn}
[T_{n}]_{p,q} = k_n^{-1/2} \sum_{i=1}^{k_n} [\widehat{\bG}_n^{(i)}(h)]_{p,q}, \quad
[\widetilde{T}_{n}]_{p,q} = k_n^{-1/2} \sum_{i=1}^{k_n} [\widetilde{\bG}_n^{(i)}(h)]_{p,q}, \quad p,q \in \{1,\dots,m\}.
\end{equation} Here, $[\widetilde{\bG}_n^{(i)}(h)]_{p,q}$ represents independent copy of $[\widehat{\bG}_n^{(i)}(h)]_{p,q}$.

Next, let $\text{vec}(T_{n}) = ([T_{n}]_{p,q})_{1 \leq p, q \leq m}$ be the vectorization of $\{[T_{n}]_{p,q}\}$ and $\text{vec}(\widetilde{T}_{n})$ is defined similarly but replaces $[T_{n}]_{p,q}$ with $[\widetilde{T}_{n}]_{p,q}$. Our goal is to show that $\text{vec}(\widehat{\bG}_{n}(h))$ and $\text{vec}(\widetilde{T}_{n})$ are asymptotically negligible, thus, having the same asymptotic distribution. To prove this, we use an intermediate random variable, $\text{vec}(T_{n})$. We first show
\begin{equation} \label{eq:pf1}
\text{vec}(\widehat{\bG}_{n}(h)) - \text{vec}(T_{n}) \Pcon 0.
\end{equation} 
To show this, we bound the first and second moments of the difference. Since $\Ex[\text{vec}(\widehat{\bG}_{n}(h))] = \Ex[\text{vec}(T_{n})] = \mathbf{0}_{m \times m}$, the first moment of the difference is zero. To bound the second moment, we will show
\begin{equation} \label{eq:varbound1}
\lim_{n\rightarrow \infty} \var \left\{ [\widehat{\bG}_{n}(h)]_{p,q} - [T_{n}]_{p,q} \right\} = 0, \quad p,q \in \{1, \dots,m\}.
\end{equation} 
Then, by Markov's inequality, we can prove (\ref{eq:pf1}). To show (\ref{eq:varbound1}), let 
$[\widehat{\bG}_{0n}(h)]_{p,q}$ be the empirical process of $\bQ(h; \btheta_0)$ without edge-correction as in (\ref{eq:G0n}). Similarly, we define $[\widehat{\bG}_{0n}^{(i)}(h)]_{p,q}$ and $[T_{0n}]_{p,q}$ as analogous no-edge-corrected estimators to (\ref{eq:Gni}) and (\ref{eq:Tn}), respectively. Then, by the Cauchy-Schwarz inequality, we have
\begin{eqnarray*} 
\var \left\{ [\widehat{\bG}_{n}(h)]_{p,q} - [T_{n}]_{p,q} \right\}
&\leq& 3 \bigg(
\var  \left\{ [\widehat{\bG}_{n}(h)]_{p,q} - [\widehat{\bG}_{0n}(h)]_{p,q} \right\}+
\var  \left\{ [\widehat{\bG}_{0n}(h)]_{p,q} - [T_{0n}]_{p,q} \right\} \nonumber \\
&& \quad +
\var  \left\{ [T_{0n}]_{p,q} - [T_{n}]_{p,q} \right\}
 \bigg), \quad p,q \in \{1,\dots,m\}.
\end{eqnarray*} 
By Theorem \ref{theorem:edge}, the first and third terms above are of order $O(n^{-1})$. The second term converges to zero as $n \rightarrow \infty$, due to the calculations in \cite{p:paw-09}, page 4200 (details omitted). Therefore, altogether, we show (\ref{eq:varbound1}) and thus, (\ref{eq:pf1}) holds.

Next, we will show
\begin{equation} \label{eq:pf2}
\text{vec}(T_{n}) - \text{vec}(\widetilde{T}_{n}) \Dcon 0.
\end{equation} 
To show this, we focus on the characteristic functions of $\text{vec}(T_{n})$ and $\text{vec}(\widetilde{T}_{n})$.
For $n \in \mathbb{N}$, let $\phi_n(\underline{t})$ and $\widetilde{\phi}_n(\underline{t})$ be the characteristic functions of 
$\text{vec}(T_{n})$ and $\text{vec}(\widetilde{T}_{n})$, respectively, where $\underline{t} \in \mathbb{R}^{m\times m}$. 
To show (\ref{eq:pf2}), it is enough to show
\begin{equation} \label{eq:pf22}
\lim_{n\rightarrow \infty} | \phi_n(\underline{t}) - \widetilde{\phi}_n(\underline{t})| = 0, \quad t \in \mathbb{R}^{m\times m}.
\end{equation}
Proof of (\ref{eq:pf22}) is standard method using telescoping sum. Let
\begin{equation*}
U_{n}^{(i)} = \exp (  i k_n^{-1/2} \langle \underline{t}, \text{vec} (\widehat{\bG}_n^{(i)})  \rangle, \quad i \in \{1, \dots,k_n\},
\end{equation*} 
where $\langle \cdot, \cdot \rangle$ denotes the dot product. Similarly, for $i \in \{1,\dots,k_n\}$, we can define $\widetilde{U}_{n}^{(i)}$ by replacing $[\widehat{\bG}_n^{(i)}(h)]_{p,q}$ with $[\widetilde{\bG}_n^{(i)}(h)]_{p,q}$ in the above definition. Then, by definition,
\begin{equation*}
\phi_n(\underline{t}) = \Ex \left[ \prod_{i=1}^{k_n} U_{n}^{(i)} \right] \quad \text{and} \quad
\widetilde{\phi}_n(\underline{t}) = \Ex \left[ \prod_{i=1}^{k_n} \widetilde{U}_{n}^{(i)} \right].
\end{equation*} Since $\widetilde{U}_{n}^{(i)}$ are jointly independent and has the same marginal distribution with $U_{n}^{(i)}$, we have $\widetilde{\phi}_n(\underline{t}) = \prod_{i=1}^{k_n} \Ex \widetilde{U}_{n}^{(i)} = \prod_{i=1}^{k_n} \Ex U_{n}^{(i)}$.
Therefore, using telescoping sum argument (cf. \cite{p:paw-09}, equation (13)), we have
\begin{equation} \label{eq:phi-diff}
| \phi_n(\underline{t}) - \widetilde{\phi}_n(\underline{t})|
\leq \sum_{j=1}^{k_n-1} \left| \cov \left\{ \prod_{i=1}^{j} U_{n}^{(i)}, U_{n}^{(j+1)}  \right\}
\right|.
\end{equation}
Now, we use the $\alpha$-mixing condition. We first note that $\text{vec} (\widehat{\bG}_n^{(i)}) \in (\mathcal{F}(D_{m(n)}^{i}))^{m\times m}$, where for $E \subseteq \mathbb{R}^d$, $\mathcal{F}(E)$ is the sigma algebra generated by $\underline{X}$ in the sampling window $E$. Therefore, for $i \in \{1,\dots,k_n-1\}$,
\begin{equation*}
\prod_{i=1}^{j} U_{n}^{(i)} \in \mathcal{F}(\cup_{i=1}^{j} D_{m(n)}^{i} ) \quad \text{and} \quad
U_{n}^{(j+1)} \in \mathcal{F}( D_{m(n)}^{j+1}).
\end{equation*} We note that
\begin{equation*}
|U_{j}| \leq 1, \quad
|D_{m(n)}^{j+1}| \leq  |\cup_{i=1}^{j} D_{m(n)}^{i}| = j m(n)^d,
 \quad \text{and} \quad
d\left(\cup_{i=1}^{j} D_{m(n)}^{i}, D_{m(n)}^{j+1}\right) \leq n^{\eta}, \quad j \in \{1, \dots,k_n-1\},
\end{equation*} where the last inequality is due to (\ref{eq:distmn}).
Therefore, using the $\alpha$-mixing coefficient defined as in (\ref{eq:alpha}) and the strong-mixing inequality (cf. \cite{p:paw-09}, equation (9)), we have
\begin{equation*} 
\left| \cov \left\{ \prod_{i=1}^{j} U_{n}^{(i)}, U_{n}^{(j+1)}  \right\} \right| \leq
4 \alpha_{ j m(n)^d } ( n^{\eta}) \leq C j m(n)^d n^{-\eta(d+\varepsilon)},
 \quad j \in \{1, \dots,k_n-1\},
\end{equation*} where the last inequality is due to \ref{assump:1}(ii). Summing the above over $j$ and using the bounds (\ref{eq:kn}) and (\ref{eq:mn}), we have
\begin{equation} \label{eq:phi-diff3}
| \phi_n(\underline{t}) - \widetilde{\phi}_n(\underline{t})|
\leq 4  \sum_{j=1}^{k_n-1}  j m(n)^d n^{-\eta (d+\varepsilon)}
\leq C k_n^2 m(n)^d n^{-\eta (d+\varepsilon)} = O(n^{2d-\beta d-\eta(d+\varepsilon)}).
\end{equation} Since $\eta \in (2d/(2d+\varepsilon), \beta)$, we have for all $t \in \mathbb{R}^{m\times m}$,
$\lim_{n\rightarrow \infty} |\phi_n(\underline{t}) - \widetilde{\phi}_n(\underline{t})| = 0$. Therefore, we show (\ref{eq:pf22}), thus, shows (\ref{eq:pf2}).

Back to our goal, combining (\ref{eq:pf1}) and (\ref{eq:pf2}), $\text{vec}(\widehat{\bG}_{n}(h))$ and $\text{vec}(\widetilde{T}_{n})$ share the same asymptotic distribution. Now, we find the asymptotic distribution of $\text{vec}(\widetilde{T}_{n})$. Recall (\ref{eq:Tn}). We have $\text{vec}(\widetilde{T}_{n}) = k_n^{-1/2} \sum_{i=1}^{k_n} \text{vec}(\widetilde{\bG}_n^{(i)}(h))$, where $\{ \text{vec}(\widetilde{\bG}_n^{(i)}(h)) \}$ are i.i.d. mean zero random vectors. Since $\{D_n\}$ is c.a.w., we let $k_n \rightarrow \infty$. Therefore, under (\ref{eq:momentQ1}), we can apply the Lyapunov Central Limit Theorem to conclude $\text{vec}(\widetilde{T}_{n})$ converges to the centered normal distribution. To calculate the asymptotic covariance matrix of $\{ \text{vec}(\widetilde{\bG}_n^{(1)}(h)) \}$, since $\{D_{m(n)}^{1}\}$ satisfies (\ref{eq:caw}), the asymptotic covariance matrix of $\{ \text{vec}(\widetilde{\bG}_n^{(1)}(h)) \}$ is the same as the asymptotic covariance matrix of $\text{vec}(\widetilde{\bG}_n(h))$, where the exact expression can be found in Appendix \ref{sec:sigma}. Altogether, we prove the theorem.
 \hfill $\Box$

\subsection{Proof of Theorem \ref{thm:normal_c}} \label{proof:thm2}

First, we will show that $\widehat{\bm{\theta}}_{n}$, the minimizer of $U_{n}(\bm{\theta})$, exists for all $n \in \mathbb{N}$. To demonstrate this, it suffices to show that $\int_{0}^{R} [\bQ(h;\bm{\theta})]_{i,j} \, dh$ is continuous with respect to $\bm{\theta}$. By assumption, $\bQ(h;\bm{\theta})$ is continuous in $\bm{\theta}$ for fixed $h \geq 0$. Moreover, since $\Theta$ is compact and $[\bQ(\cdot;\bm{\theta})]_{i,j}$ is a positive and monotonically increasing function, we have
\begin{equation*}
\int_{0}^{R} \sup_{\btheta \in \Theta} [\bQ(h;\btheta)]_{i,j} dh \leq R \sup_{\btheta \in \Theta} [\bQ(R;\btheta)]_{i,j} < \infty.
\end{equation*} 
Therefore, by the Dominated Convergence Theorem, we show $\int_{0}^{R} [\bQ(h;\bm{\theta})]_{i,j} \, dh$ is continuous with respect to $\bm{\theta}$. Consequently, $U_n(\bm{\theta})$ is continuous, ensuring that $\widehat{\bm{\theta}}_n$ exists (which may not be unique) for all $n \in \mathbb{N}$.

Next, we show $\widehat{\bm{\theta}}_n$ is uniquely determined up to a null set and satisfies (\ref{eq:theta-consist}). Let $\widehat{\bm{\theta}}_n$ be one of the minimizers of $U_n(\bm{\theta})$. Decompose
\begin{equation*}
\widehat{\bQ}^{\circ C}_{n}(h) -  \bQ^{\circ C}(h;\widehat{\btheta}_{n}) =
 \left\{ \widehat{\bQ}^{\circ C}_{n}(h) -  \bQ^{\circ C}(h;\bm{\theta}_{0}) \right\} +  
\left\{  \bQ^{\circ C}(h;\bm{\theta}_{0}) -  \bQ^{\circ C}(h;\widehat{\bm{\theta}}_{n})\right\} = A+B.
\end{equation*} Then, by definition of $\widehat{\bm{\theta}}_{n}$, we have
$\int_{0}^{R}  \Tr (A+B)(A+B)^\top W(dh) \leq  \int_{0}^{R} \Tr AA^\top W(dh)$, where $W(dh) = w(h) dh$. Expanding the above and using $\Tr BA^\top = \Tr AB^\top$, we get
\begin{equation} \label{eq:integralAB}
0 \leq  \int_{0}^{R} \Tr BB^\top W(dh)
\leq  2 \int_{0}^{R}  | \Tr A B^\top | W(dh).
\end{equation} 
For $m \times m$ symmetric matrix $C$, let $\lambda_j(C)$ be the $j$th largest eigenvalue of $C$, $j \in \{1, \dots,m\}$. Since, $\bQ^{\circ C}$ is symmetric (here, we assume that $C$ is symmetric), by \cite{fang1994inequalities}, Theorem 3, $|\Tr AB^\top|$ is bounded with
\begin{eqnarray*}
|\Tr(AB^\top)| 
&\leq&
\max \bigg( \big|\lambda_{m}(\overline{A})\Tr(B) -\lambda_{m}(B) \cdot \left\{ m \lambda_{m}(\overline{A}) - \Tr(A) \right\} \big|, \\
&&\quad \quad \big| \lambda_{1}(\overline{A})\Tr(B) -\lambda_{m}(B) \cdot \left\{m \lambda_{1}(\overline{A}) - \Tr(A) \right\} \big|\bigg) \\
&\leq& \big| \lambda_{m}(\overline{A})\Tr(B) -\lambda_{m}(B) \cdot \left\{ m \lambda_{m}(\overline{A}) - \Tr(A) \right\} \big| \\
&&+ \big| \lambda_{1}(\overline{A})\Tr(B) -\lambda_{m}(B) \cdot \left\{m \lambda_{1}(\overline{A}) - \Tr(A) \right\} \big|,
\end{eqnarray*} where $\overline{A} = (A+A^\top)/2$.
 We will focus on the first term and the second term can be treated in the same way. Integrate the first term above, we have
\begin{eqnarray}
&&  \int_{0}^{R} \big| \lambda_{m}(A)\Tr(B) -\lambda_{m}(B) \cdot \left\{ m \lambda_{m}(A) - \Tr(A) \right\} \big| W(dh) \nonumber \\
&& \quad \leq 
\int_{0}^{R}|\lambda_{m}(\overline{A})| |\Tr(B)| W(dh) +
 \int_{0}^{R}|\lambda_{m}(B)| \cdot |m \lambda_{m}(\overline{A}) - \Tr(A) | W(dh).
\label{eq:integralAB1}
\end{eqnarray}
Now, we bound both $|\lambda_{m}(\overline{A})|$ and $|m \lambda_{m}(\overline{A}) - \Tr(A)|$. Since 
$[A]_{i,j} = [\widehat{\bQ}_{n}(h)]_{i,j}^{c_{i,j}} -[\bQ_{}(h;\btheta_0)]_{i,j}^{c_{i,j}},$
from (\ref{eq:unifJ}) and the continuous mapping theorem, we have $\lim_{n\rightarrow \infty} \sup_{1\leq i,j \leq m} \sup_{0 \leq h \leq R} | [A]_{i,j}| = 0$ almost surely. Consequently, we obtain
\begin{equation} \label{eq:eigenconverge}
 \lim_{n\rightarrow \infty} \sup_{0 \leq h \leq R} \max_{1\leq j \leq m} |\lambda_{j}(\overline{A})|
= 0 \quad \text{a.s.}
\end{equation}
Moreover, since $\Tr (A) = \Tr(\overline{A}) =  \sum_{i=1}^{m} \lambda_{i}(\overline{A})$, we have
$\left| m \lambda_{m}(\overline{A}) - \Tr(A) \right| \leq m |\lambda_m(\overline{A})|+
\sum_{i=1}^{m} |\lambda_{i}(\overline{A}) | \leq 2m \max_{1\leq j \leq m} |\lambda_{j}(\overline{A})|$.
Therefore, from (\ref{eq:eigenconverge}), 
\begin{equation} \label{eq:eigenconverge2}
\lim_{n\rightarrow \infty} \sup_{0 \leq h \leq R} \left| m \lambda_{m}(\overline{A}) - \Tr(A) \right|
= 0 \quad \text{a.s.}
\end{equation}
Substitute (\ref{eq:eigenconverge}) and (\ref{eq:eigenconverge2}) into (\ref{eq:integralAB1}) and using the inequalities, $|\lambda_{m}(B)| \leq \sum_{j=1}^{m} |\lambda_j (B)|$ and $|\Tr B| \leq \sum_{j=1}^{m} |\lambda_j (B)|$,
we have
\begin{eqnarray*}
&&\int_{0}^{R}  \big| \lambda_{m}(A)\Tr(B) -\lambda_{m}(B) \cdot \left\{ m \lambda_{m}(A) - \Tr(A) \right\} \big| W(dh) \\
&& \quad \leq o_{a.s.}(1) \int_{0}^{R} \left( |\Tr(B)| + |\lambda_{m}(B)| \right) W(dh)
\leq \leq o_{a.s.}(1) \times  2 \int_{0}^{R} \left\{ \sum_{j=1}^{m} |\lambda_j (B)| \right\} W(dh),
\end{eqnarray*} where $o_{a.s.}(1)$ is a sequence of random variables that converges to zero almost surely.
Similarly, 
\begin{equation*}
\int_{0}^{R} \big| \lambda_{1}(\overline{A})\Tr(B) -\lambda_{m}(B) \cdot \left\{m \lambda_{1}(\overline{A}) - \Tr(A) \right\} \big| W(dh)
\leq o_{a.s.}(1) \times   \int_{0}^{R} \left\{ \sum_{j=1}^{m} |\lambda_j (B)| \right\} W(dh).
\end{equation*} 
Therefore, substitute the above two inequalities into (\ref{eq:integralAB}) and using the Cauchy-Schwarz inequality and Jensen's inequality, we get
\begin{eqnarray*}
0 \leq \int_{0}^{R} \Tr BB^\top W(dh) \leq 0 &\leq& \int_{0}^{R} w(h) |\Tr AB^\top| W(dh) \\
&\leq&
o_{a.s.}(1) \times  \int_{0}^{R} \left\{ \sum_{j=1}^{m} |\lambda_j (B)| \right\} W(dh) \\
&\leq&o_{a.s.}(1) \sqrt{m} \times \int_{0}^{R} \left\{ \sum_{j=1}^{m} |\lambda_j (B)|^2 \right\}^{1/2} W(dh) \\
&=& o_{a.s.}(1) \int_{0}^{R}  (\Tr BB^\top)^{1/2} W(dh) 
\leq o_{a.s.}(1) \left\{ \int_{0}^{R} \Tr BB^\top W(dh)\right\}^{1/2}.
\end{eqnarray*} 
Therefore, $\int_{0}^{R} \Tr(BB^\top W(dh)) = \int_{0}^{R} w(h) \left\{ \bQ^{\circ C}(h;\widehat{\bm{\theta}}_{n}) 
- \bQ^{\circ C}(h;\bm{\theta}_{0}) \right\}^2 dh \rightarrow 0$ almost surely as $n \rightarrow \infty$. Finally, since $\bQ^{\circ C}(\cdot; \bm{\theta})$ is uniformly continuous with respect to $\bm{\theta}$ and $\bm{\theta} \mapsto \bQ^{\circ C}(\cdot; \bm{\theta})$ is injective, by the continuous mapping theorem, we have $\widehat{\bm{\theta}}_{n} \rightarrow \bm{\theta}_{0}$ almost surely as $n \rightarrow \infty$. Therefore, we prove (\ref{eq:theta-consist}) and also show that $\widehat{\bm{\theta}}_{n}$ is uniquely determined up to a null set.

Next, we will show the asymptotic normality of $\widehat{\bm{\theta}}_n$. By using Taylor expansion, there exists $\widetilde{\bm{\theta}}_n$, a convex combination of $\widehat{\bm{\theta}}_{n}$ and $\bm{\theta}_{0}$, such that
\begin{equation}
\frac{\partial  U_{n}}{\partial \bm{\theta}} (\widehat{\bm{\theta}}_{n}) = \frac{\partial  U_{n}}{\partial \bm{\theta}} (\bm{\theta}_{0})
+ \frac{\partial^2  U_{n}}{\partial \bm{\theta} \partial \bm{\theta}^{\top}} (\widetilde{\bm{\theta}}_{n}) ( \widehat{\bm{\theta}}_{n} - \bm{\theta}_{0}) = \bm{0}.
\label{eq:taylor}
\end{equation} Now, we calculate the first and second derivatives of $U_n$ defined as in (\ref{eq:multiK_discrep_diff_power2}). 
By simple algebra, we have
\begin{eqnarray}
 -\frac{\partial U_{n}}{\partial \bm{\theta}} (\bm{\theta}_{0})  
&=& 2 \sum_{i,j=1}^{m} c_{i,j} \int_{0}^{R}  \left\{ [\widehat{\bQ}_n(h)]_{i,j}^{c_{i,j}} - [\bQ(h;\bm{\theta}_{0})]_{i,j}^{c_{i,j}} \right\} [\bQ(h;\bm{\theta}_0)]_{i,j}^{c_{i,j}-1}  [\nabla_{\btheta}\bQ (r;\bm{\theta}_{0})]_{i,j} W(dh) \nonumber \\
&=& 2 A_{n}(\bm{\theta}_{0}), 
\label{eq:Un-1st}
\end{eqnarray} 
\begin{eqnarray}
 \frac{\partial^2  U_{n}}{\partial \bm{\theta} \partial \bm{\theta}^{\top}} (\widetilde{\bm{\theta}}_{n}) &=&
2 \sum_{i,j=1}^{m} c_{i,j}^2 \int_{0}^{R} [\bQ(h;\widetilde{\bm{\theta}}_n)]_{i,j}^{2c_{i,j}-2}
\left\{[\nabla_{\btheta}\bQ (h;\widetilde{\bm{\theta}}_{n})]_{i,j}\right\} \left\{[\nabla_{\btheta}\bQ (h;\widetilde{\bm{\theta}}_{n})]_{i,j}\right\}^\top W(dh)  \nonumber \\
&& - 2\bigg[ \sum_{i,j=1}^{m} c_{i,j}(c_{i,j}-1) \int_{0}^{R} \left\{ [\widehat{\bQ}_n(h)]_{i,j}^{c_{i,j}} - [\bQ(h;\widetilde{\bm{\theta}}_n)]_{i,j}^{c_{i,j}} \right\} \nonumber \\
&&  \quad  \quad \quad \quad \times [\bQ(h;\widetilde{\bm{\theta}}_n)]_{i,j}^{c_{i,j}-2}
 \left\{[\nabla_{\btheta}\bQ (h;\widetilde{\bm{\theta}}_{n})]_{i,j}\right\} \left\{[\nabla_{\btheta}\bQ (h;\widetilde{\bm{\theta}}_{n})]_{i,j}\right\}^\top W(dh) \bigg]  \nonumber \\
&& - 2\left[ \sum_{i,j=1}^m c_{i,j}\int_0^{R} \left\{ [\widehat{\bQ}_n(h)]_{i,j}^{c_{i,j}} - [\bQ(h;\widetilde{\bm{\theta}}_n)]_{i,j}^{c_{i,j}} \right\} [\bQ(h;\widetilde{\bm{\theta}}_n)]_{i,j}^{c_{i,j}-1} [\nabla_{\btheta}^2 (h;\widetilde{\bm{\theta}}_{n})]_{i,j} W(dh) \right]. \nonumber \\
\label{eq:Un-2nd}
\end{eqnarray}
Since $\widehat{\bm{\theta}}_n \rightarrow \bm{\theta}_0$ almost surely from (\ref{eq:theta-consist}), we also have $\widetilde{\bm{\theta}}_n \rightarrow \bm{\theta}_0$ almost surely as $n \rightarrow \infty$. Therefore, using a similar approach as that used to show (\ref{eq:unifJ}), we have
\begin{equation} \label{eq:Qdiffeq1}
\lim_{n\rightarrow \infty} \max_{1\leq i,j \leq m} \int_0^{R} \left| [\widehat{\bQ}_n(h)]_{i,j}^{c_{i,j}} - [\bQ(h;\widetilde{\bm{\theta}}_n)]_{i,j}^{c_{i,j}} \right| W(dh)= 0 \quad \text{a.s.}
\end{equation} Using the above and Assumption \ref{assump:3}(ii), we can show the second and third term in (\ref{eq:Un-2nd}) are asymptotic negligible, also we have
\begin{equation}
\frac{\partial^2  U_{n}}{\partial \bm{\theta} \partial \bm{\theta}^{\top}} (\widetilde{\bm{\theta}}_{n}) 
 = 2B(\btheta_0) + o_{a.s.}( \textbf{1}_{m \times m}),
\label{eq:Un-2nd2}
\end{equation} where $B(\btheta_0)$ is defined as in (\ref{eq:Btheta})
and $o_{a.s.}( \textbf{1}_{m \times m})$ denotes sequence of $m\times m$ random matrices such that each entry converges to zero almost surely.  Substitute  (\ref{eq:Un-1st}) and (\ref{eq:Un-2nd2}) into (\ref{eq:taylor}), we have
\begin{equation}
\sqrt{|D_n|}( \widehat{\bm{\theta}}_{n} - \bm{\theta}_{0}) =
\left\{ B(\btheta_0)^{-1} + o_{a.s.}( \textbf{1}_{m \times m})\right\} \sqrt{|D_n|} A_n(\btheta_0).
\label{eq:taylor2}
\end{equation}
Next, let
\begin{eqnarray} 
 \widetilde{A}_n(\btheta_0)
&=&  \sum_{i,j=1}^{m} c_{i,j}^2 \left\{ 
[\widehat{\bQ}_{1n}(R)]_{i,j} - \Ex \left[ \widehat{\bQ}_{1n}(R)]_{i,j}\right]
\right\} \nonumber \\
&=& \sum_{i,j=1}^{m} c_{i,j}^2 \int_{0}^{R}  \left\{ [\widehat{\bQ}_n(h)]_{i,j} - [\bQ(h;\bm{\theta}_{0})]_{i,j} \right\} [\bQ(h;\bm{\theta}_0)]_{i,j}^{2c_{i,j}-2}  [\nabla_{\btheta}\bQ (r;\bm{\theta}_{0})]_{i,j} W(dh), \nonumber \\
\label{eq:Atilden}
\end{eqnarray} where $[\widehat{\bQ}_{1n}(R)]_{i,j}$ is defined as in (\ref{eq:Q1n}).
Since $\bQ(h;\btheta)$ is continuously differentiable with respect to $\btheta$, for $h \geq 0$,
 there exist $[\bQ^{*}(h)]_{i,j}$ between $[\widehat{\bQ}_n(h)]_{i,j}$ and $[\bQ(h;\bm{\theta}_{0})]_{i,j}$
such that
\begin{equation*}
[\widehat{\bQ}_n(h)]_{i,j}^{c_{i,j}} - [\bQ(h;\bm{\theta}_{0})]_{i,j}^{c_{i,j}}
= c_{i,j} \left\{ [\widehat{\bQ}_n(h)]_{i,j} - [\bQ(h;\bm{\theta}_{0})]_{i,j}\right\} [\bQ^{*}(h)]_{i,j}^{c_{i,j}-1}.
\end{equation*} Therefore, the difference $\sqrt{|D_n|}  \{A_n(\btheta_0) -  \widetilde{A}_n(\btheta_0) \}$ is bounded with
\begin{eqnarray*}
&& \sqrt{|D_n|} |A_n(\btheta_0) -  \widetilde{A}_n(\btheta_0)|_{1} \nonumber \\
&& \quad \leq
\sqrt{|D_n|} \sum_{i,j=1}^{m} c_{i,j}^2 \int_{0}^{R} \left| [\widehat{\bQ}_n(h)]_{i,j} - [\bQ(h;\bm{\theta}_{0})]_{i,j} \right|
\left| [\bQ^{*}(h)]_{i,j}^{c_{i,j}-1} - [\bQ(h;\bm{\theta}_0)]_{i,j}^{c_{i,j}-1}\right| \\
&& \quad \quad \quad  \quad \quad \times \left| [\bQ(h;\bm{\theta}_0)]_{i,j}^{c_{i,j}-1} \right| \left| [\nabla_{\btheta}\bQ (r;\bm{\theta}_{0})]_{i,j} \right|_{1} W(dh),
\end{eqnarray*} where for vector $\bx= (x_1, \dots, x_p)^\top$, $|\bx|_1 = \sum_{i=1}^{p}|x_i|$. Using similar argument as in (\ref{eq:Qdiffeq1}), it can be shown that 
$\max_{1\leq i,j\leq m}\sup_{0 \leq h \leq R} | [\bQ^{*}(h)]_{i,j} - [\bQ(h;\bm{\theta}_{0})]_{i,j}| \rightarrow 0$ almost surely as 
$n \rightarrow \infty$ and
\begin{eqnarray} 
\sqrt{|D_n|} |A_n(\btheta_0) -  \widetilde{A}_n(\btheta_0)|_1 &\leq&
\sum_{i,j=1}^{m} c_{i,j}^2 o_p(1)  \times \sqrt{|D_n|} \int_{0}^{R}   \left| [\widehat{\bQ}_n(h)]_{i,j} - [\bQ(h;\bm{\theta}_{0})]_{i,j} \right| W(dh) \nonumber \\
&=& o_p(1) \times \sum_{i,j=1}^{m} c_{i,j}^2 \int_{0}^{R} |[\widehat{\bG}_{n}(h)]_{i,j}| W(dh),
\label{eq:Anbound1}
\end{eqnarray} where $[\widehat{\bG}_{n}(h)]_{i,j}$ is defined as in (\ref{eq:Gn2}). Using (\ref{eq:momentQ1}) and Jensen's inequality, it can be shown that
\begin{equation}
 \sqrt{|D_n|}|A_n(\btheta_0) -  \widetilde{A}_n(\btheta_0)|_1
= o_p(1) O_p(1) = o_p(1).
\label{eq:Anbound2}
\end{equation} Therefore, substitute (\ref{eq:Anbound2}) into (\ref{eq:taylor2}) gives
\begin{equation}
\sqrt{|D_n|}( \widehat{\bm{\theta}}_{n} - \bm{\theta}_{0}) = B(\btheta_0)^{-1} \sqrt{|D_n|} \widetilde{A}_n(\btheta_0)
+ o_p(\textbf{1}_{m \times 1}).
\label{eq:taylor3}
\end{equation}
Next, we will show the asymptotic normality of $\sqrt{|D_n|} \widetilde{A}_n(\bm{\theta}_0)$. Recall (\ref{eq:Atilden}). We have $\mathbb{E} [\widetilde{A}_n(\bm{\theta}_0)]= 0$. Using similar techniques as those used to show the asymptotic normality of $\mathrm{vec}(\widehat{\bm{G}}_{n}(h))$ in the proof of Theorem \ref{thm:Jasymp}, we can derive the asymptotic normality of the vectorization of $\sqrt{|D_n|}\{ [\widehat{\bm{Q}}_{1n}(R)]_{i,j} - \mathbb{E} [ \widehat{\bm{Q}}_{1n}(R)]_{i,j} \}$ (details omitted). Here, we use Assumption \ref{assump:3}(iii) to apply the Lyapunov Central Limit Theorem for multivariate i.i.d. random variables. 
To calculate the asymptotic covariance matrix of $\sqrt{|D_n|} \widetilde{A}_n(\bm{\theta}_0)$, we note that
\begin{eqnarray*}
|D_n| \var \{ \widetilde{A}_n(\btheta_0) \}  &=& \sum_{i_{1},j_{1},i_{2},j_{2}=1}^{m} c_{i_1,j_1}^2 c_{i_2,j_2}^2 \int_{0}^{R} 
\int_{0}^{R}  w(s) w(h)  \cov  \left\{ [\widehat{\bG}_n(s)]_{i_{1},j_{1}}, [\widehat{\bG}_n(h)]_{i_{2},j_{2}} \right\}  \nonumber \\
 &&\quad \quad \quad \quad \times \Biggl\{ 
[\bQ(s;\bm{\theta}_0)]_{i_{1},j_{1}}^{2c_{i_1,j_1}-2} 
[\bQ(h;\bm{\theta}_0)]_{i_{2},j_{2}}^{2c_{i_2,j_2}-2}
 \Biggr\} \nonumber \\
 &&\quad \quad \quad \quad \times  
\left\{ [\nabla_{\btheta}\bQ(s;\bm{\theta}_{0})]_{i_1,j_1} \right\}
\left\{ [\nabla_{\btheta}\bQ(h;\bm{\theta}_{0})]_{i_2,j_2} \right\}^{\top} 
dsdh.
\end{eqnarray*} Therefore, using the notion $\sigma^2_{(i_{1},j_{1}: i_{2},j_{2})} (s,h)$ and $ S(\bm{\theta}_{0})$ defined as in (\ref{eq:sigma}) and (\ref{eq:Stheta}), respectively, we have $\lim_{n \rightarrow \infty}|D_n| \var \{ \widetilde{A}_n(\btheta_0) \} = S(\btheta_0)$ and thus,
\begin{equation}\label{eq:An-normal}
\sqrt{|D_n|} \widetilde{A}_n(\btheta_0) \Dcon \mathcal{N} (\textbf{0}_p, S(\btheta)).
\end{equation} 
Finally, combining (\ref{eq:An-normal}), (\ref{eq:taylor3}), and using the delta method, we have
\begin{equation*}
\sqrt{|D_n|}( \widehat{\bm{\theta}}_{n} - \bm{\theta}_{0}) \Dcon
\mathcal{N} (\textbf{0}_p, B(\btheta_0)^{-1} S(\btheta) B(\btheta_0)^{-1}).
\end{equation*}
Thus, proves (\ref{eq:theta-normal}). \hfill $\Box$

\section{Technical Lemmas} \label{sec:tech}

In this section, we prove two auxiliary lemmas. The first lemma addresses the conditions of the LGCP model in order to satisfy Assumption \ref{assump:1}.

\begin{lemma} \label{lemma:LGCP}
Let $\underline{X}= (X_1, \dots, X_m)$ be an $m$-variate stationary LGCP. For $i,j \in \{1,\dots,m\}$, the cross-covariance process of the $(i,j)$th component of the latent multivariate Gaussian random field is denoted by $C_{i,j}(\bs)$ for $\bs \in \mathbb{R}^d$. Then, the following two assertions hold:
\begin{itemize}
\item[(i)] For $i,j \in \{1,\dots,m\}$, suppose $C_{i,j}(\bs)$ is absolutely integrable. Then, $\underline{X}$ is ergodic and satisfies Assumption \ref{assump:1}(i) for all $\ell \in \mathbb{N}$. Therefore, (\ref{eq:momentQ1}) holds for all $\delta >0$.
\item[(ii)]  For $i,j \in \{1,\dots,m\}$, suppose $|C_{i,j}(\bs)| = O(\|\bs\|^{-2d-\varepsilon})$ as $\|\bs\| \rightarrow \infty$ for some $\varepsilon>0$. Therefore, \ref{assump:1}(ii) holds.
\end{itemize}
\end{lemma}
\noindent \textit{Proof}. 
Proof of (ii) is a direct consequence of \cite{b:dou-94}, Corollary 2 on page 59. To prove (i), we note that by \cite{moller1998log}, Theorem 3, $\underline{X}$ is ergodic if $|C_{i,j}(\bs)| \rightarrow 0$ as $\|\bs\| \rightarrow \infty$ for all $i,j \in \{1,\dots,m\}$. To show the (absolute) integrability of the reduced joint cumulants, for $i,j \in \{1,\dots,m\}$ and $\bs_1, \bs_2 \in \mathbb{R}^d$, let
\begin{equation*}
\lambda^{(i)}(\bs) = \lambda^{(i)}, \quad 
\lambda^{(i,j)}_{1,1}(\bs_1, \bs_2), \quad \text{and} \quad
g^{(i,j)}(\bs_1-\bs_2) = \lambda^{(i,j)}_{1,1}(\bs_1, \bs_2) / \{\lambda^{(i)}(\bs_1) \lambda^{(j)}(\bs_2)\}
\end{equation*} 
Let $\gamma_{(1,1,1)}^{(i,j,k)}(\bx_1, \bx_2, \bx_3)$ denote the first-order intensity, second-order cross intensity, and cross pair correlation function, respectively, where $i, j, k$ are distinct. We only prove Assumption \ref{assump:1}(i) for $\gamma_{(1,1,1)}^{(i,j,k)}(\bx_1, \bx_2, \bx_3)$. The general case can be treated in the same way. Using the cumulant identity in \cite{b:bri-81}, equation (2.3.1), we have
\begin{eqnarray}
\gamma_{(1,1,1)}^{(i,j,k)}(\bs_1, \bs_2, \bs_3)
&=& \lambda_{(1,1,1)}^{(i,j,k)}(\bs_1, \bs_2, \bs_3) \nonumber \\
&& - \left\{ \lambda_{}^{(i)}(\bs_1)  \lambda_{(1,1)}^{(j,k)}(\bs_2, \bs_3) +
\lambda_{}^{(j)}(\bs_2)  \lambda_{(1,1)}^{(i,k)}(\bx_1, \bs_3)+
\lambda_{}^{(k)}(\bs_3)  \lambda_{(1,1)}^{(i,j)}(\bs_1, \bs_2) \right\} \nonumber \\
&& +2 \lambda_{}^{(i)}(\bs_1)\lambda_{}^{(j)} (\bs_2) \lambda_{}^{(k)} (\bs_3).
\label{eq:cueq}
\end{eqnarray}
From \cite{moller1998log}, equations (2) and (12), the scaled joint intensities of the stationary LGCP can be written in terms of the product of $g^{(i,j)}$. For example,
\begin{equation} \label{eq:moller-1}
\frac{ \lambda_{(1,1)}^{(i,j)}(\bs_1, \bs_2)}{\lambda_{}^{(i)}\lambda_{}^{(j)}}
= g^{(i,j)}(\bs_1-\bs_2), \quad
\frac{\gamma_{(1,1,1)}^{(i,j,k)}(\bs_1, \bs_2, \bs_3)}{\lambda_{}^{(i)}\lambda_{}^{(j)} \lambda_{}^{(k)}}
 = g^{(i,j)}(\bs_1-\bs_2) g^{(j,k)}(\bs_2-\bs_3) g^{(k,i)}(\bs_3-\bs_1).
\end{equation}
Substitute (\ref{eq:moller-1}) into (\ref{eq:cueq}) and after some algebra, under stationarity, we get
\begin{eqnarray}
\frac{\gamma_{(1,1,1)}^{(i,j,k)}(\bs_1, \bs_2, \bs_3)}{\lambda_{}^{(i)}\lambda_{}^{(j)} \lambda_{}^{(k)}}
&=& g^{(i,j)}(\bs_1-\bs_2) g^{(j,k)}(\bs_2-\bs_3) g^{(k,i)}(\bs_3-\bs_1) \nonumber \\
&&-\left\{ g^{(i,j)}(\bs_1-\bs_2) + g^{(j,k)}(\bs_2-\bs_3) + g^{(k,i)}(\bs_3-\bs_1) \right\}+2
\label{eq:cueq2}
\end{eqnarray} Let $\bu_1 = \bs_2-\bs_1$ and $\bu_2 = \bs_3-\bs_1$ for $\bs_1, \bs_2, \bs_3 \in \mathbb{R}^d$. Then, the scaled reduced joint cumulant can be written as
\begin{equation}
\frac{\gamma_{(1,1,1),red}^{(i,j,k)}(\bu_1, \bu_2)}{\lambda_{}^{(i)}\lambda_{}^{(j)} \lambda_{}^{(k)}}
= g_1(\bu_1)g_2(\bu_1-\bu_2) g_3(\bu_2) - \{ g_1(\bu_1) + g_2(\bu_1-\bu_2) + g_3(\bu_2)\} +2,
\label{eq:cueq3}
\end{equation} 
where $g_1 = g^{(i,j)}$, $g_2 = g^{(j,k)}$, and $g_3 = g^{(k,i)}$. We also note from \cite{moller1998log}, Theorem 1, that $g_1(\bu) = \exp(C_{i,j}(\bu))$. Therefore, since we assume $C_{i,j}(\cdot)$ is absolutely integrable, we have $|C_{i,j}(\bu)| \rightarrow 0$ as $\|\bu\| \rightarrow \infty$. This implies that $g_1(\bu) \rightarrow 1$ as $\|\bu\| \rightarrow \infty$. Similarly, $g_2(\bu)$ and $g_3(\bu)$ converge to 1 as $\|\bu\| \rightarrow \infty$. Our goal is to express the right-hand side in (\ref{eq:cueq3}) as a function of $h_k(\bu) = g_k(\bu) - 1$. After some algebra, we have
\begin{eqnarray}
 \frac{\gamma_{(1,1,1),red}^{(i,j,k)}(\bu_1, \bu_2)}{\lambda_{}^{(i)}\lambda_{}^{(j)} \lambda_{}^{(k)}} 
&=& h_1(\bu_1)h_2(\bu_1-\bu_2) h_3(\bu_2) \nonumber \\
&&  + \{ h_1(\bu_1)h_2(\bu_1-\bu_2) +h_2(\bu_1-\bu_2) h_3(\bu_2) + h_1(\bu_1) h_3(\bu_2)
\}.
\label{eq:cueq4}
\end{eqnarray} Since $|C_{i,j}(\bu)|$ is bounded, for $k \in \{1, 2, 3\}$, there exist  $C_k, M_k>0$ such that
\begin{equation*}
|h_k(\bu)| = |\exp (C_{}(\bu)) -1| \leq C_k  \cdot |(C_{}(\bu))| < M_k, \quad \bu \in \mathbb{R}^d,
\end{equation*} where $C_{}(\cdot)$ is the cross covariance process which may varies by the value $k \in \{1,2,3\}$. Therefore, integral of the first term in (\ref{eq:cueq4}) is bounded with
\begin{eqnarray*}
&& \int_{\mathbb{R}^{2d}} |h_1(\bu_1)h_2(\bu_1-\bu_2) h_3(\bu_2)| d\bu_1 d\bu_2 
\leq M_2 \left( \int_{\mathbb{R}^{d}} |h_1(\bu_1)| d\bu_1 \right)
\left( \int_{\mathbb{R}^{d}} |h_3(\bu_2)|d\bu_2 \right) \\
&& \quad \leq M_2 C_1 C_3 \left( \int_{\mathbb{R}^{d}} |C_{i,j}(\bu_1)| d\bu_1 \right)
\left( \int_{\mathbb{R}^{d}} |C_{k,i}(\bu_2)| d\bu_2 \right) <\infty.
\end{eqnarray*} Similarly, all four terms in (\ref{eq:cueq4}) are absolutely integrable, thus, 
$\int_{\mathbb{R}^{2d}} |\gamma_{(1,1,1),red}^{(i,j,k)}(\bu_1, \bu_2)| d\bu_1 d\bu_2 <\infty$.
This proves the Lemma.
\hfill $\Box$

Next, we require the integrability of the function of joint intensities, which is used to derive the expression of the asymptotic covariance matrix $\lim_{n\rightarrow \infty} \cov \{[\widehat{\bm{G}}_{0n}(h)]_{i,j} , [\widehat{\bm{G}}_{0n}(h)]_{i,j}\}$.

\begin{lemma} \label{lemma:cuexpansion}
Let $\underline{X}=(X_1, \dots,X_m)$ be a simple multivariate stationary point process that satisfies Assumption \ref{assump:1}(i) (for $\ell=4$), and let $\{D_n\}_{n \in \mathbb{N}}$ on $\mathbb{R}^d$ be the sequence of sampling windows that satisfies (\ref{eq:caw}). For $i \neq j \in \{1, \dots,m\}$, define $\lambda^{(i,j)}_{2,2,\mathrm{red}} (\cdot, \cdot, \cdot)$ and $\lambda^{(i,j)}_{1,1,\mathrm{red}} (\cdot)$ as in (\ref{eq:joint-red}). Then, we have
\begin{equation*}
\sup_{\bu_1, \bu_2 \in \mathbb{R}^d} \int_{} \left| \lambda^{(i,j)}_{2,2,red} (\bu_2,\bu_1,\bu_2+\bu_3) -  \lambda^{(i,j)}_{1,1,red} (\bu_1) \lambda^{(i,j)}_{1,1,red} (\bu_3)\right| d\bu_2 <\infty.
\end{equation*}
\end{lemma}

\noindent \textit{Proof}. 
Let $\gamma_{\underline{n}, \mathrm{red}}^{}$ be the reduced joint cumulant intensity function defined as in (\ref{eq:cumul-red}). This proof requires a lengthy cumulant expansion. Using the cumulant expansions in \cite{p:paw-09}, page 4196, and after lengthy calculation, we have
\begin{eqnarray*}
&& \lambda^{(i,j)}_{2,2,red} (\bu_2,\bu_1,\bu_2+\bu_3) -  \lambda^{(i,j)}_{1,1,red} (\bu_1) \lambda^{(i,j)}_{1,1,red} (\bu_3) \\
&&=\gamma_{2,2,red}^{(i,j)}(\bu_2, \bu_1, \bu_2+\bu_3) + \lambda^{(i)} \big[
\gamma_{1,2,red}^{(i,j)}(\bu_1-\bu_2, \bu_3) + 
\lambda^{(i)} \gamma_{2,red}^{(j)}(\bu_2+\bu_3-\bu_1)
+
\lambda^{(i)} \gamma_{1,1,red}^{(i,j)} (\bu_1-\bu_2)\big]
\\
&&~ + \lambda^{(i)} \gamma_{1,2,red}^{(i,j)} (\bu_1, \bu_2+\bu_3)+
\lambda^{(j)} \gamma_{2,1,red}^{(i,j)} (\bu_2, \bu_2+\bu_3)
+\lambda^{(j)} \gamma_{2,1,red}^{(i,j)} (\bu_2, \bu_1) \\
&&~ + \gamma_{2,red}^{(i)}(\bu_2) \big[ \gamma_{2,red}^{(j)}(\bu_2+\bu_3-\bu_1) + (\lambda^{(i)})^2 \big]
+ \gamma_{1,1,red}^{(i,j)}(\bu_2+\bu_3) \big[ \gamma_{1,1,red}^{(i,j)}(\bu_1-\bu_2) + \lambda^{(i)}\lambda^{(j)} \big].
\end{eqnarray*}
Under Assumption \ref{assump:1}(i) (for $\ell=4$), each term above are absolutely integrable with respective to $\bu_2$ and the bound does not depend on $\bu_1$ and $\bu_3$. Thus, we get desired result.
\hfill $\Box$

\section{Additional simulation results} \label{appen:fig}

In this section, we supplement the simulation results in Section \ref{sec:simu}.

\subsection{Explicit forms} \label{appen:explicit}

Recall that $(X_1, X_2)$ is driven by the latent intensity field $(\Lambda_1(\bs), \Lambda_2(\bs))$, where the log-intensity field is Gaussian. Combining (\ref{eq:LMC}) and (\ref{eq:exp-cov}), it is easily seen that the marginal and cross covariances of $\log\Lambda_1$ and $\log\Lambda_2$ have the following expressions:
\begin{eqnarray}
    C_{11}(r;\bm{\theta}) &=&  \sigma_{Z_1}^2 \exp{(-r/\phi_{Z_1})} + \sigma_{Z_3}^2 \exp{(-r/\phi_{Z_3})}, \nonumber \\
    C_{22}(r;\bm{\theta}) &=& \sigma_{Z_2}^2 \exp{(-r/\phi_{Z_2})} + \sigma_{Z_3}^2 \exp{(-r/\phi_{Z_3})}, \nonumber \\
 C_{12}(r;\bm{\theta}) &=& C_{21}(r;\bm{\theta}) =  b \sigma_{Z_3}^2 \exp{(-r/\phi_{Z_3})},
\label{eq:Covst}
\end{eqnarray}
where the parameter of interest is $\btheta = (\sigma_{Z_1}, \phi_{Z_1}, \sigma_{Z_2}, \phi_{Z_2}, \sigma_{Z_3}, \phi_{Z_3})^\top$. The marginal and cross correlation of $\log\Lambda_1$ and $\log\Lambda_2$ has the following expressions:
 \begin{equation}
\corr_{11}(r;\btheta)=\dfrac{C_{11}(r;\bm{\theta})}{\sigma_{Z_1}^2 + \sigma_{Z_3}^2},
\corr_{22}(r;\btheta)=\dfrac{C_{22}(r;\bm{\theta})}{\sigma_{Z_2}^2 + \sigma_{Z_3}^2},
\corr_{12}(r;\btheta)=\dfrac{C_{12}(r;\bm{\theta})}{\sqrt{ (\sigma_{Z_1}^2 + \sigma_{Z_3}^2) (\sigma_{Z_2}^2 + \sigma_{Z_3}^2)}}.
 \label{eq:corr_11}
\end{equation}
Lastly, using (\ref{eq:Covst}), the cross-correlation coefficient of $X_1$ and $X_2$, denoted by \\ $\rho = \corr \{ \log \Lambda_1(\bu), \log \Lambda_2(\bu) \}$, is 
\begin{equation} \label{eq:rho}
\rho = \rho (\btheta) = b \sigma_{Z_3}^2  / \sqrt{ (\sigma_{Z_1}^2 + \sigma_{Z_3}^2) (\sigma_{Z_2}^2 + \sigma_{Z_3}^2)}.
\end{equation} 
Therefore, positive (resp., negative) sign in $b$ indicates positive (resp., negative) correlation between $X_1$ and $X_2$.  

\subsection{Additional figures and tables} \label{appen:add-fig}

Here, we provide additional figures and tables that are not displayed in Section \ref{sec:simu}. 


\begin{figure}[h!]
      \begin{subfigure}{0.5\textwidth} 
                \centering
    \includegraphics[width=0.9\linewidth]{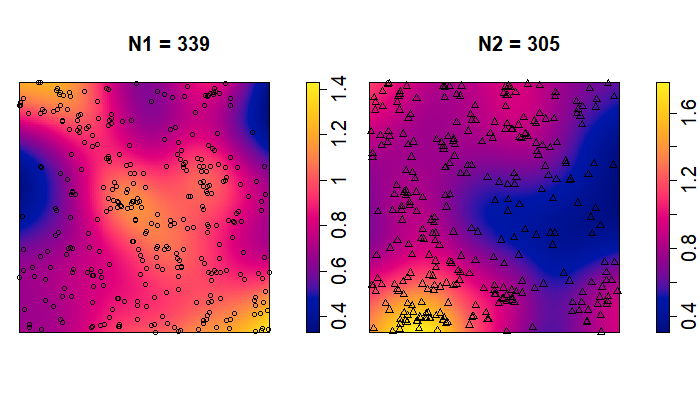}
    \vspace*{-7mm}
    \caption{(M1) with $\rho$=0.166.}
    \label{fig:P1_2_pos_density}
    \end{subfigure}
 \begin{subfigure}{0.5\textwidth}
       \centering
 \includegraphics[width=0.9\linewidth]{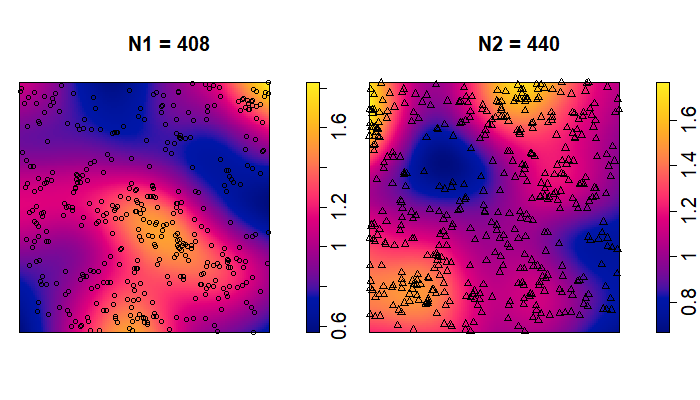}
 \vspace*{-7mm}
    \caption{(M1) with $\rho$=-0.166.}
    \label{fig:P1_2_neg_density}
    \end{subfigure}

\vspace{-1.1em} 
   
\begin{subfigure}{0.5\textwidth} 
                \centering
    \includegraphics[width=0.9\linewidth]{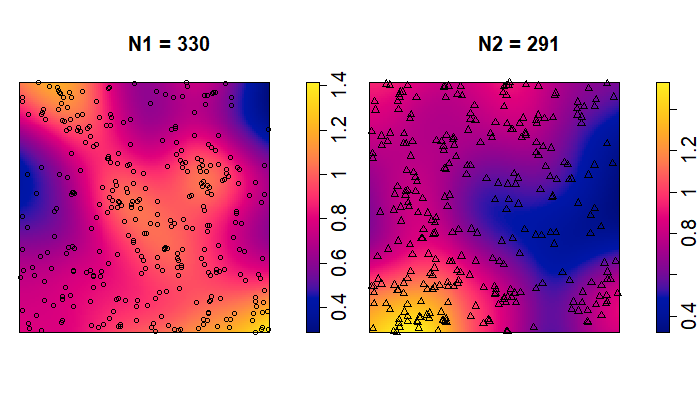}
    \vspace*{-7mm}
    \caption{ (M2) with $\rho$=0.339.}
    \label{fig:P2_2_pos_density}
    \end{subfigure}
 \begin{subfigure}{0.5\textwidth}
       \centering
 \includegraphics[width=0.9\linewidth]{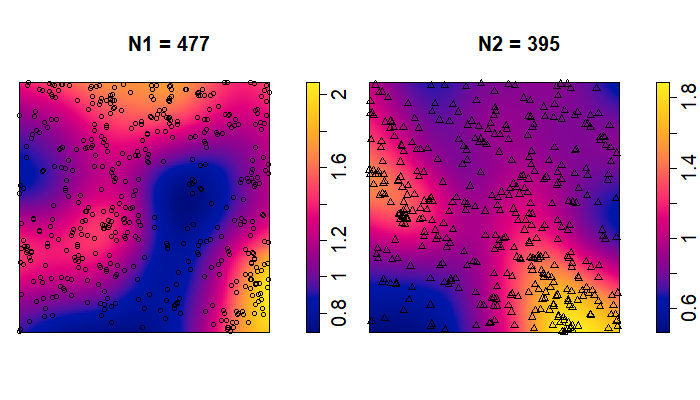}
 \vspace*{-7mm}
    \caption{ (M2) with $\rho$=-0.339.}
    \label{fig:P2_2_neg_density}
    \end{subfigure}

\vspace{-1.1em}    

\begin{subfigure}{0.5\textwidth} 
                \centering
    \includegraphics[width=0.9\linewidth]{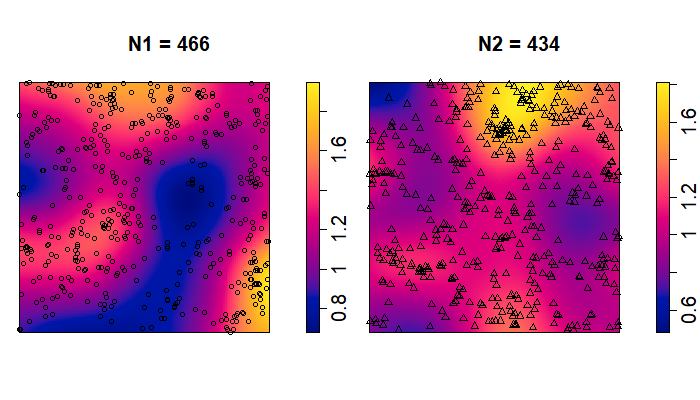}
    \vspace*{-7mm}
    \caption{ (M3) with  $\rho$=0.541.}
    \label{fig:P3_2_pos_density}
    \end{subfigure}
 \begin{subfigure}{0.5\textwidth}
       \centering
 \includegraphics[width=0.9\linewidth]{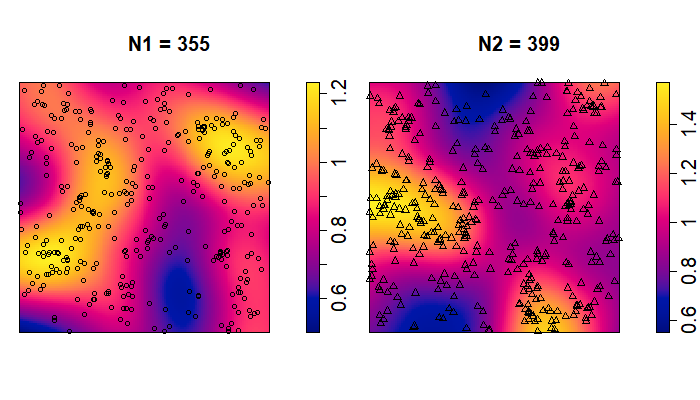}
 \vspace*{-7mm}
    \caption{ (M3) with  $\rho$=-0.541.}
    \label{fig:P3_2_neg_density}
    \end{subfigure}

\vspace{-1.1em}
    
\begin{subfigure}{0.5\textwidth} 
                \centering
    \includegraphics[width=0.9\linewidth]{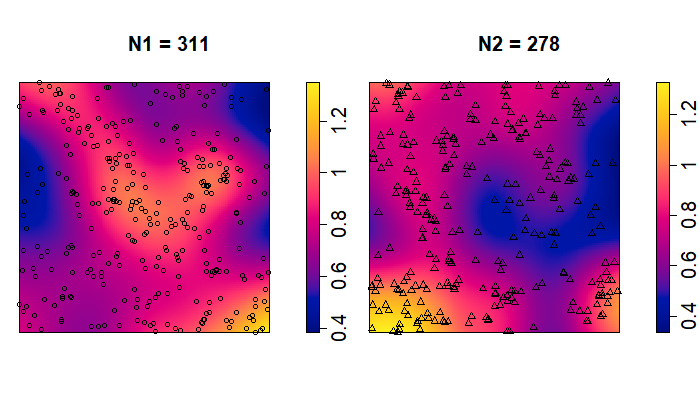}
    \vspace*{-7mm}
    \caption{(M4) with  $\rho$=0.758.}
    \label{fig:P4_2_pos_density}
    \end{subfigure}
 \begin{subfigure}{0.5\textwidth}
       \centering
 \includegraphics[width=0.9\linewidth]{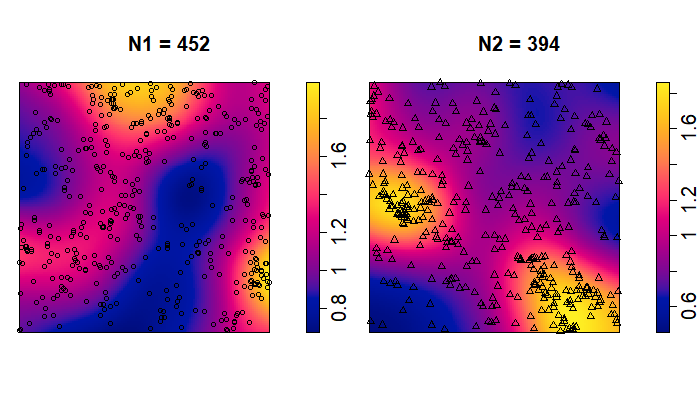}
 \vspace*{-7mm}
    \caption{ (M4) with $\rho$=-0.758.}
    \label{fig:P4_2_neg_density}
    \end{subfigure}
    
   \caption{Realization of the parametric bivariate LGCP models that are considered in Section \ref{sec:simu} ($\rho$: correlation coefficient; $\circ$: first process; $\triangle$: second process; $N_1$: total number of points of the first process; $N_2$: total number of points of the second process). The heatmap indicates the intensity function of the point process. Here, we use the sampling window $D = [-10,10]^2$ and the common first order intensities $\lambda_1^{(1)} = \lambda_1^{(2)} =  1$. }
    \label{fig:pp_simu_biv_lgcp_density} 
\end{figure}

\begin{table}[h!]
\scriptsize
\begin{center}
\begin{tabular}{c|ccccccc}
\multicolumn{4}{c}{} & 
\multicolumn{4}{c}{Number of Monte Carlo samples of $\bV_n(\widehat{\btheta}_n)$} \\
\cmidrule(lr){5-8}
Model & & Correlation & Experiment & 100 & 300 & 600 & 1000 \\ \hline \hline
\multirow{12}{*}{(M1)} & \multirow{4}{*}{Optimal $(c, R)$} & \multirow{2}{*}{Negative}  & I &(0.5, 2.00) &(0.5, 2.00) &(0.5, 2.00) &(0.5, 2.00)  \\
 &  &  & II  &(0.5, 2.00) &(0.5, 2.00) &(0.5, 2.00) &(0.5, 2.00)  \\
 & &\multirow{2}{*}{Positive} & I &(0.5, 2.50) &(0.5, 2.50) &(0.5, 2.50) &(0.5, 2.50)  \\
  &  &  & II &(0.5, 2.50) &(0.5, 2.50) &(0.5, 2.50) &(0.5, 2.50)  \\ 
\cmidrule(lr){2-8}
 &  \multirow{4}{*}{$\log \det \widehat{\Sigma}_n(\btheta_0)$} & \multirow{2}{*}{Negative} & I  &-34.03 &-34.39 &-34.46 &-34.41  \\
  & & & II  &-34.60 &-33.98 &-34.26 &-34.38  \\
 & & \multirow{2}{*}{Positive} & I  &-10.69 &-10.15 &-10.05 &-10.19  \\
  & & & II &-10.61 &-10.34 &-9.97 &-9.94  \\
\cmidrule(lr){2-8}
 & \multirow{4}{*}{Time (min)} & \multirow{2}{*}{Negative} & I &56.36 &158.03 &313.80 &514.89  \\
 & & & II  &47.64 &158.07 &310.81 &516.73  \\
 & & \multirow{2}{*}{Positive} & I &36.82 &108.59 &210.18 &370.70  \\
   & &  & II &33.29 &123.43 &216.95 &347.79  \\ \hline
\multirow{12}{*}{(M2)} & \multirow{4}{*}{Optimal $(c, R)$} & \multirow{2}{*}{Negative} & I &(0.5, 3.50) &(0.5, 3.50) &(0.5, 3.50) &(0.5, 3.50)  \\
 &  &  & II  &(0.5, 3.50) &(0.5, 3.50) &(0.5, 3.50) &(0.5, 3.50)  \\
 & & \multirow{2}{*}{Positive} & I &(0.4, 2.50) &(0.4, 2.50) &(0.4, 2.50) & (0.4, 2.50) \\
 &  &  & II &(0.4, 3.25) &(0.4, 2.50) &(0.4, 2.50) &(0.4, 2.50)  \\ 
\cmidrule(lr){2-8}
 & \multirow{4}{*}{$\log \det \widehat{\Sigma}_n(\btheta_0)$} &\multirow{2}{*}{Negative} & I  &-32.77 &-33.16 &-33.16 &-33.36  \\
  &  &  & II  &-32.33 &-32.71 &-32.83 &-32.98  \\
 & & \multirow{2}{*}{Positive} & I &-17.98 &-16.81 &-16.61 &-16.73  \\
 &  &  & II  &-16.58 &-15.93 &-16.12 &-16.20  \\
\cmidrule(lr){2-8}
& \multirow{4}{*}{Time (min)} & \multirow{2}{*}{Negative} & I &41.01 &118.57 &220.71 &386.21  \\
 &  &  & II &35.22 &110.44 &229.82 &399.69  \\
 & & \multirow{2}{*}{Positive} & I &37.72 &108.65 &211.62 &351.63  \\
  &  &  & II &34.16 &106.01 &211.83 &352.45  \\ \hline
\multirow{12}{*}{(M3)} & \multirow{4}{*}{Optimal $(c, R)$} & \multirow{2}{*}{Negative} & I &(0.5, 1.00) &(0.5, 1.00) &(0.5, 1.00) &(0.5, 1.00)  \\
 &  &  & II &(0.5, 1.00) &(0.5, 1.00) &(0.5, 1.00) &(0.5, 1.00)  \\
 & & \multirow{2}{*}{Positive} & I &(0.4, 2.50) &(0.4, 2.50) &(0.4, 2.50) & (0.4, 2.50) \\
 &  &  & II &(0.4, 3.00) &(0.4, 2.50) &(0.4, 2.50) &(0.4, 2.50)  \\
\cmidrule(lr){2-8}
 & \multirow{4}{*}{$\log \det \widehat{\Sigma}_n(\btheta_0)$} & \multirow{2}{*}{Negative} & I &-31.22 &-31.43 &-30.97 &-31.04  \\
 & & & II &-31.42 &-31.25 &-31.64 &-31.56  \\
 & & \multirow{2}{*}{Positive} & I &-21.76 &-21.60 &-21.75 &-21.44  \\
 & &  & II &-19.98 &-20.00 &-20.57 &-21.01  \\
\cmidrule(lr){2-8}
& \multirow{4}{*}{Time (min)} & \multirow{2}{*}{Negative} & I &45.76 &128.01 &253.18 & 411.64 \\
  & & & II &39.20 &126.49 &250.20 &414.76  \\
& & \multirow{2}{*}{Positive} & I &39.24 &114.95 &227.93 &362.86  \\
  & & & II &34.63 &111.82 &219.91 &354.70  \\ \hline
\multirow{12}{*}{(M4)} & \multirow{4}{*}{Optimal $(c, R)$} & \multirow{2}{*}{Negative} & I &(0.5, 3.50) &(0.5, 3.50) &(0.5, 3.50) &(0.5, 3.50)  \\
 &  &  & II &(0.5, 2.50) &(0.5, 3.50) &(0.5, 3.25) &(0.5, 3.25)  \\
 & & \multirow{2}{*}{Positive} & I &(0.5, 3.50) &(0.5, 3.50) & (0.5, 3.50)& (0.5, 3.50) \\
 &  &  & II &(0.5, 2.50) &(0.5, 3.50) &(0.5, 3.50) &(0.5, 3.50)  \\
\cmidrule(lr){2-8}
 & \multirow{4}{*}{$\log \det \widehat{\Sigma}_n(\btheta_0)$} &\multirow{2}{*}{Negative}  & I &-33.31 &-33.40 &-33.37 &-33.31  \\
 & & & II &-32.40 &-32.55 &-32.99 &-33.15  \\
 & & \multirow{2}{*}{Positive} & I &-25.69 &-23.10 &-23.44 &-23.55  \\
  & & & II &-19.63 &-24.39 &-23.63 &-23.48  \\
\cmidrule(lr){2-8}
& \multirow{4}{*}{Time (min)} & \multirow{2}{*}{Negative} & I &38.01 &104.84 &210.83 &349.03  \\
  & & & II &33.22 &107.95 &223.86 &350.76  \\
& & \multirow{2}{*}{Positive} & I &38.06 &107.68 &212.14 &350.64  \\
  & & & II &33.81 &108.43 &218.71 &347.56  \\ \hline
\end{tabular}
\caption{The optimal control parameter $(c,R)$, the log determinant of the estimated asymptotic covariance matrix based on the optimal $(c,R)$, and the total computing time of grid search based on the two realizations (Experiment I and II) for different models from bivariate LGCP in Table \ref{tab:true_par_LMC} and different Monte Carlo simulations to calculate asymptotic covariance matrix. Here, the sampling window is set to $D=[-5,5]^2$. } \label{tab:selection-add-Q} 
\end{center}
\end{table}

\begin{figure}[h!]
    \centering
    \includegraphics[width=\linewidth]{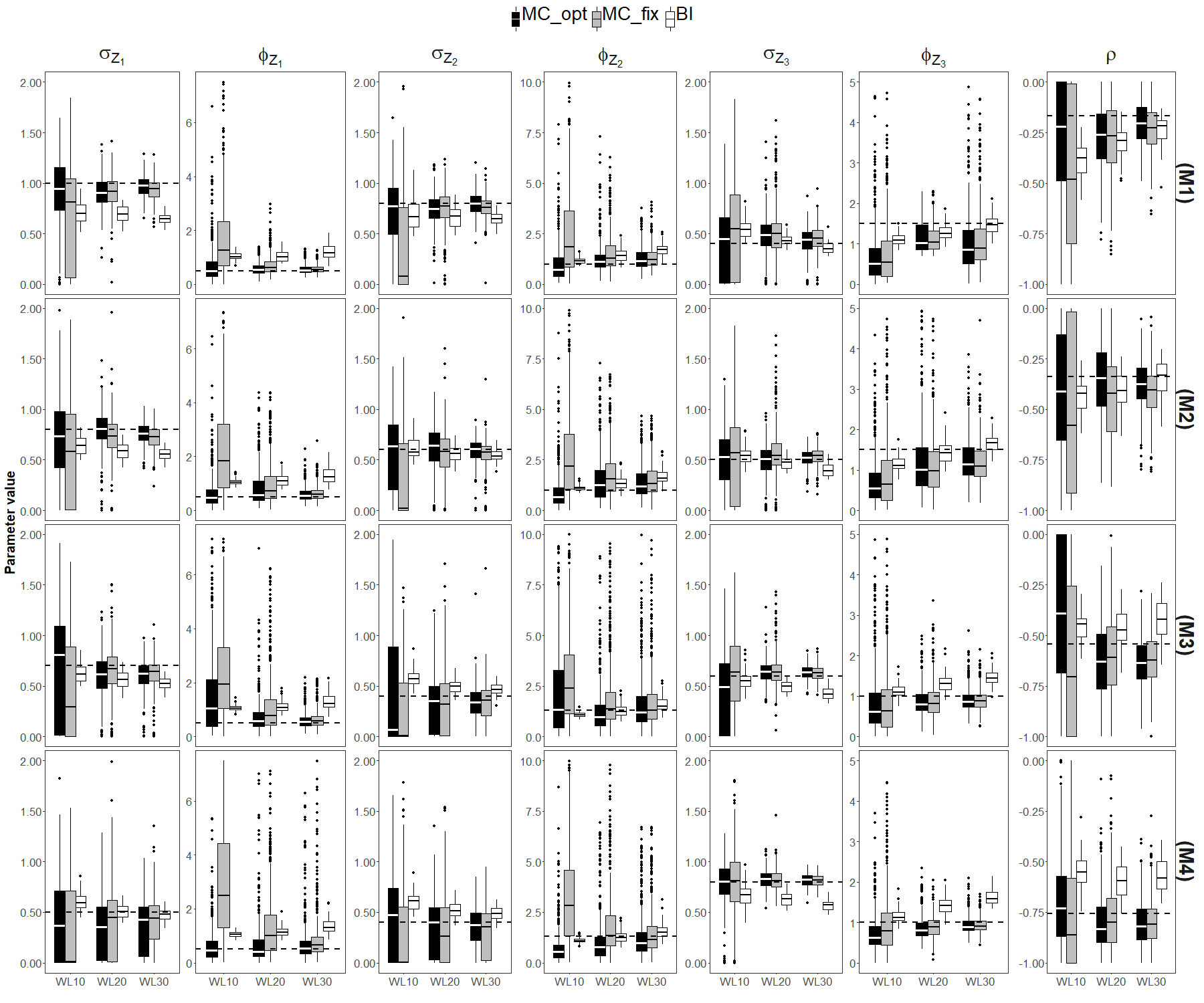}
    \caption{
Similar to Figure \ref{fig:boxplot_MCM_BI_diffWL_pos}, but for negatively correlated models.}
\label{fig:boxplot_MCM_BI_diffWL_neg}
\end{figure}

\begin{landscape}
\begin{table}[h!]
\tiny
\begin{center}
\begin{tabular}{c|c|c|c|c|c|c|c|c|c|c}
Window & Correlation & Estimator & (c,R) & $\sigma_{Z_1}$ &  $\phi_{Z_1}$ & $\sigma_{Z_2}$ & $\phi_{Z_2}$ & $\sigma_{Z_3}$ & $\phi_{Z_3}$ & $\rho$  \\
 \hline \hline
\multirow{6}{*}{WL=10}  & \multirow{3}{*}{Negative} &\multirow{2}{*}{(MC)} & Opt &0.27 \quad 0.35 \quad 0.36 &0.58 \quad 1.70 \quad 1.74 &0.29 \quad 0.37 \quad 0.39 &0.70 \quad 1.07 \quad 1.07 &0.29 \quad 0.33 \quad 0.33 &1.99 \quad 7.75 \quad 7.75 &0.24 \quad 0.28 \quad 0.30  \\ 
                        \cmidrule(lr){4-11}
                        &                           &    & Fix &0.44 \quad 0.51 \quad 0.59 &2.08 \quad 3.56 \quad 4.08 &0.52 \quad 0.44 \quad 0.61 &2.03 \quad 2.70 \quad 3.23 &0.40 \quad 0.45 \quad 0.48 &2.36 \quad 7.63 \quad 7.67 &0.39 \quad 0.38 \quad 0.48  \\ 
                        \cmidrule(lr){3-11}
                        &                           &(BI)& -- &0.29 \quad 0.11 \quad 0.31 &0.55 \quad 0.16 \quad 0.57 &0.15 \quad 0.14 \quad 0.18 &0.18 \quad 0.16 \quad 0.22 &0.16 \quad 0.10 \quad 0.18 &0.39 \quad 0.14 \quad 0.42 &0.22 \quad 0.09 \quad 0.24  \\ 
                       \cmidrule(lr){2-11}
                        & \multirow{3}{*}{Positive} &\multirow{2}{*}{(MC)} & Opt &0.27 \quad 0.36 \quad 0.37 &0.43 \quad 1.18 \quad 1.20 &0.28 \quad 0.36 \quad 0.37 &0.74 \quad 1.32 \quad 1.32 &0.41 \quad 0.45 \quad 0.47 &3.91 \quad 6.16 \quad 6.92 &0.29 \quad 0.30 \quad 0.35  \\ 
                        \cmidrule(lr){4-11}
                        &                           &    & Fix & 0.32 \quad 0.42 \quad 0.44&1.39 \quad 3.98 \quad 4.18 &0.43 \quad 0.47 \quad 0.54 &2.05 \quad 3.79 \quad 4.14 &0.52 \quad 0.60 \quad 0.60 &2.9e2 \quad 3.3e3 \quad 3.3e3 &0.29 \quad 0.35 \quad 0.36  \\ 
                        \cmidrule(lr){3-11}
                        &                           &(BI)& -- &0.28 \quad 0.14 \quad 0.31 &0.54 \quad 0.15 \quad 0.55 &0.14 \quad 0.13 \quad 0.16 &0.20 \quad 0.17 \quad 0.25 &0.17 \quad 0.08 \quad 0.18 &0.36 \quad 0.17 \quad 0.39 &0.22 \quad 0.09 \quad 0.24  \\ 
                        \cmidrule(lr){1-11}
\multirow{6}{*}{WL=20}  & \multirow{3}{*}{Negative} &\multirow{2}{*}{(MC)} & Opt &0.14 \quad 0.15 \quad 0.18 &0.17 \quad 0.23 \quad 0.24 &0.12 \quad 0.15 \quad 0.16 &0.50 \quad 0.80 \quad 0.85 &0.17 \quad 0.20 \quad 0.21 &0.47 \quad 0.42 \quad 0.52 &0.15 \quad 0.16 \quad 0.19 \\ 
                        \cmidrule(lr){4-11}
                        &                           &    & Fix &0.14 \quad 0.16 \quad 0.18 &0.27 \quad 0.36 \quad 0.41 &0.14 \quad 0.19 \quad 0.19 &0.72 \quad 0.94 \quad 1.09 &0.21 \quad 0.27 \quad 0.28 &0.48 \quad 0.41 \quad 0.52 &0.17 \quad 0.19 \quad 0.22  \\ 
                        \cmidrule(lr){3-11}
                        &                           &(BI)& -- &0.31 \quad 0.08 \quad 0.32 &0.55 \quad 0.21 \quad 0.59 &0.14 \quad 0.10 \quad 0.17 &0.46 \quad 0.32 \quad 0.55 &0.05 \quad 0.06 \quad 0.07 &0.26 \quad 0.21 \quad 0.30 &0.13 \quad 0.08 \quad 0.15  \\ 
                       \cmidrule(lr){2-11}
                        & \multirow{3}{*}{Positive} &\multirow{2}{*}{(MC)} & Opt &0.15 \quad 0.21 \quad 0.22 &0.21 \quad 0.33 \quad 0.33 &0.14 \quad 0.19 \quad 0.19 &0.42 \quad 0.60 \quad 0.61 &0.20 \quad 0.23 \quad 0.25 &1.96 \quad 3.89 \quad 4.03 &0.15 \quad 0.17 \quad 0.19  \\ 
                        \cmidrule(lr){4-11}
                        &                           &    & Fix &0.12 \quad 0.14 \quad 0.15 &0.16 \quad 0.21 \quad 0.22 &0.12 \quad 0.15 \quad 0.15 &0.43 \quad 0.68 \quad 0.69 &0.34 \quad 0.36 \quad 0.39 &4.82 \quad 5.98 \quad 7.45 & 0.23 \quad 0.22 \quad 0.27 \\ 
                        \cmidrule(lr){3-11}
                        &                           &(BI)& -- &0.33 \quad 0.08 \quad 0.33 &0.52 \quad 0.20 \quad 0.56 &0.15 \quad 0.09 \quad 0.17 &0.44 \quad 0.26 \quad 0.50 &0.07 \quad 0.05 \quad 0.09 &0.28 \quad 0.30 \quad 0.31 &0.17 \quad 0.07 \quad 0.18  \\ 
                        \cmidrule(lr){1-11}
\multirow{6}{*}{WL=30}  & \multirow{3}{*}{Negative} &\multirow{2}{*}{(MC)} & Opt &0.08 \quad 0.10 \quad 0.10 &0.13 \quad 0.17 \quad 0.18 &0.09 \quad 0.11 \quad 0.11 &0.45 \quad 0.55 \quad 0.61 &0.12 \quad 0.15 \quad 0.16 &0.80 \quad 0.85 \quad 0.95 &0.09 \quad 0.11 \quad 0.11  \\ 
                        \cmidrule(lr){4-11}
                        &                           &    & Fix &0.09 \quad 0.10 \quad 0.12 &0.12 \quad 0.15 \quad 0.16 &0.08 \quad 0.10 \quad 0.11 &0.44 \quad 0.55 \quad 0.63 &0.11 \quad 0.13 \quad 0.14 &0.73 \quad 0.75 \quad 0.85 &0.10 \quad 0.11 \quad 0.13  \\ 
                        \cmidrule(lr){3-11}
                        &                           &(BI)& -- &0.35 \quad 0.05 \quad 0.35 &0.69 \quad 0.28 \quad 0.75 &0.15 \quad 0.07 \quad 0.16 &0.72 \quad 0.35 \quad 0.80  &0.06 \quad 0.06 \quad 0.07 &0.19 \quad 0.23 \quad 0.24 &0.07 \quad 0.07 \quad 0.10  \\ 
                       \cmidrule(lr){2-11}
                        & \multirow{3}{*}{Positive} &\multirow{2}{*}{(MC)} & Opt &0.10 \quad 0.14 \quad 0.14 &0.15 \quad 0.21 \quad 0.21 &0.09 \quad 0.12 \quad 0.12 &0.31 \quad 0.48 \quad 0.48 &0.11 \quad 0.15 \quad 0.15 &2.03 \quad 4.07 \quad 4.29 &0.09 \quad 0.11 \quad 0.11  \\ 
                        \cmidrule(lr){4-11}
                        &                           &    & Fix &0.08 \quad 0.09 \quad 0.09 &0.10 \quad 0.12 \quad 0.12 &0.08 \quad 0.10 \quad 0.10 &0.25 \quad 0.38 \quad 0.38 &0.17 \quad 0.22 \quad 0.23 &3.55 \quad 5.96 \quad 6.68 &0.12 \quad 0.13 \quad 0.15  \\ 
                        \cmidrule(lr){3-11}
                        &                           &(BI)& -- &0.39 \quad 0.06 \quad 0.39 &0.60 \quad 0.21 \quad 0.63 &0.19 \quad 0.06 \quad 0.20 &0.67 \quad 0.32 \quad 0.74 &0.05 \quad 0.05 \quad 0.06 &0.40 \quad 0.40 \quad 0.50 &0.17 \quad 0.08 \quad 0.19  \\ 
\hline

\end{tabular}
\end{center}

\caption{The MAE, SD and RMSE (3 columns from left to right for each parameter) of two MC estimators and BI estimator from (M1) for different sampling windows. Here, ``Opt'' stands for the MC estimator using optimal control parameters and ``Fix'' stands for the MC estimator using fixed control parameter $(c,R)=(0.2,0.15\text{WL})$, where $\text{WL}$ is the window length. 
} 
\label{tab:bias-M1}
\end{table}
\end{landscape}

\begin{landscape}
\begin{table}[h!]
\tiny
\begin{center}
\begin{tabular}{c|c|c|c|c|c|c|c|c|c|c}
Window & Correlation & Estimator & (c,R) & $\sigma_{Z_1}$ &  $\phi_{Z_1}$ & $\sigma_{Z_2}$ & $\phi_{Z_2}$ & $\sigma_{Z_3}$ & $\phi_{Z_3}$ & $\rho$  \\
 \hline \hline
\multirow{6}{*}{WL=10}  & \multirow{3}{*}{Negative} &\multirow{2}{*}{(MC)} & Opt &0.34 \quad 0.42 \quad 0.44 &0.46 \quad 0.97 \quad 1.00 &0.31 \quad 0.38 \quad 0.38 &0.76 \quad 1.68 \quad 1.68 &0.24 \quad 0.30 \quad 0.30 &1.07 \quad 1.37 \quad 1.50 & 0.27 \quad 0.31 \quad 0.32 \\ 
                        \cmidrule(lr){4-11}
                        &                           &    & Fix &0.47 \quad 0.51 \quad 0.57 &2.90 \quad 3.84 \quad 4.76 &0.49 \quad 0.60 \quad 0.65 &9.56 \quad 21.6 \quad 23.5 &0.36 \quad 0.43 \quad 0.43 &1.93 \quad 4.18 \quad 4.21 &0.39 \quad 0.39 \quad 0.43  \\ 
                        \cmidrule(lr){3-11}
                        &                           &(BI)& -- &0.16 \quad 0.08 \quad 0.18 &0.57 \quad 0.14 \quad 0.59 &0.09 \quad 0.11 \quad 0.11 &0.13 \quad 0.13 \quad 0.18 &0.08 \quad 0.09 \quad 0.10 &0.36 \quad 0.18 \quad 0.39 &0.10 \quad 0.08 \quad 0.12  \\ 
                       \cmidrule(lr){2-11}
                        & \multirow{3}{*}{Positive} &\multirow{2}{*}{(MC)} & Opt &0.31 \quad 0.39 \quad 0.41 &0.75 \quad 3.51 \quad 3.54 &0.32 \quad 0.40 \quad 0.40 &0.90 \quad 1.99 \quad 2.00 &0.40 \quad 0.45 \quad 0.48 &5.27 \quad 7.63 \quad 8.97 &0.32 \quad 0.33 \quad 0.36  \\ 
                        \cmidrule(lr){4-11}
                        &                           &    & Fix &0.35 \quad 0.42 \quad 0.46 &1.99 \quad 5.72 \quad 6.01 &0.41 \quad 0.43 \quad 0.48 &2.47 \quad 4.86 \quad 5.28 &0.55 \quad 0.60 \quad 0.60 &7.7e2 \quad 6.2e2 \quad 6.2e2 &0.39 \quad 0.41 \quad 0.41  \\ 
                        \cmidrule(lr){3-11}
                        &                           &(BI)& -- &0.17 \quad 0.11 \quad 0.19 &0.56 \quad 0.14 \quad 0.58 &0.09 \quad 0.13 \quad 0.13 &0.15 \quad 0.16 \quad 0.21 &0.08 \quad 0.08 \quad 0.10 &0.34 \quad 0.19 \quad 0.38 &0.11 \quad 0.10 \quad 0.14  \\ 
                        \cmidrule(lr){1-11}
\multirow{6}{*}{WL=20}  & \multirow{3}{*}{Negative} &\multirow{2}{*}{(MC)} & Opt &0.14 \quad 0.19 \quad 0.18 &0.47 \quad 0.69 \quad 0.77 &0.19 \quad 0.25 \quad 0.25 &0.89 \quad 1.22 \quad 1.33 &0.13 \quad 0.18 \quad 0.18 &0.81 \quad 1.07 \quad 1.09 &0.16 \quad 0.20 \quad 0.20  \\ 
                        \cmidrule(lr){4-11}
                        &                           &    & Fix &0.16 \quad 0.22 \quad 0.23 &0.59 \quad 0.79 \quad 0.93 &0.20 \quad 0.26 \quad 0.27 &1.06 \quad 1.29 \quad 1.53 &0.16 \quad 0.22 \quad 0.23 &0.73 \quad 0.81 \quad 0.89 &0.19 \quad 0.21 \quad 0.23  \\ 
                        \cmidrule(lr){3-11}
                        &                           &(BI)& -- &0.21 \quad 0.09 \quad 0.23 &0.64 \quad 0.23 \quad 0.68 &0.08 \quad 0.09 \quad 0.10 &0.41 \quad 0.34 \quad 0.51 &0.05 \quad 0.06 \quad 0.07 &0.23 \quad 0.29 \quad 0.29 &0.09 \quad 0.10 \quad 0.11  \\ 
                       \cmidrule(lr){2-11}
                        & \multirow{3}{*}{Positive} &\multirow{2}{*}{(MC)} & Opt &0.16 \quad 0.22 \quad 0.22 &0.22 \quad 0.37 \quad 0.37 &0.15 \quad 0.20 \quad 0.20 &0.49 \quad 0.71 \quad 0.71 &0.14 \quad 0.18 \quad 0.20 &2.16 \quad 3.50 \quad 3.84 &0.15 \quad 0.18 \quad 0.19  \\ 
                        \cmidrule(lr){4-11}
                        &                           &    & Fix &0.14 \quad 0.18 \quad 0.19 &0.27 \quad 0.47 \quad 0.49 &0.16 \quad 0.22 \quad 0.23 &0.60 \quad 0.93 \quad 0.94 &0.28 \quad 0.29 \quad 0.36 &4.35 \quad 5.01 \quad 6.39 &0.23 \quad 0.21 \quad 0.27  \\ 
                        \cmidrule(lr){3-11}
                        &                           &(BI)& -- &0.23 \quad 0.08 \quad 0.24 &0.60 \quad 0.21 \quad 0.63 &0.08 \quad 0.08 \quad 0.10 &0.34 \quad 0.26 \quad 0.42 &0.05 \quad 0.06 \quad 0.06 &0.28 \quad 0.33 \quad 0.34 &0.12 \quad 0.08 \quad 0.14  \\ 
                        \cmidrule(lr){1-11}
\multirow{6}{*}{WL=30}  & \multirow{3}{*}{Negative} &\multirow{2}{*}{(MC)} & Opt &0.09 \quad 0.11 \quad 0.12 &0.20 \quad 0.27 \quad 0.28 &0.09 \quad 0.13 \quad 0.13 &0.66 \quad 0.86 \quad 0.96 &0.07 \quad 0.08 \quad 0.09 &0.51 \quad 0.56 \quad 0.61 &0.10 \quad 0.12 \quad 0.12  \\ 
                        \cmidrule(lr){4-11}
                        &                           &    & Fix &0.11 \quad 0.11 \quad 0.14 &0.21 \quad 0.27 \quad 0.30 &0.09 \quad 0.13 \quad 0.13 &0.73 \quad 0.90 \quad 1.06 &0.07 \quad 0.08 \quad 0.09 &0.50 \quad 0.52 \quad 0.60 &0.11 \quad 0.12 \quad 0.14  \\ 
                        \cmidrule(lr){3-11}
                        &                           &(BI)& -- &0.24 \quad 0.06 \quad 0.25 &0.81 \quad 0.35 \quad 0.88 &0.07 \quad 0.06 \quad 0.08 &0.69 \quad 0.44 \quad 0.82 &0.10 \quad 0.06 \quad 0.12 &0.24 \quad 0.25 \quad 0.29 &0.08 \quad 0.09 \quad 0.09  \\ 
                       \cmidrule(lr){2-11}
                        & \multirow{3}{*}{Positive} &\multirow{2}{*}{(MC)} & Opt &0.09 \quad 0.11 \quad 0.11 &0.14 \quad 0.20 \quad 0.20 &0.09 \quad 0.11 \quad 0.11 &0.39 \quad 0.57 \quad 0.57&0.11 \quad 0.14 \quad 0.15 &0.60 \quad 0.75 \quad 0.76 &0.10 \quad 0.12 \quad 0.13  \\ 
                        \cmidrule(lr){4-11}
                        &                           &    & Fix &0.08 \quad 0.10 \quad 0.10 &0.13 \quad 0.18 \quad 0.19 &0.09 \quad 0.11 \quad 0.11 &0.35 \quad 0.50 \quad 0.50 &0.11 \quad 0.15 \quad 0.16 &0.62 \quad 0.77 \quad 0.82 &0.10 \quad 0.12 \quad 0.13  \\ 
                        \cmidrule(lr){3-11}
                        &                           &(BI)& -- &0.30 \quad 0.06 \quad 0.31 &0.63 \quad 0.26 \quad 0.68 &0.13 \quad 0.05 \quad 0.14 &0.52 \quad 0.30 \quad 0.59 &0.04 \quad 0.05 \quad 0.05 &0.60 \quad 0.40 \quad 0.70 &0.16 \quad 0.08 \quad 0.18  \\ 
\hline

\end{tabular}
\end{center}

\caption{Similar to Table \ref{tab:bias-M1}, but for (M2).
} 
\label{tab:bias-M2}
\end{table}
\end{landscape}

\begin{landscape}
\begin{table}[h!]
\tiny
\begin{center}
\begin{tabular}{c|c|c|c|c|c|c|c|c|c|c}
Window & Correlation & Estimator & (c,R) & $\sigma_{Z_1}$ &  $\phi_{Z_1}$ & $\sigma_{Z_2}$ & $\phi_{Z_2}$ & $\sigma_{Z_3}$ & $\phi_{Z_3}$ & $\rho$  \\
 \hline \hline
\multirow{6}{*}{WL=10}  & \multirow{3}{*}{Negative} &\multirow{2}{*}{(MC)} & Opt &0.44 \quad 0.51 \quad 0.51 &1.61 \quad 2.52 \quad 2.91 &0.46 \quad 0.51 \quad 0.52 &7.14 \quad 20.7 \quad 21.7 &0.32 \quad 0.35 \quad 0.39 &1.44 \quad 3.50 \quad 3.57 &0.34 \quad 0.36 \quad 0.39  \\ 
                        \cmidrule(lr){4-11}
                        &                           &    & Fix &0.48 \quad 0.50 \quad 0.55 &2.68 \quad 3.06 \quad 4.03 &0.42 \quad 0.47 \quad 0.49 &10.5 \quad 24.7 \quad 26.7 &0.34 \quad 0.41 \quad 0.41 &1.29 \quad 3.29 \quad 3.33 &0.35 \quad 0.38 \quad 0.39  \\ 
                        \cmidrule(lr){3-11}
                        &                           &(BI)& -- &0.10 \quad 0.09 \quad 0.12 &0.56 \quad 0.12 \quad 0.58 &0.18 \quad 0.09 \quad 0.20 &0.22 \quad 0.14 \quad 0.24 &0.09 \quad 0.10 \quad 0.10 &0.16 \quad 0.18 \quad 0.22 &0.09 \quad 0.07 \quad 0.11  \\ 
                       \cmidrule(lr){2-11}
                        & \multirow{3}{*}{Positive} &\multirow{2}{*}{(MC)} & Opt &0.32 \quad 0.40 \quad 0.40 &0.44 \quad 0.81 \quad 0.83 &0.33 \quad 0.38 \quad 0.38 &9.17 \quad 1.5e2 \quad 1.5e2 &0.34 \quad 0.38 \quad 0.42 &4.84 \quad 6.44 \quad 7.87 &0.25 \quad 0.28 \quad 0.29  \\ 
                        \cmidrule(lr){4-11}
                        &                           &    & Fix &0.38 \quad 0.45 \quad 0.47 &0.81 \quad 1.38 \quad 1.52 &0.38 \quad 0.46 \quad 0.47 &31.3 \quad 3.3e2 \quad 3.3e2 &0.62 \quad 0.67 \quad 0.67 &12.9 \quad 12.5 \quad 17.9 &0.42 \quad 0.43 \quad 0.44  \\ 
                        \cmidrule(lr){3-11}
                        &                           &(BI)& -- &0.12 \quad 0.10 \quad 0.13 &0.55 \quad 0.12 \quad 0.56 &0.19 \quad 0.10 \quad 0.21 &0.22 \quad 0.16 \quad 0.25 &0.07 \quad 0.10 \quad 0.09 &0.19 \quad 0.19 \quad 0.24 &0.09 \quad 0.10 \quad 0.11  \\ 
                        \cmidrule(lr){1-11}
\multirow{6}{*}{WL=20}  & \multirow{3}{*}{Negative} &\multirow{2}{*}{(MC)} & Opt &0.19 \quad 0.26 \quad 0.28 &0.39 \quad 0.67 \quad 0.71 &0.22 \quad 0.27 \quad 0.28 &0.86 \quad 1.42 \quad 1.42 &0.09 \quad 0.11 \quad 0.12 &0.30 \quad 0.35 \quad 0.37 &0.16 \quad 0.18 \quad 0.19  \\ 
                        \cmidrule(lr){4-11}
                        &                           &    & Fix &0.18 \quad 0.25 \quad 0.26 &0.79 \quad 1.17 \quad 1.35 &0.27 \quad 0.32 \quad 0.33 &1.21 \quad 1.99 \quad 2.10 &0.11 \quad 0.15 \quad 0.16 &0.36 \quad 0.46 \quad 0.47 &0.17 \quad 0.19 \quad 0.20  \\ 
                        \cmidrule(lr){3-11}
                        &                           &(BI)& -- &0.14 \quad 0.08 \quad 0.16 &0.62 \quad 0.24 \quad 0.66 &0.11 \quad 0.08 \quad 0.13 &0.25 \quad 0.33 \quad 0.32 &0.11 \quad 0.05 \quad 0.12 &0.33 \quad 0.25 \quad 0.40 &0.10 \quad 0.09 \quad 0.12  \\ 
                       \cmidrule(lr){2-11}
                        & \multirow{3}{*}{Positive} &\multirow{2}{*}{(MC)} & Opt &0.10 \quad 0.09 \quad 0.12 &0.56 \quad 0.12 \quad 0.58 &0.18 \quad 0.09 \quad 0.20 &0.22 \quad 0.14 \quad 0.24 &0.09 \quad 0.10 \quad 0.10 &0.16 \quad 0.18 \quad 0.22 &0.09 \quad 0.07 \quad 0.11  \\ 
                        \cmidrule(lr){4-11}
                        &                           &    & Fix &0.16 \quad 0.21 \quad 0.21 &0.35 \quad 0.59 \quad 0.62 &0.23 \quad 0.29 \quad 0.29 &0.90 \quad 1.31 \quad 1.32 &0.20 \quad 0.24 \quad 0.29 &1.36 \quad 1.84 \quad 2.19 &0.17 \quad 0.18 \quad 0.20  \\ 
                        \cmidrule(lr){3-11}
                        &                           &(BI)& -- &0.16 \quad 0.07 \quad 0.18 &0.56 \quad 0.20 \quad 0.60 &0.08 \quad 0.07 \quad 0.10 &0.21 \quad 0.25 \quad 0.25 &0.07 \quad 0.05 \quad 0.08 &0.45 \quad 0.28 \quad 0.52 &0.06 \quad 0.08 \quad 0.08  \\ 
                        \cmidrule(lr){1-11}
\multirow{6}{*}{WL=30}  & \multirow{3}{*}{Negative} &\multirow{2}{*}{(MC)} & Opt &0.13 \quad 0.15 \quad 0.17 &0.21 \quad 0.29 \quad 0.30 &0.15 \quad 0.18 \quad 0.20 &0.94 \quad 1.47 \quad 1.52 &0.06 \quad 0.07 \quad 0.08 &0.23 \quad 0.27 \quad 0.29 &0.13 \quad 0.12 \quad 0.15  \\ 
                        \cmidrule(lr){4-11}
                        &                           &    & Fix &0.12 \quad 0.14 \quad 0.16 &0.24 \quad 0.33 \quad 0.36 &0.16 \quad 0.20 \quad 0.21 &0.99 \quad 1.64 \quad 1.72 &0.06 \quad 0.07 \quad 0.08 &0.22 \quad 0.25 \quad 0.27 &0.12 \quad 0.13 \quad 0.15  \\ 
                        \cmidrule(lr){3-11}
                        &                           &(BI)& -- &0.17 \quad 0.07 \quad 0.19 &0.79 \quad 0.35 \quad 0.86 &0.08 \quad 0.06 \quad 0.09 &0.35 \quad 0.38 \quad 0.48 &0.18 \quad 0.05 \quad 0.18 &0.47 \quad 0.22 \quad 0.52 &0.13 \quad 0.09 \quad 0.15  \\ 
                       \cmidrule(lr){2-11}
                        & \multirow{3}{*}{Positive} &\multirow{2}{*}{(MC)} & Opt &0.09 \quad 0.11 \quad 0.11 &0.15 \quad 0.21 \quad 0.22 &0.11 \quad 0.14 \quad 0.14 &0.62 \quad 1.05 \quad 1.05 &0.08 \quad 0.07 \quad 0.10 &0.39 \quad 0.52 \quad 0.61 &0.11 \quad 0.12 \quad 0.13  \\ 
                        \cmidrule(lr){4-11}
                        &                           &    & Fix &0.09 \quad 0.12 \quad 0.12 &0.17 \quad 0.23 \quad 0.24 &0.12 \quad 0.16 \quad 0.16 &0.77 \quad 1.24 \quad 1.25 &0.06 \quad 0.08 \quad 0.08 &0.48 \quad 0.65 \quad 0.76 &0.09 \quad 0.11 \quad 0.11  \\ 
                        \cmidrule(lr){3-11}
                        &                           &(BI)& -- &0.24 \quad 0.06 \quad 0.25 &0.61 \quad 0.23 \quad 0.65 &0.04 \quad 0.05 \quad 0.05 &0.23 \quad 0.28 \quad 0.30 &0.09 \quad 0.04 \quad 0.10 &0.85 \quad 0.35 \quad 0.91 &0.07 \quad 0.07 \quad 0.09  \\ 
\hline

\end{tabular}
\end{center}

\caption{Similar to Table \ref{tab:bias-M1}, but for (M3).
} 
\label{tab:bias-M3}
\end{table}
\end{landscape}

\begin{landscape}
\begin{table}[h!]
\tiny
\begin{center}
\begin{tabular}{c|c|c|c|c|c|c|c|c|c|c}
Window & Correlation & Estimator & (c,R) & $\sigma_{Z_1}$ &  $\phi_{Z_1}$ & $\sigma_{Z_2}$ & $\phi_{Z_2}$ & $\sigma_{Z_3}$ & $\phi_{Z_3}$ & $\rho$  \\
 \hline \hline
\multirow{6}{*}{WL=10}  & \multirow{3}{*}{Negative} &\multirow{2}{*}{(MC)} & Opt &0.37 \quad 0.42 \quad 0.43 &0.51 \quad 1.18 \quad 1.20 &0.35 \quad 0.39 \quad 0.39 &0.91 \quad 1.23 \quad 1.32 &0.16 \quad 0.22 \quad 0.22 &0.49 \quad 0.91 \quad 0.93 &0.19 \quad 0.24 \quad 0.25  \\ 
                        \cmidrule(lr){4-11}
                        &                           &    & Fix &0.48 \quad 0.50 \quad 0.53 &5.53 \quad 7.36 \quad 9.17 &0.42 \quad 0.45 \quad 0.46 &9.25 \quad 20.65 \quad 22.50 &0.26 \quad 0.35 \quad 0.35 &0.81 \quad 1.72 \quad 1.73 &0.26 \quad 0.32 \quad 0.32  \\ 
                        \cmidrule(lr){3-11}
                        &                           &(BI)& -- &0.11 \quad 0.09 \quad 0.13 &0.56 \quad 0.11 \quad 0.57 &0.21 \quad 0.09 \quad 0.23 &0.22 \quad 0.13 \quad 0.25 &0.15 \quad 0.12 \quad 0.17 &0.19 \quad 0.20 \quad 0.25 &0.21 \quad 0.09 \quad 0.23  \\ 
                       \cmidrule(lr){2-11}
                        & \multirow{3}{*}{Positive} &\multirow{2}{*}{(MC)} & Opt &0.39 \quad 0.41 \quad 0.42 &0.50 \quad 2.84 \quad 2.84 &0.35 \quad 0.39 \quad 0.39 &1.48 \quad 5.34 \quad 5.34 &0.23 \quad 0.31 \quad 0.33 &1.98 \quad 3.61 \quad 3.96 &0.19 \quad 0.23 \quad 0.23  \\ 
                        \cmidrule(lr){4-11}
                        &                           &    & Fix &0.38 \quad 0.44 \quad 0.45 &4.88 \quad 26.1 \quad 26.5 &0.38 \quad 0.47 \quad 0.47 &1.81 \quad 5.57 \quad 5.61 &0.66 \quad 0.71 \quad 0.74 &6.85 \quad 5.17 \quad 8.46 &0.29 \quad 0.38 \quad 0.39  \\ 
                        \cmidrule(lr){3-11}
                        &                           &(BI)& -- &0.10 \quad 0.10 \quad 0.14 &0.56 \quad 0.11 \quad 0.57 &0.22 \quad 0.11 \quad 0.25 &0.23 \quad 0.17 \quad 0.26 &0.12 \quad 0.12 \quad 0.15 &0.21 \quad 0.21 \quad 0.27 &0.18 \quad 0.11 \quad 0.22  \\ 
                        \cmidrule(lr){1-11}
\multirow{6}{*}{WL=20}  & \multirow{3}{*}{Negative} &\multirow{2}{*}{(MC)} & Opt &0.28 \quad 0.29 \quad 0.33 &0.50 \quad 0.94 \quad 0.97 &0.22 \quad 0.26 \quad 0.27 &0.96 \quad 1.78 \quad 1.78 &0.08 \quad 0.09 \quad 0.09 &0.26 \quad 0.29 \quad 0.31 &0.12 \quad 0.14 \quad 0.14  \\ 
                        \cmidrule(lr){4-11}
                        &                           &    & Fix &0.28 \quad 0.36 \quad 0.37 &1.33 \quad 2.27 \quad 2.58 &0.29 \quad 0.33 \quad 0.34 &1.35 \quad 2.68 \quad 2.81 &0.08 \quad 0.10 \quad 0.10 &0.23 \quad 0.27 \quad 0.29 &0.14 \quad 0.17 \quad 0.17  \\ 
                        \cmidrule(lr){3-11}
                        &                           &(BI)& -- &0.06 \quad 0.07 \quad 0.07 &0.67 \quad 0.21 \quad 0.70 &0.13 \quad 0.08 \quad 0.15 &0.22 \quad 0.29 \quad 0.29 &0.17 \quad 0.06 \quad 0.18 &0.43 \quad 0.23 \quad 0.48 &0.16 \quad 0.09 \quad 0.19  \\ 
                       \cmidrule(lr){2-11}
                        & \multirow{3}{*}{Positive} &\multirow{2}{*}{(MC)} & Opt &0.23 \quad 0.33 \quad 0.33 &0.41 \quad 0.65 \quad 0.67 &0.21 \quad 0.31 \quad 0.31 &0.71 \quad 0.79 \quad 0.86 &0.08 \quad 0.11 \quad 0.11 &0.55 \quad 1.02 \quad 1.11 &0.12 \quad 0.15 \quad 0.16  \\ 
                        \cmidrule(lr){4-11}
                        &                           &    & Fix &0.23 \quad 0.32 \quad 0.32 &0.42 \quad 0.62 \quad 0.65 &0.23 \quad 0.31 \quad 0.31 &0.75 \quad 0.82 \quad 0.90 &0.13 \quad 0.20 \quad 0.21 &1.52 \quad 2.84 \quad 3.17 &0.11 \quad 0.14 \quad 0.14  \\ 
                        \cmidrule(lr){3-11}
                        &                           &(BI)& -- &0.05 \quad 0.06 \quad 0.07 &0.61 \quad 0.17 \quad 0.64 &0.07 \quad 0.07 \quad 0.09 &0.18 \quad 0.19 \quad 0.21 &0.12 \quad 0.06 \quad 0.13 &0.52 \quad 0.28 \quad 0.59 &0.08 \quad 0.07 \quad 0.10  \\ 
                        \cmidrule(lr){1-11}
\multirow{6}{*}{WL=30}  & \multirow{3}{*}{Negative} &\multirow{2}{*}{(MC)} & Opt &0.22 \quad 0.25 \quad 0.28 &0.42 \quad 0.85 \quad 0.88 &0.17 \quad 0.21 \quad 0.22 &0.78 \quad 1.14 \quad 1.14 &0.05 \quad 0.06 \quad 0.07 &0.19 \quad 0.23 \quad 0.24 &0.10 \quad 0.11 \quad 0.12  \\ 
                        \cmidrule(lr){4-11}
                        &                           &    & Fix &0.20 \quad 0.25 \quad 0.27 &0.55 \quad 1.04 \quad 1.12 &0.19 \quad 0.23 \quad 0.24 &0.80 \quad 1.18 \quad 1.19 &0.05 \quad 0.06 \quad 0.06 &0.16 \quad 0.18 \quad 0.20 &0.10 \quad 0.12 \quad 0.12  \\ 
                        \cmidrule(lr){3-11}
                        &                           &(BI)& -- &0.05 \quad 0.06 \quad 0.07 &0.88 \quad 0.32 \quad 0.93 &0.10 \quad 0.06 \quad 0.11 &0.34 \quad 0.41 \quad 0.49 &0.23 \quad 0.05 \quad 0.24 &0.61 \quad 0.19 \quad 0.64 &0.18 \quad 0.08 \quad 0.20  \\ 
                       \cmidrule(lr){2-11}
                        & \multirow{3}{*}{Positive} &\multirow{2}{*}{(MC)} & Opt &0.13 \quad 0.17 \quad 0.17 &0.27 \quad 0.40 \quad 0.41 &0.11 \quad 0.14 \quad 0.14 &0.58 \quad 0.80 \quad 0.81 &0.06 \quad 0.07 \quad 0.08 &0.22 \quad 0.29 \quad 0.31 &0.08 \quad 0.10 \quad 0.11 \\ 
                        \cmidrule(lr){4-11}
                        &                           &    & Fix &0.13 \quad 0.18 \quad 0.18 &0.31 \quad 0.45 \quad 0.47 &0.12 \quad 0.16 \quad 0.16 &0.66 \quad 0.88 \quad 0.89 &0.06 \quad 0.07 \quad 0.07 &0.24 \quad 0.30 \quad 0.34 &0.08 \quad 0.11 \quad 0.11  \\ 
                        \cmidrule(lr){3-11}
                        &                           &(BI)& -- &0.12 \quad 0.05 \quad 0.13 &0.67 \quad 0.24 \quad 0.71 &0.05 \quad 0.06 \quad 0.06 &0.20 \quad 0.25 \quad 0.26 &0.14 \quad 0.04 \quad 0.15 &0.87 \quad 0.28 \quad 0.91 &0.04 \quad 0.06 \quad 0.06  \\ 
\hline

\end{tabular}
\end{center}

\caption{Similar to Table \ref{tab:bias-M1}, but for (M4).
} 
\label{tab:bias-M4}
\end{table}
\end{landscape}


\begin{landscape}
\begin{table}[h!]
\scriptsize
\begin{center}
\begin{tabular}{c|c|c|c|c|c|c|c|c|c}
\multicolumn{4}{c}{} & 
\multicolumn{6}{c}{MAE, SD and RMSE of estimators} \\
\cmidrule(lr){5-10}
Window & Estimator & $(c,R)$ & time & $\kappa$ & $\eta$ & $\alpha_{1}$ & $\alpha_{2}$ & $\alpha_{3}$ & $\alpha_{4}$ \\
 \hline \hline
\multirow{3}{*}{WL = 10} & \multirow{2}{*}{MC} & Opt &2.65 &0.08 \quad 0.12 \quad 0.14 &0.10 \quad 0.23 \quad 0.24 &0.33 \quad 0.39 \quad 0.43 &0.19 \quad 0.23 \quad 0.24 &0.32 \quad 0.39 \quad 0.39 &0.31 \quad 0.37 \quad 0.39 \\
 & & Fix &1.66  &0.09 \quad 0.13 \quad 0.14 &0.11 \quad 0.23 \quad 0.25 &0.33 \quad 0.39 \quad 0.43 &0.18 \quad 0.21 \quad 0.22 &0.31 \quad 0.38 \quad 0.39 &0.31 \quad 0.37 \quad 0.38 \\
\cmidrule(lr){2-10}
& WCL & -- & 7.21 &0.08 \quad 0.14 \quad 0.16 &0.06 \quad 0.07 \quad 0.08 &0.41 \quad 0.53 \quad 0.53 &0.21 \quad 0.26 \quad 0.26 &0.36 \quad 0.44 \quad 0.45 &0.36 \quad 0.45 \quad 0.45 \\
\cmidrule(lr){1-10} 
\multirow{3}{*}{WL = 20} & \multirow{2}{*}{MC} & Opt &1.92 &0.03 \quad 0.04 \quad 0.04 &0.04 \quad 0.05 \quad 0.06 &0.17 \quad 0.20 \quad 0.20 &0.10 \quad 0.12 \quad 0.12 &0.15 \quad 0.19 \quad 0.19 &0.18 \quad 0.21 \quad 0.22 \\
 &  & Fix & 2.75 &0.03 \quad 0.03 \quad 0.04 &0.04 \quad 0.05 \quad 0.05 &0.18 \quad 0.21 \quad 0.22 &0.11 \quad 0.13 \quad 0.13 &0.17 \quad 0.22 \quad 0.22 &0.19 \quad 0.23 \quad 0.24 \\
\cmidrule(lr){2-10}
& WCL & -- & 6.09 &0.03 \quad 0.04 \quad 0.04 &0.03 \quad 0.05 \quad 0.05 &0.20 \quad 0.25 \quad 0.25 &0.12 \quad 0.15 \quad 0.15 &0.18 \quad 0.23 \quad 0.23 &0.19 \quad 0.23 \quad 0.24 \\
\cmidrule(lr){1-10}
\multirow{3}{*}{WL = 30} & \multirow{2}{*}{MC} & Opt &2.17 &0.02 \quad 0.02 \quad 0.02 &0.02 \quad 0.02 \quad 0.03 &0.13 \quad 0.16 \quad 0.16 &0.07 \quad 0.09 \quad 0.09 &0.10 \quad 0.13 \quad 0.13 &0.12 \quad 0.14 \quad 0.14 \\
 &  & Fix &7.42 &0.02 \quad 0.02 \quad 0.02 &0.03 \quad 0.04 \quad 0.04 &0.15 \quad 0.18 \quad 0.18 &0.08 \quad 0.10 \quad 0.10 &0.12 \quad 0.15 \quad 0.16 &0.14 \quad 0.16 \quad 0.17 \\
\cmidrule(lr){2-10}
& WCL & -- & 7.61 &0.02 \quad 0.02 \quad 0.02 &0.02 \quad 0.03 \quad 0.03 &0.14 \quad 0.18 \quad 0.18 &0.09 \quad 0.10 \quad 0.10 &0.12 \quad 0.15 \quad 0.15 &0.13 \quad 0.16 \quad 0.16 \\
\hline
\end{tabular}
\end{center}

\caption{Average computing time per simulation (unit: minute), MAE, SD and RMSE (3 columns from left to right for each parameter) for MC and WCL estimators from multivariate PSNCP Models for different sampling windows. Here, ``Opt'' stands for the MC estimator using optimal control parameters and ``Fix'' stands for the MC estimator using fixed control parameter $(c,R)=(0.2,0.15\text{WL})$, where $\text{WL}$ is the window length. 
} 
\label{tab:comp-time-est-PSNCP}
\end{table}
\end{landscape}

\section{Application: Terrorism in Nigeria} \label{sec:applic} 

\begin{figure}[h!]
\centering
\includegraphics[width=0.4\textwidth]{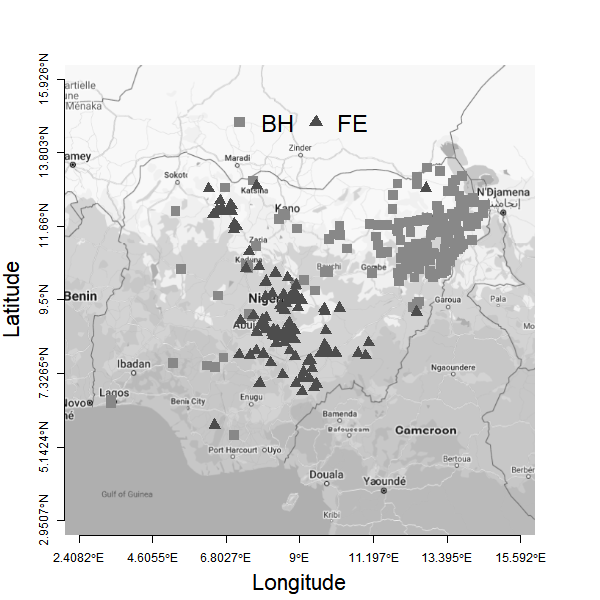}
\caption{Point pattern of terror attacks by Boko Haram (BH; 436 terror attacks) and Fulani Extremists (FE; 156 terror attacks) in Nigeria during 2014.}
\label{fig:pointpattern}
\end{figure}

In this section, we apply our methods to the bivariate point pattern data of the 2014 terrorism attacks in Nigeria. 
Figure \ref{fig:pointpattern} plots the point pattern of terror attacks by Boko Haram (BH; 436 terror attacks) and Fulani Extremists (FE; 156 terror attacks) in Nigeria during 2014 which is obtained from the Global Terrorism Database \citep[GTD;][]{start}.
In the raw data, there are several events with identical spatial coordinates which occur at different times. Thus, we add random Gaussian noise with a standard deviation of $10^{-3}$ degrees (${}^\circ$) to both coordinates to distinguish these events. We observe in Figure \ref{fig:pointpattern} that the BH attacks are mostly concentrated in the northeast corner of the country's border, while the majority of the FE attacks are located in the middle of the countryside. This indicates repulsiveness between the two sources of terror attacks.

Now, we fit our data using the bivariate LGCP model with negative cross-correlation. As referees criticized, since the point pattern data may indicate that the first-order intensities are not homogeneous over the domain, it is more appropriate to fit inhomogeneous point process models (e.g., inhomogeneous bivariate LGCP model) to describe the nature of the given point pattern data. 
However, the likelihood-ratio based inhomogeneity test for both marginal point patterns of BH and FE did not reject the null of homogeneity and one can still obtain meaningful information on the second-order interactions within and between BH and FE by fitting the (homogeneous) bivariate LGCP model as in Section \ref{sec:sim-model}.
To elaborate, intriguingly, under the SOIRS framework, we show that the non-edge corrected $Q$-function estimator as in (\ref{eq:Q0}) estimates the second-order structure of the (stationary) intensity-reweighted point process. Therefore, the fitted scale and range parameter values in Section \ref{sec:sim-model} may have interpretation even in the case when the process is inhomogeneous. Details on this work will be reported in future research.


Back to our bivariate terrorist attack data, let $\btheta = (\sigma_{\text{BH}}, \phi_{\text{BH}}, \sigma_{\text{FE}}, \phi_{\text{FE}}, \sigma_{\text{Common}}, \phi_{\text{Common}})^\top$ be the parameter of interest, where the indices ``BH'', ``FE'', and ``Common'' correspond to $Z_1$, $Z_2$, and $Z_3$ in (\ref{eq:exp-cov}), respectively. Since it is apparent from the figure above that BH and FE repel each other, we set $b=-1$ to indicate negative cross-correlation. Using the methods in Section \ref{sec:select_tune_par} (see also Section \ref{sec:opt_control_par}), we search for the optimal control parameters over the grids $c \in \{0.1, 0.2, 0.3, 0.4, 0.5\}$ and $R \in \{60 \text{km}, 80 \text{km}, \dots, 500 \text{km}\}$, where the sampling window is approximately $D = 1,347 \times 1,088 \text{km}^2$. Consequently, we select the optimal parameters $(c, R) = (0.1, 420 \text{km})$.

In Table \ref{tab:est_par_biv_terrorism_LMC_0.1_420}, we report the parameter estimation, asymptotic standard error estimator, and two 95\% confidence intervals (CIs): asymptotic and simulation-based, as described in Section \ref{sec:CRs}. When reporting parameter estimates, we also include the cross-correlation coefficient $\rho$ between BH and FE (see, (\ref{eq:rho})). From the results in Table \ref{tab:est_par_biv_terrorism_LMC_0.1_420}, we observe that both the 95\% asymptotic and simulation-based CIs suggest no significant difference in the scale ($\sigma_{\text{BH}}$ and $\sigma_{\text{FE}}$) and range ($\phi_{\text{BH}}$ and $\phi_{\text{FE}}$) parameters associated with BH and FE. The appearance of negative values in the lower bound of the asymptotic confidence interval is due to the large standard error. However, both 95\% CIs of the cross-correlation coefficient $\widehat{\rho}$ lie on the negative side, suggesting significant repulsion between BH and FE attacks.
\begin{table}[h!]
    \centering
\small
    \begin{tabular}{c|cccc}
\hline
       Param. & EST & Asymptotic SE 
        & 95\% asymptotic CI
        & 95\% simulation-based CI
        \\  \hline \hline
    $\sigma_{\text{BH}}$ &1.28 &0.41 & (-0.47, 2.09) & (0.22, 1.58) \\
    $\phi_{\text{BH}}$ &63.99 &58.04 & (-49.77, 177.75) & (5.32, 158.62) \\
    $\sigma_{\text{FE}}$ &1.94 &0.46 & (1.05, 2.83) & (1.15, 2.47) \\
    $\phi_{\text{FE}}$ &12.68 &24.72 & (-35.77, 61.13) & (1.68, 45.48) \\
    $\sigma_{\text{Common}}$ &1.33 &0.44 & (0.48, 2.19) & (0.58, 1.54) \\
    $\phi_{\text{Common}}$ &370.43 &305.90 & (-229.12, 969.99) & (46.46, 400.95) \\
    $\rho$ &-0.41 &0.17 & (-0.74, -0.08) & (-0.62, -0.11) \\
      \hline
    \end{tabular}
\caption{
Estimated parameter value (EST), asymptotic standard error (SE), and 95\% asymptotic and simulation-based confidence intervals (CI) for the minimum contrast method to fit the terror attacks in Nigeria in 2014 by BH and FE. Units of $\phi_{x}$ ($x \in \{\text{BH}, \text{FE}, \text{Common}\}$) are in kilometers.
}
    \label{tab:est_par_biv_terrorism_LMC_0.1_420}
\end{table}


\bibliographystyle{plainnat}
\bibliography{mybib}

\begin{thebibliography}{47}
\providecommand{\natexlab}[1]{#1}
\providecommand{\url}[1]{\texttt{#1}}
\expandafter\ifx\csname urlstyle\endcsname\relax
  \providecommand{\doi}[1]{doi: #1}\else
  \providecommand{\doi}{doi: \begingroup \urlstyle{rm}\Url}\fi

\bibitem[Baddeley et~al.(2014)Baddeley, Jammalamadaka, and Nair]{p:bad-14}
A.~Baddeley, A.~Jammalamadaka, and G.~Nair.
\newblock Multitype point process analysis of spines on the dendrite network of
  a neuron.
\newblock \emph{J. R. Stat. Soc. Ser. C. Appl. Stat.}, 63\penalty0
  (5):\penalty0 673--694, 2014.

\bibitem[Baddeley et~al.(2000)Baddeley, M{\o}ller, and Waagepetersen]{p:bad-00}
A.~J. Baddeley, J.~M{\o}ller, and R.~Waagepetersen.
\newblock Non-and semi-parametric estimation of interaction in inhomogeneous
  point patterns.
\newblock \emph{Stat. Neerl.}, 54\penalty0 (3):\penalty0 329--350, 2000.

\bibitem[Biscio and Svane(2022)]{p:bis-22}
C.~A.~N. Biscio and A.~M. Svane.
\newblock A functional central limit theorem for the empirical {R}ipley’s
  {K}-function.
\newblock \emph{Electron. J. Stat.}, 16\penalty0 (1):\penalty0 3060--3098,
  2022.

\bibitem[Biscio and Waagepetersen(2019)]{p:bis-19}
C.~A.~N. Biscio and R.~Waagepetersen.
\newblock A general central limit theorem and a subsampling variance estimator
  for $\alpha$-mixing point processes.
\newblock \emph{Scand. J. Stat.}, 46\penalty0 (4):\penalty0 1168--1190, 2019.

\bibitem[Brillinger(1981)]{b:bri-81}
D.~R. Brillinger.
\newblock \emph{Time series: {D}ata {A}nalysis and {T}heory}, volume~36.
\newblock {SIAM}, 1981.

\bibitem[Choiruddin et~al.(2020)Choiruddin, Cuevas-Pacheco, Coeurjolly, and
  Waagepetersen]{p:cho-20}
A.~Choiruddin, F.~Cuevas-Pacheco, J.-F. Coeurjolly, and R.~Waagepetersen.
\newblock Regularized estimation for highly multivariate log {G}aussian {C}ox
  processes.
\newblock \emph{Stat. Comput.}, 30\penalty0 (3):\penalty0 649--662, 2020.

\bibitem[Chu et~al.(2022)Chu, Guan, Waagepetersen, and Xu]{p:chu-22}
T.~Chu, Y.~Guan, R.~Waagepetersen, and G.~Xu.
\newblock Quasi-likelihood for multivariate spatial point processes with
  semiparametric intensity functions.
\newblock \emph{Spat. Stat.}, 50:\penalty0 100605, 2022.

\bibitem[Cressie(1993)]{b:cre-93}
N.~Cressie.
\newblock \emph{Statistics for {S}patial {D}ata}.
\newblock John Wiley \& Sons, Hokoben, NJ., 1993.
\newblock Revised edition.

\bibitem[Cui et~al.(2018)Cui, Wang, and Haenggi]{cui2018vehicle}
Q.~Cui, N.~Wang, and M.~Haenggi.
\newblock Vehicle distributions in large and small cities: {S}patial models and
  applications.
\newblock \emph{IEEE Trans. Veh. Technol.}, 67\penalty0 (11):\penalty0
  10176--10189, 2018.

\bibitem[Daley and Vere-Jones(2003)]{b:dal-03}
D.~J. Daley and D.~Vere-Jones.
\newblock \emph{An introduction to the theory of point processes: volume {I}:
  elementary theory and methods}.
\newblock Springer, New York City, NY., 2003.
\newblock second edition.

\bibitem[Davies and Hazelton(2013)]{p:dav-haz-13}
T.~M. Davies and M.~L. Hazelton.
\newblock Assessing minimum contrast parameter estimation for spatial and
  spatiotemporal log-{G}aussian {C}ox processes.
\newblock \emph{Stat. Neerl.}, 67\penalty0 (4):\penalty0 355--389, 2013.

\bibitem[Deng et~al.(2014)Deng, Waagepetersen, and Guan]{p:deng-14}
C.~Deng, R.~P. Waagepetersen, and Y.~Guan.
\newblock A combined estimating function approach for fitting stationary point
  process models.
\newblock \emph{Biometrika}, 101\penalty0 (2):\penalty0 393--408, 2014.

\bibitem[Deng et~al.(2017)Deng, Guan, Waagepetersen, and Zhang]{p:deng-17}
C.~Deng, Y.~Guan, R.~P. Waagepetersen, and J.~Zhang.
\newblock Second-order quasi-likelihood for spatial point processes.
\newblock \emph{Biometrics}, 73\penalty0 (4):\penalty0 1311--1320, 2017.

\bibitem[Diggle(2003)]{diggle2003statistical}
P.~J. Diggle.
\newblock \emph{Statistical {A}nalysis of {S}patial {P}oint {P}atterns}.
\newblock Hodder education publishers, London, UK., 2003.
\newblock Second edition.

\bibitem[Doguwa and Upton(1989)]{p:dog-89}
S.~I. Doguwa and G.~J.~G. Upton.
\newblock Edge-corrected estimators for the reduced second moment measure of
  point processes.
\newblock \emph{Biom. J.}, 31\penalty0 (5):\penalty0 563--575, 1989.

\bibitem[Doukhan(1994)]{b:dou-94}
P.~Doukhan.
\newblock \emph{Mixing: properties and examples}.
\newblock Springer, New York City, NY., 1994.

\bibitem[Fang et~al.(1994)Fang, Loparo, and Feng]{fang1994inequalities}
Y.~Fang, K.~A. Loparo, and X.~Feng.
\newblock Inequalities for the trace of matrix product.
\newblock \emph{IEEE Trans. Automat. Control}, 39\penalty0 (12):\penalty0
  2489--2490, 1994.

\bibitem[Guan(2009)]{p:gua-09}
Y.~Guan.
\newblock A minimum contrast estimation procedure for estimating the
  second-order parameters of inhomogeneous spatial point processes.
\newblock \emph{Stat. Interface}, 2\penalty0 (1):\penalty0 91--99, 2009.

\bibitem[Guan and Sherman(2007)]{p:gua-07}
Y.~Guan and M.~Sherman.
\newblock On least squares fitting for stationary spatial point processes.
\newblock \emph{J. R. Stat. Soc. Ser. B. Stat. Methodol.}, 69\penalty0
  (1):\penalty0 31--49, 2007.

\bibitem[Guan et~al.(2004)Guan, Sherman, and Calvin]{guan2004nonparametric}
Y.~Guan, M.~Sherman, and J.~A. Calvin.
\newblock A nonparametric test for spatial isotropy using subsampling.
\newblock \emph{J. Amer. Statist. Assoc.}, 99\penalty0 (467):\penalty0
  810--821, 2004.

\bibitem[Hanisch and Stoyan(1979)]{hanisch1979formulas}
K.~H. Hanisch and D.~Stoyan.
\newblock Formulas for the second-order analysis of marked point processes.
\newblock \emph{Statistics}, 10\penalty0 (4):\penalty0 555--560, 1979.

\bibitem[Heinrich(1992)]{p:hei-92}
L.~Heinrich.
\newblock Minimum contrast estimates for parameters of spatial ergodic point
  processes.
\newblock In \emph{Transactions of the 11th Prague conference on random
  processes, information theory and statistical decision functions}, pages
  479--492. Academic Publishing House, 1992.

\bibitem[Heinrich and Pawlas(2008)]{p:hei-08}
L.~Heinrich and Z.~Pawlas.
\newblock Weak and strong convergence of empirical distribution functions from
  germ-grain processes.
\newblock \emph{Statistics}, 42\penalty0 (1):\penalty0 49--65, 2008.

\bibitem[Helmers and Zitikis(1999)]{p:hel-99}
R.~Helmers and R.~Zitikis.
\newblock On estimation of poisson intensity functions.
\newblock \emph{Ann. Inst. Statist. Math.}, 51\penalty0 (2):\penalty0 265--280,
  1999.

\bibitem[Hessellund et~al.(2022)Hessellund, Xu, Guan, and
  Waagepetersen]{p:hes-22}
K.~B. Hessellund, G.~Xu, Y.~Guan, and R.~Waagepetersen.
\newblock Second-order semi-parametric inference for multivariate log
  {G}aussian {C}ox processes.
\newblock \emph{J. R. Stat. Soc. Ser. C. Appl. Stat.}, 71\penalty0
  (1):\penalty0 244--268, 2022.

\bibitem[Jalilian et~al.(2015)Jalilian, Guan, Mateu, and
  Waagepetersen]{p:jal-15}
A.~Jalilian, Y.~Guan, J.~Mateu, and R.~Waagepetersen.
\newblock Multivariate product-shot-noise {C}ox point process models.
\newblock \emph{Biometrics}, 71\penalty0 (4):\penalty0 1022--1033, 2015.

\bibitem[Jolivet(1978)]{p:jol-78}
E.~Jolivet.
\newblock Central limit theorem and convergence of empirical processes for
  stationary point processes.
\newblock In \emph{Point processes and queuing problems: Colloquia Mathematica
  Societatis J\'{a}nos Bolyai}, volume~24, pages 117--161, 1978.

\bibitem[Jun and Cook(2024)]{p:jun-coo-22}
M.~Jun and S.~Cook.
\newblock Flexible multivariate spatiotemporal {H}awkes process models of
  terrorism.
\newblock \emph{Ann. Appl. Stat.}, 18\penalty0 (2):\penalty0 1378--1403, 2024.

\bibitem[Jun et~al.(2019)Jun, Schumacher, and Saravanan]{jun_et_al19}
M.~Jun, C.~Schumacher, and R.~Saravanan.
\newblock Global multivariate point pattern models for rain type occurrence.
\newblock \emph{Spat. Stat.}, 31, 2019.
\newblock Article 100355.

\bibitem[Kallenberg(2021)]{b:kal-21}
O.~Kallenberg.
\newblock \emph{Foundations of modern probability}.
\newblock Springer, New York City, NY, 2021.
\newblock third edition.

\bibitem[Lawson(2012)]{lawson2012bayesian}
A.~B. Lawson.
\newblock Bayesian point event modeling in spatial and environmental
  epidemiology.
\newblock \emph{Stat. Methods Med. Res.}, 21\penalty0 (5):\penalty0 509--529,
  2012.

\bibitem[M{\o}ller(2003)]{p:mol-03}
J.~M{\o}ller.
\newblock Shot noise {C}ox processes.
\newblock \emph{Adv. in Appl. Probab.}, 35\penalty0 (3):\penalty0 614--640,
  2003.

\bibitem[M{\o}ller and D{\'\i}az-Avalos(2010)]{moller2010structured}
J.~M{\o}ller and C.~D{\'\i}az-Avalos.
\newblock Structured {S}patio-{T}emporal {S}hot-{N}oise {C}ox {P}oint {P}rocess
  {M}odels, with a {V}iew to {M}odelling {F}orest {F}ires.
\newblock \emph{Scand. J. Stat.}, 37\penalty0 (1):\penalty0 2--25, 2010.

\bibitem[M{\o}ller and Waagepetersen(2007)]{p:mol-07}
J.~M{\o}ller and R.~P. Waagepetersen.
\newblock Modern spatial point process modelling and inference (with
  discussion).
\newblock \emph{Scand. J. Stat.}, 34:\penalty0 643--711, 2007.

\bibitem[M{\o}ller et~al.(1998)M{\o}ller, Syversveen, and
  Waagepetersen]{moller1998log}
J.~M{\o}ller, A.~R. Syversveen, and R.~P. Waagepetersen.
\newblock Log {G}aussian {C}ox processes.
\newblock \emph{Scand. J. Stat.}, 25\penalty0 (3):\penalty0 451--482, 1998.

\bibitem[Nguyen and Zessin(1979)]{p:ngu-79}
X.~X. Nguyen and H.~Zessin.
\newblock Ergodic theorems for spatial processes.
\newblock \emph{Z. Wahrscheinlichkeitstheorie verw. Gebiete (Probab. Theory
  Related Fields)}, 48\penalty0 (2):\penalty0 133--158, 1979.

\bibitem[Ogata(1978)]{p:oga-78}
Y.~Ogata.
\newblock The asymptotic behaviour of maximum likelihood estimators for
  stationary point processes.
\newblock \emph{Ann. Inst. Statist. Math.}, 30\penalty0 (1):\penalty0 243--261,
  1978.

\bibitem[Pawlas(2009)]{p:paw-09}
Z.~Pawlas.
\newblock Empirical distributions in marked point processes.
\newblock \emph{Stochastic Process. Appl.}, 119\penalty0 (12):\penalty0
  4194--4209, 2009.

\bibitem[Rajala et~al.(2018)Rajala, Murrell, and Olhede]{p:raj-18}
T.~Rajala, D.~J. Murrell, and S.~C. Olhede.
\newblock Detecting multivariate interactions in spatial point patterns with
  {G}ibbs models and variable selection.
\newblock \emph{J. R. Stat. Soc. Ser. C. Appl. Stat.}, 67\penalty0
  (5):\penalty0 1237--1273, 2018.

\bibitem[Ripley(1976)]{p:rip-76}
B.~D. Ripley.
\newblock The second-order analysis of stationary point processes.
\newblock \emph{J. Appl. Probab.}, 13\penalty0 (2):\penalty0 255--266, 1976.

\bibitem[Rosenblatt(1956)]{p:ros-56}
M.~Rosenblatt.
\newblock A central limit theorem and a strong mixing condition.
\newblock \emph{Proc. Natl. Acad. Sci. USA}, 42\penalty0 (1):\penalty0 43--47,
  1956.

\bibitem[Rue et~al.(2009)Rue, Martino, and Chopin]{rue2009approximate}
H.~Rue, S.~Martino, and N.~Chopin.
\newblock Approximate {B}ayesian inference for latent {G}aussian models by
  using integrated nested {L}aplace approximations.
\newblock \emph{J. R. Stat. Soc. Ser. B. Stat. Methodol.}, 71\penalty0
  (2):\penalty0 319--392, 2009.

\bibitem[START(2022)]{start}
START.
\newblock Global terrorism database 1970 - 2020 [data file], 2022.
\newblock National Consortium for the Study of Terrorism and Responses to
  Terrorism, \url{https://www.start.umd.edu/gtd}.

\bibitem[Taylor et~al.(2015)Taylor, Davies, Rowlingson, and
  Diggle]{taylor2015bayesian}
B.~M. Taylor, T.~Davies, B.~S. Rowlingson, and P.~J. Diggle.
\newblock Bayesian {I}nference and {D}ata {A}ugmentation {S}chemes for
  {S}patial, {S}patiotemporal and {M}ultivariate {L}og-{G}aussian {C}ox
  {P}rocesses in {R}.
\newblock \emph{J. Stat. Softw.}, 63\penalty0 (7):\penalty0 1--48, 2015.

\bibitem[Waagepetersen and Guan(2009)]{p:waa-09}
R.~Waagepetersen and Y.~Guan.
\newblock Two-step estimation for inhomogeneous spatial point processes.
\newblock \emph{J. R. Stat. Soc. Ser. B. Stat. Methodol.}, 71\penalty0
  (3):\penalty0 685--702, 2009.

\bibitem[Waagepetersen et~al.(2016)Waagepetersen, Guan, Jalilian, and
  Mateu]{p:was-16}
R.~Waagepetersen, Y.~Guan, A.~Jalilian, and J.~Mateu.
\newblock Analysis of multispecies point patterns by using multivariate
  log-{G}aussian {C}ox processes.
\newblock \emph{J. R. Stat. Soc. Ser. C. Appl. Stat.}, 65\penalty0
  (1):\penalty0 77--96, 2016.

\bibitem[Yang and Guan(2024)]{p:yang-24}
J.~Yang and Y.~Guan.
\newblock Fourier analysis of spatial point processes.
\newblock \emph{arXiv preprint arXiv:2401.06403}, 2024.

\end{thebibliography}

\end{document}